\documentclass[]{article}
\usepackage[utf8]{inputenc}
\usepackage{amsmath}
\usepackage{amsthm}
\usepackage{amsfonts}
\usepackage{braket}
\usepackage{tikz}
\usepackage{graphicx}
\usepackage[title]{appendix}
\usepackage[export]{adjustbox}[2011/08/13]
\usepackage{fullpage}
\usepackage{cite}
\usepackage{authblk}
\usepackage{xcolor}

\usepackage{caption}
\usepackage{subcaption}
\usepackage{multibib}
\newcites{appref}{References}

\usepackage{algorithm}
\usepackage[noend]{algpseudocode}
\usepackage[colorlinks=true]{hyperref}

\usetikzlibrary{quantikz}

\DeclareMathOperator{\Tr}{Tr}

\algdef{SE}[SUBALG]{Indent}{EndIndent}{}{\algorithmicend\ }%
\algtext*{Indent}
\algtext*{EndIndent}

\makeatletter
\algnewcommand{\LineComment}[1]{\Statex \hskip\ALG@thistlm \(\triangleright\) #1}
\newtheorem{prop}{Proposition}
\newtheorem*{prop*}{Proposition}

\theoremstyle{remark}

\makeatother

\title{Learning quantum data with the quantum Earth Mover's distance}
\author[1,2]{Bobak Toussi Kiani}
\author[2,3,4,5]{Giacomo De Palma}
\author[6]{Milad Marvian}
\author[7]{Zi-Wen Liu}
\author[2,3]{Seth Lloyd}
\affil[1]{Department of Electrical Engineering and Computer Science, Massachusetts Institute of Technology, Cambridge, MA, USA}
\affil[2]{Research Laboratory of Electronics, Massachusetts Institute of Technology, Cambridge, MA, USA}
\affil[3]{Department of Mechanical Engineering, Massachusetts Institute of Technology, Cambridge, MA, USA}
\affil[4]{Scuola Normale Superiore, Pisa, Italy}
\affil[5]{Department of Mathematics, University of Bologna, Bologna, Italy}
\affil[6]{Department of Electrical and Computer Engineering, University of New Mexico, Albuquerque, NM, USA}
\affil[7]{Perimeter Institute for Theoretical Physics, Waterloo, Canada}
\date{}

\begin{document}

\maketitle
\begin{abstract}
    Quantifying how far the output of a learning algorithm is from its target is an essential task in machine learning. However, in quantum settings, the loss landscapes of commonly used distance metrics often produce undesirable outcomes such as poor local minima and exponentially decaying gradients. To overcome these obstacles, we consider here the recently proposed quantum earth mover's (EM) or Wasserstein-1 distance as a quantum analog to the classical EM distance. We show that the quantum EM distance possesses unique properties, not found in other commonly used quantum distance metrics, that make quantum learning more stable and efficient. We propose a quantum Wasserstein generative adversarial network (qWGAN) which takes advantage of the quantum EM distance and provides an efficient means of performing learning on quantum data. We provide examples where our qWGAN is capable of learning a diverse set of quantum data with only resources polynomial in the number of qubits.
\end{abstract}

\section{Introduction}

A fundamental task in quantum machine learning is designing efficient algorithms for learning quantum states \cite{benedetti2019adversarial,dallaire2018quantum,torlai2020machine,gao2018experimental,aaronson2007learnability,rocchetto2019experimental,lloyd2018quantum,carrasquilla2019reconstructing,chakrabarti2019quantum,beer2020training}, transformations \cite{kiani2020learning,mitarai2018quantum,bisio2010optimal,quintino2019reversing,lloyd2020quantum,carolan2020variational,beer2020training,sharma2020noise}, and classical data stored as or generated by quantum states \cite{benedetti2019generative,liu2018differentiable,coyle2020born}. In the general setup, one is given a target quantum object, say a target quantum state, and aims to generate or approximate that target object by efficiently learning parameters in a quantum circuit. For example, quantum generative adversarial networks (qGAN) are parameterized sets of quantum circuits and quantum operators designed to learn target states or transformations via optimization over the parameters of a quantum generator and discriminator \cite{lloyd2018quantum,chakrabarti2019quantum}. 

A crucial component of a quantum machine learning algorithm is an objective function (often a distance metric) which determines how close a generated object is to its target. This choice of metric is important not only as a measure of performance but also as a means for optimization. For example, certain metrics lead to efficient algorithms for calculating gradients with respect to that metric, allowing algorithms to perform optimization via gradient based optimizers (\textit{e.g.,} gradient descent). Naturally, for learning pure states, a common choice for the distance metric is a function of the inner product between quantum states. Similarly, for learning density matrices, researchers commonly choose distance metrics which simplify to a function of the inner product when measuring the distance between pure states. 

Previous approaches to learning quantum data have typically suffered from the presence of vanishing gradients \cite{mcclean2018barren,wang2020noise,cerezo2020cost} and poor local minima \cite{pechen2011there,moore2012exploring,cerezo2020variational} in the loss landscape induced by the choice of distance metric. Intuitively, these ``barren plateaus'' and traps arise due to the fact that random quantum states have inner product that diminishes exponentially with the number of qubits. Our approach helps surmount these challenges by formulating an algorithm which provides an efficient means for learning pure and mixed states using the recently proposed quantum earth mover's (EM) distance, also known as the quantum Wasserstein distance of order 1 \cite{de2020quantum}. As we will demonstrate, the quantum EM distance is a natural distance metric for optimization over local operations and avoids common pitfalls faced by other distance metrics which reduce to functions of the inner product. This is in agreement with results in classical machine learning, where algorithms employing the earth mover's distance are often more stable and avoid issues with vanishing or exploding gradients \cite{arjovsky2017wasserstein,chen2018adversarial,rubner1998metric,villani2008optimal,gulrajani2017improved} (see \autoref{app:rw} for a more complete discussion of the literature and \autoref{app:clEM} for a presentation of the classical EM distance). Intuitively, the quantum EM distance can be interpreted as a continuous version of a quantum ``hamming distance", which allows local gates to optimize over the few qubits on which they act instead of over some global distance metric which often decays exponentially in the number of qubits.

In this work, we study the quantum EM distance from an applied setting and make the following contributions. First, we overview the construction of the quantum EM distance and analyze the properties of different quantum distance metrics and the loss landscapes they produce in quantum machine learning settings. Here, we show that the quantum EM distance has unique advantages over other common distance metrics. Then, to operationalize the quantum EM distance, we devise a new heuristic method to approximate the quantum EM distance efficiently given copies of quantum states. In learning settings, this leads to our development of a quantum Wasserstein generative adversarial network (qWGAN) which is a quantum analog to the classical Wasserstein generative adversarial network \cite{arjovsky2017wasserstein} (see also \cite{chakrabarti2019quantum}). Importantly, like its classical analog, our qWGAN employs an earth mover's distance in its cost function. Numerical results show that our qWGAN is efficient at learning quantum data with shallow circuits in various settings. Finally, we discuss near term applications of our qWGAN for both classical and quantum problems.

\section{Quantum distance metrics and quantum EM distance}

To approximate or reconstruct a target probability distribution with a machine learning algorithm, the choice of distance metric, measuring how well the approximating distribution matches the target distribution, is crucial to the performance of the algorithm. Classically, generative adversarial networks (GAN) provide a neural network approach for learning a target probability distribution and generating new samples from the approximate distribution \cite{goodfellow2014generative,arjovsky2017wasserstein}. The choice of loss metric for a GAN is a distance or divergence metric which is minimized when the target and generated distributions coincide. 

In the quantum setting, distance metrics between states or density matrices are employed in the implementation of quantum generative adversarial networks (qGAN) \cite{hu2019quantum,chakrabarti2019quantum,lloyd2018quantum,cerezo2020variational}. As in the classical setting, the choice of distance metric is crucial to the runtime and performance of the quantum machine learning algorithm. Here, we consider common distance metrics and show that the quantum earth mover's (EM) distance recently defined in \cite{de2020quantum} possesses desirable properties that are not found in the other metrics.

For a brief overview of the notation used in quantum mechanics, we refer the reader to \autoref{app:QM}. 
Let $\rho, \sigma \in \mathbb{C}^{N \times N}$ be the density matrices corresponding to two quantum states, \textit{e.g.,} $\rho$ can be the quantum state generated by a GAN and $\sigma$ is the target state.
Until now, common distance metrics employed to train quantum GANs have been unitarily invariant, \emph{i.e.}, invariant with respect to the conjugation of both quantum states with the same unitary matrix and reducing to a function of the inner product for pure states (\textit{i.e.,} orthogonal projectors with rank one). Commonly used distance metrics in prior works include:

\begin{itemize}
    \item \textbf{Trace Distance: } The simplest and most common choice (\textit{e.g.,} see \cite{hu2019quantum,benedetti2019adversarial}) is the trace distance:
    \begin{equation}
        D_1 (\rho,\sigma) = \frac{1}{2}\left\| \rho - \sigma \right\|_1\,,
    \end{equation}
    where $\|\cdot\|_1$ denotes the trace norm, \emph{i.e.}, the sum of the singular values.
    \item \textbf{Quantum Fidelity:} Another common choice (\textit{e.g.,} see \cite{beer2020training}) is the maximum absolute value squared of the inner product between purifications of $\rho$ and $\sigma$:
    \begin{equation}
        F(\rho,\sigma) = \left\|\sqrt{ \rho\vphantom{ \sigma}} \, \sqrt{\sigma\vphantom{ \rho} }\right\|_1^2\,.
    \end{equation}
    $F(\rho,\sigma)$ is often modified to $\arccos\sqrt{F(\rho,\sigma)}$ to construct a proper distance metric.
    
    \item \textbf{Quantum Wasserstein Semimetric:} Introduced in \cite{chakrabarti2019quantum} as a quantum generalization of the Wasserstein distance, this distance, denoted $qW(\rho, \sigma)$, is calculated by forming a coupling between quantum states $\rho$ and $\sigma$ in $\mathbb{C}^{N \times N}$. The coupling is a quantum state in $\left(\mathbb{C}^{N \times N}\right)^{\otimes2}$ whose marginal states are equal to $\rho$ and $\sigma$, respectively. The quantum Wasserstein semimetric is the minimum of the expectation value of the projector onto the symmetric subspace of $\left(\mathbb{C}^{N}\right)^{\otimes2}$. %(\emph{i.e.}, the invariant subspace with respect to the permutation of the two copies of $\mathbb{C}^N$) with respect to all the couplings between $\rho$ and $\sigma$.
$qW$ does not satisfy the triangle inequality, hence the name semimetric. Importantly, $qW$ is unitarily invariant, and for pure states, it reduces to a function of their inner product: for any $\ket{u},\,\ket{v}$ unit vectors in $\mathbb{C}^N$, $qW(\ket{u} \! \bra{u}, \ket{v} \! \bra{v}) = (1-\lvert \braket{v|u} \rvert ^2)/2 $.
Further details can be found in \autoref{sec:QW}.
 \end{itemize}

\paragraph{The quantum EM distance} 
 In this paper we consider the case of $n$ qubits, where $N=2^n$, and employ the quantum generalization of the Wasserstein distance of order 1 to the states of $n$ qubits recently proposed in \cite{de2020quantum} and also known as the earth mover's (EM) distance. 
    We adopt the latter terminology as it is more prevalent in the machine learning community, hereby denoting the quantum EM distance with $D_{EM}$.
    Unlike all the previously employed distances, the quantum EM distance is not unitarily invariant.
    We will show that, similar to its classical counterpart \cite{arjovsky2017wasserstein}, $D_{EM}$ possesses several properties that are desirable when learning quantum data.
    
    The quantum EM distance of \cite{de2020quantum} is based on the notion of neighboring states.
    Two quantum states of $n$ qubits are neighboring if they differ in only one qubit, \emph{i.e.}, if they coincide after one qubit is discarded.
    The quantum EM distance is the distance that is induced by the maximum norm that assigns distance at most one to any couple of neighboring states.
    We denote with $\|\cdot\|_{EM}$ the corresponding norm, whose analytical expression can be found below.
    This definition enforces the continuity of the distance with respect to local operations, \emph{i.e.}, any quantum operation acting on a single qubit can displace a state by at most one unit with respect to the quantum EM distance.
    Indeed, for the quantum states of the computational basis the quantum EM distance recovers the classical Hamming distance (equal to the number of elements that differ between two strings), \emph{i.e.}, for any two strings of $n$ bits $x$ and $y$ we have $D_{EM}(\ket{x} \! \bra{x}, \ket{y} \! \bra{y}) = h(x,y)$.
    More generally, for quantum states diagonal in the computational basis, the quantum EM distance recovers the classical EM distance.
    The quantum EM distance admits a dual formulation \cite{de2020quantum}, based on the quantum generalization of the Lipschitz constant, which is more suitable for implementation of quantum GANs.
   
    We denote with $\mathcal{O}_n$ the set of $n$-qubit observables, \emph{i.e.}, the set of the $2^n\times 2^n$ Hermitian matrices.
    The quantum Lipschitz constant of the observable $H\in\mathcal{O}_n$ is
    \begin{equation}
        \|H\|_L = 2\max_{i=1,\,\ldots,\,n}\min\left\{\|H - \tilde{H}_{\bar{i}}\|_\infty:\tilde{H}_{\bar{i}}\in\mathcal{O}_n\textnormal{ acts as identity on the $i$-th qubit}\right\}\,.
    \end{equation}
    The quantum Lipschitz constant defined above is a generalization of the Lipschitz constant for the functions on strings of $n$ bits, and coincides with the classical Lipschitz constant for the observables that are diagonal in the computational basis \cite{de2020quantum}.
    The quantum EM distance between the quantum states $\rho$ and $\sigma$ is equal to the maximum difference between the expectation values on $\rho$ and $\sigma$ of a quantum observable with Lipschitz constant at most one:
    \begin{equation}
    D_{EM}(\rho,\sigma) = \max \left\{ \Tr\left[\left(\rho-\sigma\right)H\right] : H \in \mathcal{O}_n,\; \|H\|_L \leq 1 \right\}\,.
    \label{eq:qWass1}
    \end{equation}
When the quantum EM distance plays the role of a cost function in a machine learning algorithm, it can be considered as an energy associated to the parameter configuration. For this reason, we may refer to the observables $H$ in \eqref{eq:qWass1} as Hamiltonians.

Equivalently, the quantum EM distance can be also defined by its primal formulation \cite[Definition 6]{de2020quantum}:
\begin{equation}
    D_{EM}(\rho, \sigma) = \frac{1}{2}\min\left\{\sum_{i=1}^n\|X_i\|_1:X_i\in\mathcal{O}_n,\;\mathrm{Tr}_iX_i=0\;\forall\,i=1,\,\ldots,\,n,\;\sum_{i=1}^nX_i = \rho - \sigma\right\}.
    \label{eq:primal}
\end{equation}

To show why $D_{EM}$ possesses desirable properties, we first consider the case where both the target $\sigma$ and the generated $\rho$ are pure states in a simple toy model. Here, as we will show, undesirable critical points are clearly present and endemic to the loss landscapes for metrics which are a function of the inner product between two pure states. In contrast, the quantum EM distance $D_{EM}$ avoids these undesirable critical points. Finally, we generalize the findings of this toy model to a larger class of quantum machine learning settings.

\subsection{A simple toy model}
\label{sec:toy_model}
 In this section, we consider an intuitive example which shows the advantages of using the quantum EM distance  when the learning is performed over local quantum gates. Namely, we show that the commonly used distance metrics which are a function of the inner product between states feature two key issues in learning via local gates. First, the inner product between a generated and target state fails to show improvement when gates are optimized one by one or layer-wise \cite{campos2021abrupt}. Second, when parameters are initialized randomly, gradients in this example decay exponentially when the distance metric is a function of the inner product. The quantum EM distance avoids both of these drawbacks and allows for efficient learning in this scenario. 

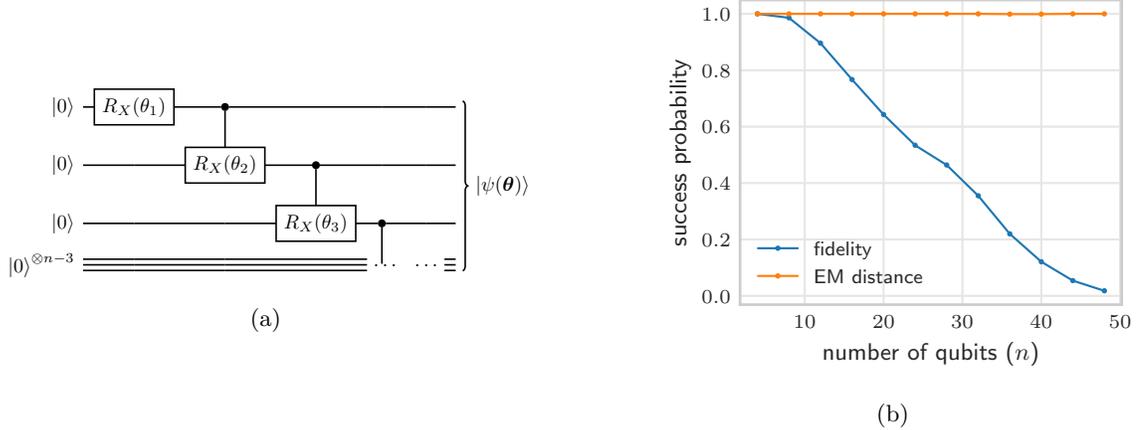
\begin{figure}
    \centering
    \begin{subfigure}[]{0.6\textwidth}
        \centering
        \begin{tikzpicture}
        \node[scale=0.75] {
        \begin{quantikz}[row sep=0.4cm, column sep=0.2cm]
            \lstick{$\ket{0}$} & \gate{R_X(\theta_1)} & \ctrl{1} & \qw & \qw & \qw & \qw \rstick[wires = 4]{$\ket{\psi(\boldsymbol{\theta})}$}  \\     
            \lstick{$\ket{0}$} & \qw & \gate{R_X(\theta_2)} & \ctrl{1} & \qw & \qw & \qw  \\
            \lstick{$\ket{0}$} & \qw & \qw & \gate{R_X(\theta_3)} & \ctrl{1} & \qw & \qw  \\
            \lstick{$\ket{0}^{\otimes n-3}$} & \qwbundle[alternate]{} & \qwbundle[alternate]{} & \qwbundle[alternate]{} & \ \ldots \qwbundle[alternate]{} & \ \ldots \ & \qwbundle[alternate]{} 
        \end{quantikz}
        };
        \end{tikzpicture}
        \caption{}
        \label{fig:simple_circuit}
    \end{subfigure}
    \hfill
    \begin{subfigure}[]{0.39\textwidth}
         \centering
         \includegraphics[]{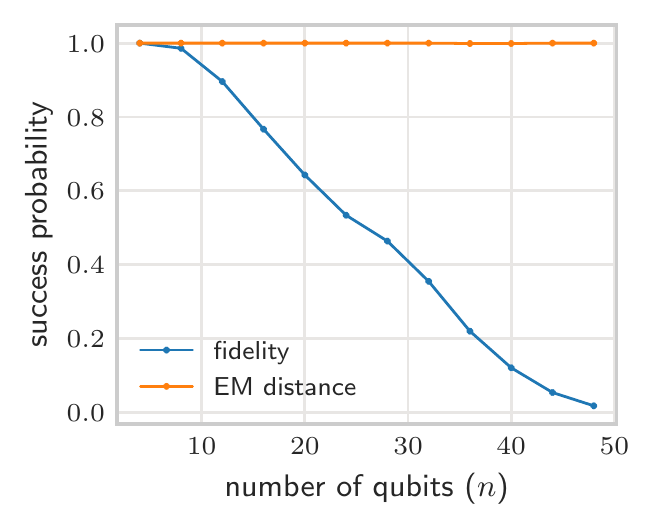}
         \caption{}
         \label{fig:GHZ_success_rate}
     \end{subfigure}
     \caption{(a) Simple quantum circuit which can generate the GHZ state when parameter values are appropriately chosen. (b) In learning settings, this circuit presents challenges for loss metrics that are a function of the inner product between target and initial states. Simulations show that, with a loss function such as fidelity (blue line), gradient based optimizers eventually fail to find global optimum in virtually all instances when circuit contains many qubits, instead converging to the state $\ket{0_n}$. In contrast, the use of the quantum EM distance (orange line) results in convergence to the global optimum in virtually all instances tested. Here, optimization is performed for up to $10^5$ steps using the Adam optimizer (simulations with EM distance generally achieved convergence within about $1000$ steps). Experiments are repeated 1000 times for each $n$ to estimate success probability. See \autoref{app:toy_model_details} for full details of experiments. }
\end{figure}

In this toy model, the task at hand is to learn the correct values of parameters in the circuit in \autoref{fig:simple_circuit} to generate the GHZ state of $n$ qubits $\ket{GHZ_n} = \left(\ket{0_n} + \ket{1_n}\right)/\sqrt{2}$. This circuit consists of a parameterized Pauli X rotation on the first qubit and controlled parameterized Pauli X rotations on later qubits (see \autoref{app:QM} for description of Pauli operators). When $n$ is a multiple of $4$, setting  $\theta_1$ equal to $\pi/2$ and all other parameters equal to $\pi$ will construct the target GHZ state.

Consider the case where one aims to maximize the fidelity $F$ between the generated state $\ket{\psi(\boldsymbol{\theta})}$ and the GHZ state $\ket{GHZ_n}$. Given our circuit, $F$ takes a simple form:
\begin{equation}
\begin{split}
    F & =  |\braket{GHZ_n| \psi(\boldsymbol{\theta})} |^2 = \left( \frac{\cos(\theta_1)}{\sqrt{2}} + \frac{\prod_{i=1}^n \sin(\theta_i)}{\sqrt{2}} \right)^2 .
\end{split}
\label{eq:toy_loss_function}
\end{equation}

The first problem associated with learning via $F$ is that the loss landscape has undesirable local minima associated with the state $\ket{0_n}$. Note that fixing any $\theta_i=0$ will force a learning algorithm (\textit{e.g.,} gradient descent) to optimize the $\cos (\theta_1 )$ term, converging to the state $\ket{0_n}$. In other words, if any algorithm aims to optimize the gates in a layer-wise fashion (\textit{e.g.,} optimizing  $\theta_1, \theta_2, \dots$ in order as in \cite{skolik2020layerwise}), that algorithm will get stuck in the local optimum at $\ket{0_n}$.

Of course, in practice, parameters $\boldsymbol{\theta}$ are typically initialized randomly so it is unlikely that any $\theta_i=0$; however, even here, we have the second issue that gradients with respect to $\theta_2, \theta_3, \dots, \theta_n$ all decay exponentially with $n$ since the product of sine functions with random parameter values decays exponentially to zero. 

\begin{equation}
        \frac{\partial F}{\partial \theta_i} =
        \left\{ \begin{array}{ll}
           -\cos \theta_1 \sin \theta_1 + \cos 2 \theta_1 \prod_{k=2}^n \sin \theta_k  &\qquad i = 1 \\
           + \sin \theta_1 \cos \theta_1 \bigl( \prod_{k=2}^n \sin \theta_k \bigr)^2 & \\ & \\
            \cos \theta_i \prod_{k\neq i} \sin(\theta_k) \bigl[ \cos(\theta_1) + \prod_{k=1}^n \sin(\theta_k) \bigr] & \qquad i>1
        \end{array} \right.\,.
        \label{eq:toy_gradients}
\end{equation}

Notably, $\frac{\partial F}{\partial \theta_i} = O(\frac{1}{2^n})$ for $i>1$, but $\frac{\partial F}{\partial \theta_i} = O(1)$ for $i=1$. For large $n$, $\frac{\partial F}{\partial\theta_1} \approx -\cos \theta_1 \sin \theta_1$ indicating that gradient optimizers will converge to a poor local minimum outputting the state $\ket{0_n}$ ($\theta_1 = 0 \mod 2 \pi$). In fact, since the loss function $F$ (equation \eqref{eq:toy_loss_function}) takes a simple form, gradient descent on the parameters can be efficiently performed classically, and results shown in \autoref{fig:GHZ_success_rate} show that gradient descent converges to the undesirable local optimum associated with the $\ket{0_n}$ state even as more qubits are added (simulations stopped after 100000 steps of optimization). 

The challenges described above are encapsulated by the feature of the inner product that, for example, the states $\ket{0000}$, $\ket{1000}$, and even $\ket{1110}$ are equally distant (orthogonal) to the state $\ket{1111}$ -- \textit{i.e.,} local updates induce no change in the inner product distance metric. Using a loss function with the quantum EM distance $D_{EM}$ naturally avoids these challenges. Since the quantum EM distance recovers the Hamming distance between two computational basis states, local operations on one and two qubits can reduce $D_{EM}$ as long as those operations reduce the Hamming distance between the target and generated states. For example, unlike the inner product distance metric, $\ket{1000}$ and $\ket{1110}$ are three and one units away respectively from the state $\ket{1111}$ in the quantum EM distance. In our toy model, local gates, even when applied in isolation, result in changes to single qubits which either reduce or increase the quantum EM distance.  

If we set $\theta_1=\pi/2$, $\theta_2 = \ldots = \theta_k = \pi$  and $\theta_{k+1} = \ldots = \theta_n = 0$ in the quantum circuit of \autoref{fig:simple_circuit}, we obtain the quantum state $\ket{\Psi_k} = \left(\ket{0_k}+(-i)^k\ket{1_k}\right)\ket{0_{n-k}}/\sqrt{2}$.
The following \autoref{prop:GHZ} intuitively explains this success and shows that the sequence of states 
$\ket{\Psi_0} = \ket{0_n},\,\ket{\Psi_1},\,\ldots,\,\ket{\Psi_{n}} = \ket{GHZ_n}$ gets closer and closer to the target state $\ket{GHZ_n}$, with a guaranteed improvement every two steps.
The proof is in \autoref{app:GHZ}.
\begin{prop}\label{prop:GHZ}
For any $k=0,\,\ldots,\,n$, let $D_k = D_{EM}\left(|\Psi_k\rangle\langle \Psi_k|,\,|GHZ_n\rangle\langle GHZ_n|\right)$.
We have $n/2 \le D_0 \le (n+1)/2$, $D_n = 0$, and $(n-k)/2 \le D_k \le \left(n-k+\sqrt{2}\right)/2$ for any $k=1,\,\ldots,\,n-1$.
In particular, we have $D_{k+2}<D_k$ for any $k=0,\,\ldots,\,n-2$.
\end{prop}

Furthermore, as shown in \autoref{fig:GHZ_success_rate}, optimization using the quantum EM distance is, in virtually all cases, successful at learning the GHZ state. In the simulations for \autoref{fig:GHZ_success_rate}, the quantum EM distance is efficiently estimated (lower bounded) using the dual formulation by considering the expectation of the generated state over a subset of $O(n)$ Hermitian operators $H_i$ all with Lipschitz constant equal to one. 

\begin{equation}
    \begin{split}
        \tilde{D}_{EM} & = \max_{H_i} \bigl|  \bra{ \psi(\boldsymbol{\theta})} H_i \ket{ \psi(\boldsymbol{\theta})}  - \bra{GHZ_n} H_i \ket{GHZ_n}  \bigr| \\
        & \leq D_{EM} \bigl(\ket{\psi(\boldsymbol{\theta})} \bra{\psi(\boldsymbol{\theta})}, \ket{GHZ_n} \bra{GHZ_n} \bigr), 
    \end{split}
\end{equation}
where $\tilde{D}_{EM}$ is the approximation which lower bounds $D_{EM}$ by taking the maximum over the $O(n)$ operators $H_i$ chosen for optimization of the circuit (see \autoref{app:toy_model_details} for list of operators). Using $\tilde{D}_{EM}$, we can successfully learn and generate the GHZ state regardless of the size of the system. Given the simplified form of our circuit, calculating $\tilde{D}_{EM}$ can be efficiently performed using a classical computer and the methodology is detailed in \autoref{app:toy_model_details}. Interestingly, though the subset of Hamiltonians considered in calculating $\tilde{D}_{EM}$ is significantly less than the total space of Hamiltonians available needed to exactly calculate $D_{EM}$, the simplified form of $\tilde{D}_{EM}$ still suffices to completely learn the GHZ state. This perhaps surprising fact is one motivation for our qWGAN, discussed later, which uses similar techniques to construct a general algorithm for learning quantum data in more complex settings.

\subsection{Properties of quantum EM distance in learning settings}
For the EM distance over probability distributions, the classic work of \cite{arjovsky2017wasserstein} showed that the EM distance has a number of properties that confer advantages in learning settings over other distance metrics such as the total variational distance. Here, we show that our quantum EM distance offers corresponding analogous properties when learning in quantum settings. These properties provide intuitive explanations for why learning was so successful in the toy model analyzed earlier. First, the quantum EM distance is super-additive with respect to the tensor product \cite[Proposition 4]{de2020quantum}:
\begin{prop}
For any two quantum states $\rho$, $\sigma$ of $n$ qubits and any $k=1,\,\ldots,\,n-1$,
\begin{equation}
    D_{EM} (\rho,\sigma) \ge D_{EM}(\rho_{1\ldots k}, \sigma_{1\ldots k}) + D_{EM}(\rho_{k+1\ldots n}, \sigma_{k+1\ldots n})\,,
\end{equation}
where $\rho_{1\ldots k}$ and $\rho_{k+1\ldots n}$ are the marginal states of $\rho$ over the first $k$ and the last $n-k$ qudits, respectively, and analogously for $\sigma$. 

In the case of product states where $\rho = \rho_{1\ldots k} \otimes \rho_{k+1\ldots n}$ and $\sigma = \sigma_{1\ldots k} \otimes \sigma_{k+1\ldots n}$, then the above is an equality:
\begin{equation}
    D_{EM} (\rho,\sigma) = D_{EM}(\rho_{1\ldots k}, \sigma_{1\ldots k}) + D_{EM}(\rho_{k+1\ldots n}, \sigma_{k+1\ldots n})\,,
\end{equation}
\end{prop}
Intuitively, the Proposition above implies that operations which reduce the distance between two states over a portion of their qubits will proportionally reduce the total distance over all of the qubits. Note that no unitarily invariant distance can have this property. For example, to learn a target state $\ket{GHZ_2}\ket{1}$, updating the state $\ket{000}$ to $\ket{GHZ_2}\ket{0}$ results in a significant improvement in the quantum EM distance but, since the updated state is still orthogonal to the target state, no unitarily invariant distance will show any improvement. As an aside, super-additivity is relevant in noisy contexts as it implies that if noise only effects a small number of qubits, then the change in the EM distance is correspondingly bounded by the number of qubits on which the noise acts. 

A second useful property of the quantum EM distance is that it recovers the classical earth mover's distance for quantum states diagonal in the canonical basis, and in particular, it recovers the classical Hamming distance for the quantum states of the computational basis \cite[Proposition 6]{de2020quantum}:
\begin{prop}
Let $p,\,q$ be probability distributions on $\{0,1\}^n$, and let
\begin{equation}\label{eq:defpq}
\rho = \sum_{x\in\{0,1\}^n}p(x)\,|x\rangle\langle x|\,,\qquad \sigma = \sum_{y\in\{0,1\}^n}q(y)\,|y\rangle\langle y|\,.
\end{equation}
Then, $D_{EM}(\rho,\sigma) = D_{EM}(p,q)$.
In particular, the quantum EM distance between vectors of the canonical basis coincides with the Hamming distance: $D_{EM}(|x\rangle\langle x|,|y\rangle\langle y|) = h(x,y)$ for any $x,\,y\in\{0,1\}^n$.
\end{prop}

The above proposition implies that advantages conferred in classical machine learning algorithms when using the classical EM distance directly translate into quantum settings when using the quantum EM distance. Finally, the quantum EM distance is always contained between the trace distance and $n$ times the trace distance \cite[Proposition 2]{de2020quantum}:
\begin{prop}
For any two quantum states $\rho,\,\sigma$,
\begin{equation}\label{eq:1T}
D_1(\rho,\sigma) \le D_{EM}(\rho,\sigma) \le n\,D_1(\rho,\sigma)\,.
\end{equation}
\end{prop}
In particular, a small quantum EM distance guarantees that the trace distance is also small and vice-versa. Thus, convergence in the quantum EM distance necessarily implies convergence in more conventional quantum distance metrics such as fidelity or trace distance.

\subsection{EM Distance Evaluation}
\label{sec:EM_evaluation}

As in the classical Wasserstein GAN \cite{arjovsky2017wasserstein}, approximations to the EM distance are required to construct learning algorithms that have efficient runtimes. Note, the quantum EM distance can be exactly evaluated using algorithms for semidefinite programs \cite{brandao2017quantum,van2020quantum} which run in time polynomial in the dimension of the quantum state and the number of constraints. Such an exact approach would require algorithmic runtimes that are exponential in the number of qubits and furthermore, do not lead to obvious methods for calculating the gradient of the quantum EM distance. Instead, we provide a procedure below to estimate the quantum EM distance between two distributions of quantum states using its dual formulation \eqref{eq:qWass1}.
To avoid cumbersome computation of Lipschitz constants, we construct a parameterized family of functions which preserve a quantum Lipschitz constraint upon optimization.
Let
\begin{equation}
    H = \sum_{\mathcal{I} \subseteq \{1,\,\ldots,\,n\}} H_{\mathcal{I}}\,,
\end{equation}
where each $H_{\mathcal{I}}$ acts non-trivially only on the qubits in the corresponding set $\mathcal{I}$. Then, Proposition 10 of \cite{de2020quantum} provides an upper bound to the quantum Lipschitz constant of a Hamiltonian in terms of its local structure.
\begin{equation}
    \| H \|_L \leq 2 \max_{i=1,\,\ldots,\,n} \left\|  \sum_{i\in \mathcal{I}\subseteq\{1,\,\ldots,\,n\}} H_{\mathcal{I}}   \right\|_{\infty} ,
    \label{eq:quant_lipsh_bound}
\end{equation}
where the maximum is taken over the qubits. The notation $i\in \mathcal{I}\subseteq\{1,\,\ldots,\,n\}$ indicates that the sum is taken only over the set of operators which act non-trivially (\textit{i.e.,} not the identity) on qubit $i$. Intuitively, since the Lipschitz constant bounds the change in a Hamiltonian induced by changes to a single qubit, the bound above can be viewed as a bound on the maximum singular value of nontrivial operators thus also bounding the corresponding Lipschitz constant. A natural choice for the operators $H_{\mathcal{I}}$ are a subset of the Pauli operators which we explore in our construction of a quantum generative adversarial network next.

\section{qWGAN Algorithm}

Our quantum Wasserstein generative adversarial net (qWGAN) consists of a discriminator and generator which approximates a target distribution over states $\rho_{\mathrm{tar}}$ by ``playing" a min-max game. Here, the generator sets its parameters $\theta$ outputting a state $G(\theta)$, and the discriminator $H(W)$ is a parameterized sum of Hermitian operators with weights $W$. In each iteration of optimization, the discriminator first sets its operator weights, outputting a Hamiltonian $H_{\max}$ which is the Hamiltonian maximizing our dual formulation estimate of $D_{EM}(G(\theta), \rho_{\mathrm{tar}})$. Then, a gradient update is performed on the parameters of the generator $\theta$. This iterative process is repeated either until convergence in the generator parameters $\theta$ or until a stopping criterion is reached. We detail the forms of the discriminator and generator as well as the steps of the algorithm in this section.

\subsection{Form of the discriminator}
In an optimal scenario, a discriminator explores the complete set of Hamiltonians which have Lipschitz constant less than or equal to one. However, this ideal case does not lend itself to efficient algorithms, and we instead construct a discriminator which efficiently estimates (lower bounds) the quantum EM distance. The discriminator we choose is a parameterized sum of strings of Pauli operators:
\begin{equation}\label{eq:H(W)}
    H(W) = \sum_{P_1,\,\ldots,\,P_n \in \{ I, X, Y, Z \}} w_{P_1\ldots P_n} \,\sigma_{P_1}^{(1)} \otimes \sigma_{P_2}^{(2)} \otimes \hdots \otimes \sigma_{P_n}^{(n)}\,,
\end{equation}
where $\sigma_I$ is the $2\times2$ identity matrix, $\sigma_X$, $\sigma_Y$ and $\sigma_Z$ are the Pauli matrices, superscripts specify the qubit on which the corresponding Pauli matrix acts and each $w_{P_1\ldots P_n}$ is the trainable parameter for the corresponding Pauli string $\sigma_{P_1}^{(1)} \otimes \sigma_{P_2}^{(2)} \otimes \hdots \otimes \sigma_{P_n}^{(n)}$. To simplify notation, we denote the set of all trainable parameters as $W$.
Finding the exact Lipschitz constant of the Hamiltonian \eqref{eq:H(W)} can be computationally expensive.
However, noting that Pauli operators have infinity norm of $1$ and applying the triangle equality, \eqref{eq:quant_lipsh_bound} provides an easily computable upper bound to $\left\|H(W)\right\|_L$, which we denote with $\left\|H(W)\right\|_{\tilde{L}}$:
\begin{equation}\label{eq:Ltilde}
    \left\|H(W)\right\|_{\tilde{L}} = 2\max_{i=1,\,\ldots,\,n}\sum_{P_1,\,\ldots,\,P_n \in \{ I, X, Y, Z \}:P_i\neq I} \left|w_{P_1\ldots P_n}\right| \ge \left\|H(W)\right\|_L\,.
\end{equation}

The Hamiltonian \eqref{eq:H(W)} has $4^n$ parameters and is impractical to train.
For this reason, we restrict optimization to operators that contain only terms acting on few qubits.
One option is to choose the set of $k$-local (\textit{i.e.,} acting on $k$ qubits and not necessarily geometrically local) Pauli operators as the discriminator.
We denote with $\mathcal{O}_n^{(k)}$ the linear span of such operators.
For example, the most general element of $\mathcal{O}_n^{(2)}$ is
\begin{equation}
\begin{split}
    H(W) =  w_I + \sum_{i=1}^n\sum_{P\in\{X,Y,Z\}}w_P^{(i)}\,\sigma_P^{(i)} + \sum_{i=1}^{n-1} \sum_{j=i+1}^n\sum_{P,\,Q \in \{X, Y, Z \}} w_{P,Q}^{(i,j)}\,\sigma_{P}^{(i)} \otimes \sigma_{Q}^{(j)},
    \label{eq:discriminator_k2}
\end{split}
\end{equation}
where each $w_I$, $w_P^{(i)}$ and $w_{P,\,Q}^{(i,j)}$ is the trainable parameter for the corresponding Pauli operator.
For $k \ll n$, there are $O(n^k)$ total terms in the above summation, polynomial in the number of qubits.

We can now define an approximated EM distance by restricting the optimization in \eqref{eq:qWass1} to $k$-local Hamiltonians and replacing the exact Lipschitz constant with the approximated Lipschitz constant \eqref{eq:Ltilde}:
    \begin{equation}\label{eq:EMtilde}
    D_{EM}^{(k)}(\rho,\sigma) = \max \left\{ \Tr\left[\left(\rho-\sigma\right)H\right] : H \in \mathcal{O}^{(k)}_n,\; \|H\|_{\tilde{L}} \leq 1 \right\}\,.
    \end{equation}
The approximated quantum EM distance is increasing with respect to $k$ and provides a lower bound to the exact quantum EM distance, \emph{i.e.}, for any two quantum states $\rho$ and $\sigma$,
\begin{equation}
    D_{EM}^{(1)}(\rho,\sigma) \le D_{EM}^{(2)}(\rho,\sigma) \ldots \le D_{EM}^{(n)}(\rho,\sigma) \le D_{EM}(\rho,\sigma)\,.
\end{equation}
The order $k$ of Pauli operators can be tuned to the complexity of a given problem. To learn truly random states, access to the full set of Hamiltonian operators is needed, but in many cases, especially when learning data generated by shallow circuits or data that can be discriminated via reduced density matrices, learning using only lower order Pauli operators can be effective.

The approximated EM distance \eqref{eq:EMtilde} can be computed with the following linear program, which can be efficiently solved. To simplify notation, we assume all parameters are enumerated in a list $W=\{w_1, w_2, \dots, w_{|W|}\}$. For each parameter $w_i$, we let $\mathcal{I}_i$ be equal to the set of qubits which the corresponding Pauli string acts on. Thus, with $|W|$ parameters and $n$ qubits, one maximizes the following linear program:
\begin{equation}
\begin{array}{ll@{}ll}
\text{maximize}  & \displaystyle\sum\limits_{j=1}^{|W|} c_{j}w_{j} &\\
\text{subject to}& \displaystyle\sum\limits_{j:i \in \mathcal{I}_j}   |w_{j}| \leq 1,  & \qquad i=1 ,..., n
\end{array}
\label{eq:linear_program}
\end{equation}
where $c_{j}$ is the trace of the product between the $j$-th Pauli string and $G(\theta)-\rho_{\mathrm{tar}}$. \textit{i.e.,} assuming $w_j$ is associated to Pauli string $\sigma_{P_a}^{(a)}\sigma_{P_b}^{(b)}\dots\sigma_{P_k}^{(k)}$, then $c_j=\Tr \left[ (G-\rho_{\mathrm{tar}}) \sigma_{P_a}^{(a)}\,\sigma_{P_b}^{(b)}\dots\sigma_{P_k}^{(k)} \right] $. In the above formulation, there exists a constraint for each qubit $i$ limiting the sum of magnitudes of operators acting on that qubit to less than or equal to one.

The linear program in \eqref{eq:linear_program} can be transformed into a standard form linear program with $n$ constraints (one for each qubit), which outputs a sparse set of at most $n$ non-zero weights \cite{bertsimas1997introduction} (the number of non-zero variables in linear programs in standard form is at most the number of constraints). Specifically, the linear program will output $n_{\mathrm{active}} \leq n$ operators with non-zero weights, called active operators, constructing a Hamiltonian $H_{\max}$ which is passed onto the generator for optimization:
\begin{equation}
    H_{\max} = \sum_{i=0}^{n_{\mathrm{active}}} w_i^* H_i^* ,
    \label{eq:discriminator_max_hamiltonian}
\end{equation}
where $w_{i}^*$ and $H_i^*$ are the weights and active operators respectively. As an example, we show in \autoref{app:correlations} that for single qubit states, the linear program outputs an optimal Hamiltonian composed of single qubit Pauli terms chosen by selecting the single qubit Pauli term which has the greatest contribution on each qubit (thus $n_{active} = n$ in this setting). Since $n_{\mathrm{active}} \leq n$, gradient updates on the generator which aims to minimize the expectation of $H_{\max}$ can be performed. However, this efficiency comes at the cost of potentially missing certain active operators in the optimal Hamiltonian. \autoref{eq:discriminator_max_hamiltonian} is naturally biased towards including low order operators in its set of active operators. Especially for problems where the optimization needs to access higher order operators, this methodology may fail to perform effectively. 

\autoref{app:correlations} compares values of the estimated distance $D_{EM}^{(2)}$ with the actual distance $D_{EM}$ for random states generated by shallow circuits showing that there is a correlation between the two measures, although the approximation may introduce an unwanted bias. As we show in \autoref{app:bias}, restricting the optimization over operators to terms acting on few qubits does not affect the value of the distance, since the optimal unconstrained Hamiltonian for the maximization problem \eqref{eq:linear_program} contains only these terms whenever the coefficients $c_j$ associated to single Pauli operators are all $\Omega(1)$. Furthermore, since operators only act on few qubits, efficient algorithms from \cite{huang2020predicting,huang2021efficient} can be applied to calculate the expectations of $|W|$ Pauli operators in \eqref{eq:discriminator_k2} using $O(\log |W|)$ copies of the generated and target states. If a large number of higher order Pauli terms are included in the decomposition above, such logarithmic dependence may no longer hold.

We stress that the approximation employed here is the central limitation in successfully applying our qWGAN model. Improving this approximation is crucial to expanding the scope of application of our qWGAN beyond the assorted examples provided in \autoref{sec:simulations}. Generally, there are two paths by which one can improve the approximation. The first path consists of identifying a more optimal relaxation to the semidefinite program in comparison to the linear relaxation of \autoref{eq:linear_program}. Existing quantum algorithms for solving semidefinite programs \cite{brandao2017quantum,van2020quantum} may help in realizing this goal. The second path consists of finding methods to more optimally choose the subset of Hamiltonians or Pauli operators included in the approximation. In some cases, higher order Pauli operators are needed to distinguish states and avoiding these altogether may produce sub-optimal results. In light of this, we describe below one technique that empirically helps in finding a good subset of Pauli operators.

\paragraph{Optional cycling of operators}
\label{par:cycling}
Over a small number of steps of optimization, changes to the expectations of operators $c_j$ are expected to be very small. Therefore, if the expectation of a given operator in the discriminator is small, it is unlikely that the operator will be chosen as an active operator over the course of optimization. Therefore, one has the option of removing these ``bad" operators and including new operators (here, we choose the new operators uniformly randomly from the set of all Pauli operators) into the set of operators over which the discriminator optimizes. Many choices exist for cycling operators; here, we opt for a simple choice where operators are cycled out when the expectation of an operator is below a threshold equal to $P(\min_i c_{i}^*)$ (\textit{i.e.,} minimum taken over all active operators) where $0 < P \leq 1$. When an operator is cycled out, a random Pauli operator is then included in the discriminator's set of operators including potentially Pauli operators that were removed in earlier cycles. Cycling operators may, of course, be detrimental if operators are cycled out that end up being useful during later phases of training. Nevertheless, in our experiments, we often find that the amount of cycling can serve as another tunable hyperparameter for improving the performance of our qWGAN.

\subsection{Form of the generator}
In its most general form, a generator is an object or function, that when given an input (potentially a sample from a random variable), outputs a state which approximates or produces a sample drawn from a distribution close to the target distribution. Similar to classical machine learning where neural networks are customized to given settings -- \textit{e.g.,} convolutional neural networks optimized for image analysis \cite{krizhevsky2017imagenet,gu2018recent,lecun2015deep} and transformer networks optimized for text analysis \cite{vaswani2017attention,brown2020language,devlin2018bert} -- the form of the generator in our quantum algorithm can and should be customized to the specific problem setting. Many options exist for constructing a generator including parameterized quantum circuits \cite{benedetti2019parameterized,du2018expressive} and quantum neural networks \cite{sharma2020trainability,schuld2014quest,killoran2019continuous,cong2019quantum}. The form of the generator determines the space of functions which a generator can access, and ideally this space should overlap with the function of the target object. Given we can only cover a limited class of generators in our analysis, we focus here on a single, though generic, form for the generator, encouraging future research to construct and analyze generators customized to specific applications in quantum machine learning.

In this generic formulation, the generator $G(\theta)$ is a function which maps a starting state $\ket{\psi_0} \bra{\psi_0}$ to a density matrix $\rho$ representing the distribution over quantum states that one aims to reconstruct. As in \cite{chakrabarti2019quantum}, our generator is constructed by a set of probabilities and associated parameterized unitaries $\{(p_1,U_1), \dots, (p_r, U_r) \}$: 
\begin{equation}
    G(\theta) = \sum_{i=1}^r p_i U_i \ket{\psi_0} \bra{\psi_0} U_i^\dagger  ,
    \label{eq:generic_generator}
\end{equation}
where we use $\theta$ to denote the set of all parameters for the generator which includes the probabilities $p_i$ and parameters for each unitary $U_i$. $r$ is the maximum rank of the output density matrix which can be tuned as a hyperparameter. Later, we consider $U_i$ constructed by parameterized quantum circuits with one and two qubit gates. The choice of these parameterized circuits depends on the nature of the problem (see \autoref{sec:simulations} for examples).

\subsection{qWGAN optimization procedure}

The algorithm for the qWGAN detailed in \autoref{alg:qWGAN} iteratively optimizes parameters of the generator and discriminator, consistent with methods used in classical GANs \cite{arjovsky2017wasserstein}. The following two steps are repeated until convergence in the parameters of the generator $\theta$. First, the parameters $w$ are updated using the linear program \eqref{eq:linear_program} to maximize the quantum EM distance $D_{EM}$ in equation \eqref{eq:qWass1}. Then, a gradient update is performed on the parameters of the generator $\theta$.

\begin{algorithm}
\caption{qWGAN with quantum earth mover's distance}
\begin{algorithmic}[1]

\Require initial discriminator operators: $H_i^{[0]}$ \Comment{\textit{e.g.,} set of $2$-local Paulis}
\Require initialization of generator parameters: $p_i^{[0]}$ and $\theta_i^{[0]}$
\Require hyperparameters for generator optimizer (\textit{e.g.,} learning rate $\alpha$) 
    \While{$\theta$, $p$ have not converged}  \Comment{alternatively, stop after $T$ steps}
        \Indent
        \LineComment  \textit{discriminator optimization: }
        \State measure operator expectations: $c_i \leftarrow \Tr{ \left[ H_i(G(\theta)-\rho_{\mathrm{tar}}) \right]}$
        \State find $w_i^*$, $H_i^*$ (linear program, equation \eqref{eq:linear_program}) \Comment{$H_{\max} = \sum_i w_i^* H_i^*$}
        \State \textbf{optional:} cycle operators
        \EndIndent
        \Indent
        \LineComment  \textit{generator optimization: }
        \State find gradients $g_p$, $g_\theta$ of $\Tr{[G(\theta)H_{\max}]}$ \Comment{see \autoref{app:grad}} \label{alg:calc_grad}
        \State perform gradient update on $\theta$ and $p$ \Comment{\textit{e.g.,} $\theta \leftarrow \theta - \alpha g_\theta$}
        \EndIndent
        
    \EndWhile  

\end{algorithmic}
\label{alg:qWGAN}
\end{algorithm}

\subsection{Properties of the gradient}

As stated earlier, one can calculate the gradients with respect to the parameters of the unitary operator implemented by the circuit (step \ref{alg:calc_grad} in \autoref{alg:qWGAN}) via the parameter shift rule \cite{schuld2019evaluating}. As an example, let $G(\theta)$ be generated by the quantum circuit that implements the unitary operator
\begin{equation}
    U(\theta) = \prod_{k}U_k\,e^{-i\theta_kP_k}\,,
\end{equation}
where each $U_k$ is a unitary operator that does not contain any parameter and each $P_k$ is a generalized Pauli operator.
Then, given an optimal Hamiltonian $H_{\max}$, one can calculate the gradient as follows for a given parameter $\theta_k$:
\begin{equation}
    \frac{\partial}{\partial \theta_k} \Tr{[G(\theta)H_{\max}]} = \Tr{[G(\theta^+)H_{\max}]} -  \Tr{[G(\theta^-)H_{\max}]},
\end{equation}
where $\theta^+$ and $\theta^-$ are the values of the parameters shifted by $\frac{\pi}{4}$ in either direction for the entry corresponding to $\theta_k$ \cite{schuld2019evaluating}. This parameter-shift rule has the benefit that the equation for the gradient is in fact exact (not an approximation). Gradients for each individual parameter must be calculated using separate circuit evaluations.

Prior work has shown that local cost functions avoid barren plateaus up to poly-logarithmic depth in the circuit \cite{cerezo2020cost}. Due to the super-additivity property of the quantum EM distance and the construction of the linear relaxation in \eqref{eq:linear_program}, the optimal Hamiltonian is heavily biased towards local terms and thus fits into this regime. We formalize this below by showing that high order Pauli strings acting nontrivially on $k$ qubits must have magnitude greater than $k$ times the smallest single qubit Pauli contribution to the optimal Hamiltonian $H_{\max}$.
\begin{prop}
Let $w^*:\{I,X,Y,Z\}^n\to\mathbb{R}$ be the set of parameters that achieve the maximum in \eqref{eq:linear_program}, and let
\begin{equation}
a = \min_{i=1,\,\ldots,\,n}\max_{P=X,Y,Z}\left|\mathrm{Tr}\left[\left(G(\theta) -\rho_{\mathrm{tar}}\right)\sigma_P^{(i)}\right]\right|\,.
\end{equation}
Then, $w^*_{P_1\ldots P_n} = 0$ for any $P_1,\,\ldots,\,P_n\in\{I,X,Y,Z\}$ such that
\begin{equation}
|c_{P_1\ldots P_n}| < a\left|\left\{i=1,\,\ldots,\,n:P_i\neq I\right\}\right|\,.
\end{equation}
In particular, $w^*_{P_1\ldots P_n} = 0$ for any Pauli string that acts nontrivially on more than $2/a$ qubits.
\end{prop}
The proof of the above is deferred to \autoref{app:bias}. As a corollary of the above, any global $k$-qubit Pauli term that appears in the optimal Hamiltonian must exceed the sums of the maximum single qubit Pauli terms on the $k$ qubits on which it acts. Thus, the optimal Hamiltonian $H_{\max}$ will be at most $2/a$-local which is constant when the expectation of single qubit Paulis is $\Omega(1)$. In these settings, the qWGAN will avoid barren plateaus whenever the generator has depth that does not grow faster than logarithmically in the number of qubits (see Theorem 2 of \cite{cerezo2020cost}).

\section{Experiments}
\label{sec:simulations}

Our qWGAN can efficiently learn quantum data of various forms. Here, we apply the qWGAN directly to the toy model of \autoref{sec:toy_model} and also consider a more general scenario where the qWGAN learns states generated by a mixing circuit previously known to suffer from barren plateaus \cite{mcclean2018barren,cerezo2020cost,huembeli2020characterizing}. In \autoref{app:other_simulations}, we include results for the qWGAN in two other scenarios: one where the generator is a circuit formed by a quantum alternating operator ansatz (QAOA) \cite{fingerhuth2018quantum,hadfield2019quantum,farhi2014quantum} and one where the qWGAN is tasked with learning mixed states. The Adam optimizer with a default learning rate of 0.01 was used to train the qWGAN \cite{kingma2014adam}. Details on the structure of the quantum circuits and on how the simulations were performed are provided in \autoref{app:circuits} and \autoref{app:computational_details}, respectively.

\paragraph{Learning the GHZ state}
\label{subsec:GHZ_simulations}
The $n$-qubit GHZ state is an entangled state which requires a simple circuit of depth $n$ to construct. However, as noted in \autoref{sec:toy_model}, the correct parameters of this circuit are hard to learn when using cost metrics that are a function of the inner product between the generated and target GHZ state. Continuing our analysis, we show that our qWGAN is especially efficient and effective at learning the correct parameters of a circuit to generate the GHZ state. 

\begin{figure}[ht]
    \centering
    \includegraphics[center]{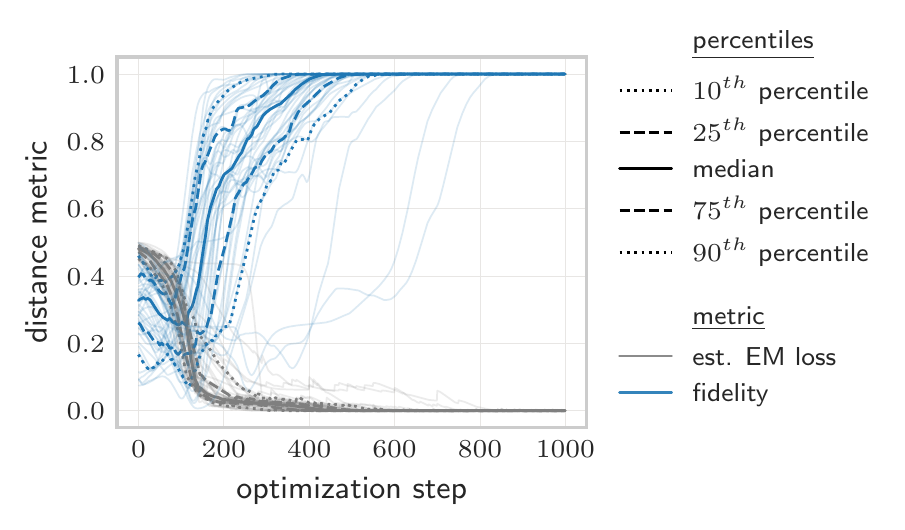}
    \caption{The qWGAN consistently generates the 8 qubit GHZ state. Estimated EM loss (quantum EM distance estimated by active operators) is plotted in grey alongside the fidelity in blue. EM distance is normalized to a maximum of one by dividing by the number of qubits. Percentiles are calculated across 50 simulations. Individual simulations are plotted as transparent lines.}
    \label{fig:GHZ_loss_percentiles}
\end{figure}

As shown in \autoref{fig:GHZ_loss_percentiles} for $10$ random simulations, our qWGAN efficiently generates the $n=8$ qubit GHZ state in around $500$ steps of optimization-- results for circuits of different size are also consistent with this analysis and detailed in \autoref{app:other_simulations}. The generator circuit has $n+2$ parameters and depth $n$. Simulations are repeated $50$ times across random initializations. The discriminator starts with access to $k=2$ local Pauli operators, cycling out ``bad'' operators every five steps. Estimated EM loss is also plotted in \autoref{fig:GHZ_loss_percentiles} (normalized by dividing by the number of qubits), which is equal to the quantum EM distance as measured by the active operators in the discriminator (lower bounding the actual quantum EM distance). Jumps in the estimated EM loss can be observed when operators are cycled and later become active, highlighting the importance of randomly cycling operators in these simulations. It is interesting to note that during the early phases of learning, the qWGAN often optimizes the EM distance while temporarily decreasing the fidelity. This learning profile is typically associated with transitions from the state $\ket{0_n}$ to the GHZ state. As our toy model indicated, this transition characterized by a temporary decrease in the fidelity is needed to reach the global optimum.

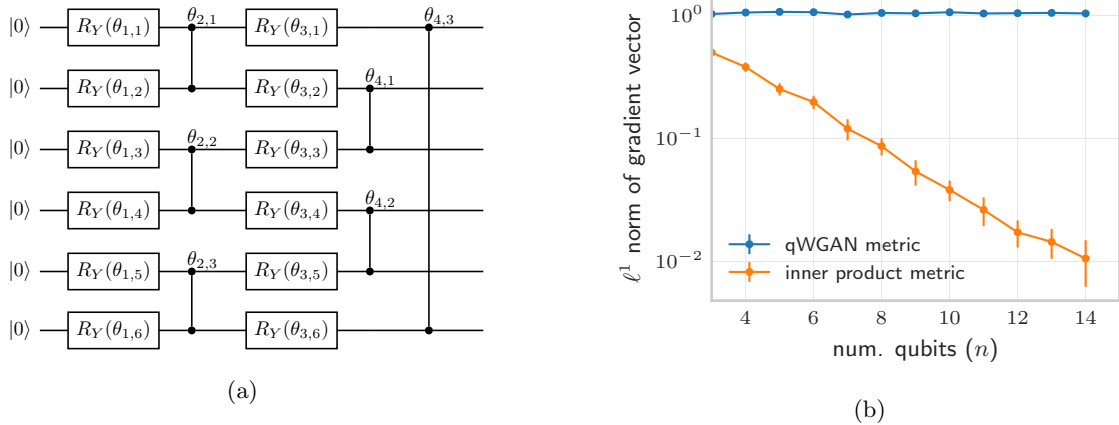
\begin{figure}[ht]
    \centering
    \begin{subfigure}[]{0.5\textwidth}
        \centering
        \begin{tikzpicture}
        \node[scale=0.75] {
        \begin{quantikz}[row sep=0.4cm, column sep=0.5cm]
            \lstick{$\ket{0}$} & \gate{R_Y(\theta_{1,1})} & \phase{\theta_{2,1}} & 
            \gate{R_Y(\theta_{3,1})} & \qw & \phase{\theta_{4,3}} & \qw \\     
            \lstick{$\ket{0}$} & \gate{R_Y(\theta_{1,2})} & \ctrl{-1} &
            \gate{R_Y(\theta_{3,2})} & \phase{\theta_{4,1}} & \qw & \qw  \\
            \lstick{$\ket{0}$} & \gate{R_Y(\theta_{1,3})} & \phase{\theta_{2,2}} &
            \gate{R_Y(\theta_{3,3})} &  \ctrl{-1} & \qw & \qw \\
            \lstick{$\ket{0}$} & \gate{R_Y(\theta_{1,4})} & \ctrl{-1} &
            \gate{R_Y(\theta_{3,4})} & \phase{\theta_{4,2}} & \qw & \qw \\
            \lstick{$\ket{0}$} & \gate{R_Y(\theta_{1,5})} & \phase{\theta_{2,3}} &
            \gate{R_Y(\theta_{3,5})} & \ctrl{-1} & \qw & \qw \\
            \lstick{$\ket{0}$} & \gate{R_Y(\theta_{1,6})} & \ctrl{-1} &  
            \gate{R_Y(\theta_{3,6})} & \qw & \ctrl{-5} & \qw
        \end{quantikz}
        };
        \end{tikzpicture}
        \caption{}
        \label{fig:mixing_circuit}
    \end{subfigure}
    \hfill
    \begin{subfigure}[]{0.49\textwidth}
         \centering
         \includegraphics[]{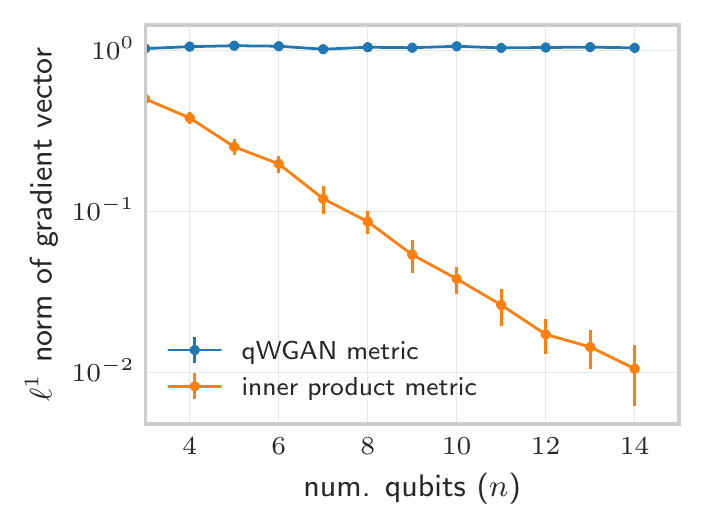}
         \caption{}
         \label{fig:mixing_gradients}
     \end{subfigure}
     \caption{(a) Single layer of mixing circuit consisting of alternating layers of parameterized Pauli Y rotations and parameterized Pauli Z-Z rotations applied to pairwise qubits. Here, the form of the circuit is shown for six qubits. (b) Learning using two layers of mixing circuit (shallow, constant depth) results in exponentially decaying gradients for conventional loss metrics. Gradients of the qWGAN remain constant while gradients with respect to a loss metric as a function of the inner product decay exponentially in the number of qubits. Gradients are calculated at first step of optimization and $\ell^1$ norm is divided by $n$ to normalize to the number of parameters in the circuit. Findings are consistent when the average is taken for the $\ell^2$ norm or of individual gradient entries as shown in \autoref{app:other_simulations}. Results are averaged across 100 simulations for each data point.  }
\end{figure}

\paragraph{Teacher-student learning}
\label{subsec:teach_student}
To analyze our qWGAN in a more general setting, we consider a ``teacher-student" setup where the circuit used to generate the target state and perform learning are both of the form shown in \autoref{fig:mixing_circuit}. This circuit is a generic mixing circuit also studied in \cite{mcclean2018barren,cerezo2020cost,huembeli2020characterizing} where barren plateaus in the loss landscape are observed. For our simulations, the target state $\phi_{\mathrm{tar}}$ is generated by a depth 2 circuit (\textit{i.e.} gates shown in \autoref{fig:mixing_circuit} repeated twice) with parameters drawn i.i.d. from the standard normal distribution. As a point of comparison, we compare our qWGAN to a quantum GAN equipped with the loss function $F = 1- \left| \braket{ \phi_{\mathrm{tar}} | \phi(\boldsymbol{\theta}) } \right|^2 $ which is a function of the inner product between the target and generated state. \autoref{fig:mixing_gradients} shows that when a circuit of the same form is used to learn the target state, gradients of $F$ (function of the inner product) decay exponentially with more qubits whereas gradients of the quantum EM loss function remain constant. Note that the exponentially decaying gradients for the inner product loss metrics are observed here for constant depth shallow circuits. This result further confirms that loss landscapes for the quantum EM distance avoid common pitfalls faced by conventional distance metrics including the Wasserstein semi-metric proposed in \cite{chakrabarti2019quantum}.

\begin{figure}[ht]
    \centering
    \includegraphics[center]{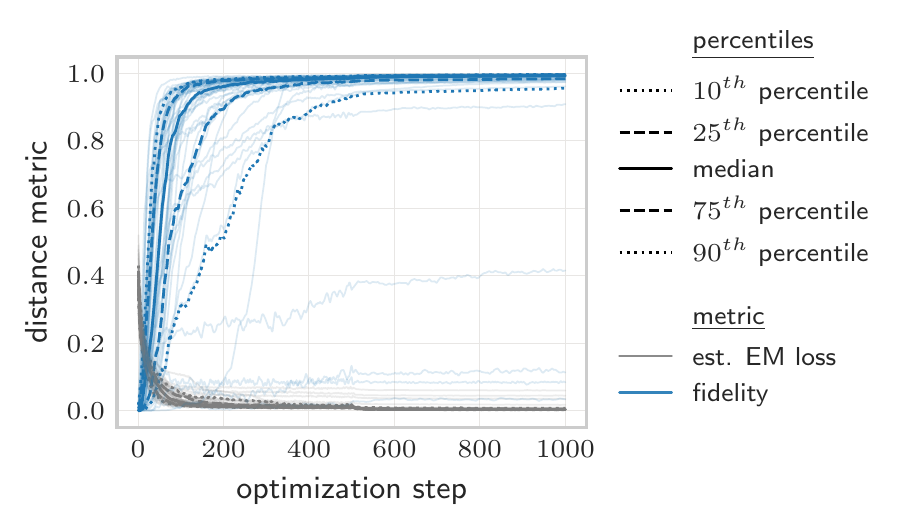}
    \caption{The student circuit is able to approximate well the state generated by the teacher circuit. Here, the target is constructed by randomly setting the parameters of a depth $2$ mixing circuit (teacher circuit). The qWGAN, equipped with a depth $4$ generator circuit, successfully learns the target state by optimizing over the quantum EM distance. Estimated EM loss (quantum EM distance estimated by active operators) is plotted in grey alongside the fidelity in blue. EM distance is normalized to a maximum of one by dividing by the number of qubits. Percentiles are calculated across 50 simulations. Individual simulations are plotted as transparent lines. }
    \label{fig:mixing_loss_plot}
\end{figure}

Furthermore, as shown in \autoref{fig:mixing_loss_plot}, the qWGAN successfully learns the states constructed by $n=8$ qubit teacher circuits using student circuits of depth $4$. This circuit has $96$ trainable parameters. Simulations are repeated $50$ times across random initializations in \autoref{fig:mixing_loss_plot}. Target states are generated by drawing the parameters of the teacher circuit i.i.d. from the standard normal distribution. Learning is typically achieved within a few hundred steps of optimization. In these simulations, the discriminator for the qWGAN contains all order $2$ Pauli operators and no cycling of the operators was performed. We find that, in general, learning in the teacher-student setting is best achieved when the student circuit is deeper than the teacher circuit. Furthermore, due to the approximations made in calculating the quantum EM distance, we find that our algorithm struggles to learn especially deep 8-qubit teacher circuits which are four layers or more in depth. For these deeper target circuits, we suspect that higher order Paulis are needed to efficiently estimate the quantum EM distance, and further improvements to the optimization must be made to incorporate these higher order Paulis in the estimation procedure. Additional simulations for different circuit sizes are shown in \autoref{app:other_simulations}.

\section{Discussion}
As interest in quantum machine learning algorithms has flourished, recent research has highlighted the challenges associated with learning using quantum computers. At the root of these challenges are adverse properties of loss landscapes in quantum machine learning settings, perhaps most notably that loss landscapes have poor local minima and exponentially decaying gradients. In this work, we show that the loss landscape induced by the quantum EM distance can potentially confer advantages in machine learning settings, especially when optimization is performed over local gates in shallow circuits. Our results provide a new approach to constructing loss landscapes which can avoid common quantum machine learning roadblocks.  

For the specific application of learning quantum data, we have proposed a qWGAN which leverages the quantum EM distance to produce an efficient learning algorithm. In accord with its classical counterpart \cite{arjovsky2017wasserstein}, we show that our qWGAN can potentially improve convergence and stability in learning quantum data. Nevertheless, the qWGAN struggles in learning more ``random" data or data generated from deep circuits. These challenges stem from two approximations made in the learning procedure. First, to ensure runtime efficiency, the discriminator was restricted to measuring a subset of local Pauli operators. Second, the optimization statement to calculate the quantum EM distance was relaxed into a linear program to ensure it can be effficiently calculated classically. Though this estimated distance is well correlated to the true distance when learning certain structured states like the GHZ state or states generated by shallow circuits, it failed to tightly bound the quantum EM distance in more challenging settings. Looking forward, improving the bounds given by relaxations of the quantum EM distance can potentially allow for application of our qWGAN in these more challenging settings.

Beyond the test cases studied here, the qWGAN has many potential applications. In quantum controls, one can use it to search for robust or optimal control parameters \cite{ge2020robust,palittapongarnpim2017learning}. For unsupervised learning, the qWGAN provides a framework and approach to quantum circuit compression, data encoding, and sampling \cite{schuld2020circuit,romero2017quantum,jones2018quantum}. For quantum error correction, one can use a qWGAN to develop new techniques for constructing quantum error codes or assisting error correction procedures \cite{nautrup2019optimizing,baireuther2018machine,bausch2020quantum,johnson2017qvector}.

\section*{Acknowledgement}
This work was funded by AFOSR, ARO under the Blue Sky Initiative, DARPA, and NSF. MM is partially supported by the NSF Grant No. CCF-1954960.
GDP is a member of the ``Gruppo Nazionale per la Fisica Matematica (GNFM)'' of the ``Istituto Nazionale di Alta Matematica ``Francesco Severi'' (INdAM)''.

\section*{Author Contributions}
BTK developed the structure of the qWGAN algorithm and performed simulations under the supervision of SL. All authors contributed to the design of the qWGAN algorithm. The EM distance was motivated by prior work from GDP, MM, and SL. BTK and GDP wrote the manuscript. GDP developed mathematical proofs in the appendix. All authors reviewed the results. 

\bibliographystyle{naturemag}
\bibliography{main.bib}

\begin{thebibliography}{100}
\expandafter\ifx\csname url\endcsname\relax
  \def\url#1{\texttt{#1}}\fi
\expandafter\ifx\csname urlprefix\endcsname\relax\def\urlprefix{URL }\fi
\providecommand{\bibinfo}[2]{#2}
\providecommand{\eprint}[2][]{\url{#2}}

\bibitem{benedetti2019adversarial}
\bibinfo{author}{Benedetti, M.}, \bibinfo{author}{Grant, E.},
  \bibinfo{author}{Wossnig, L.} \& \bibinfo{author}{Severini, S.}
\newblock \bibinfo{title}{Adversarial quantum circuit learning for pure state
  approximation}.
\newblock \emph{\bibinfo{journal}{New Journal of Physics}}
  \textbf{\bibinfo{volume}{21}}, \bibinfo{pages}{043023}
  (\bibinfo{year}{2019}).

\bibitem{dallaire2018quantum}
\bibinfo{author}{Dallaire-Demers, P.-L.} \& \bibinfo{author}{Killoran, N.}
\newblock \bibinfo{title}{Quantum generative adversarial networks}.
\newblock \emph{\bibinfo{journal}{Physical Review A}}
  \textbf{\bibinfo{volume}{98}}, \bibinfo{pages}{012324}
  (\bibinfo{year}{2018}).

\bibitem{torlai2020machine}
\bibinfo{author}{Torlai, G.} \& \bibinfo{author}{Melko, R.~G.}
\newblock \bibinfo{title}{Machine-learning quantum states in the nisq era}.
\newblock \emph{\bibinfo{journal}{Annual Review of Condensed Matter Physics}}
  \textbf{\bibinfo{volume}{11}}, \bibinfo{pages}{325--344}
  (\bibinfo{year}{2020}).

\bibitem{gao2018experimental}
\bibinfo{author}{Gao, J.} \emph{et~al.}
\newblock \bibinfo{title}{Experimental machine learning of quantum states}.
\newblock \emph{\bibinfo{journal}{Physical review letters}}
  \textbf{\bibinfo{volume}{120}}, \bibinfo{pages}{240501}
  (\bibinfo{year}{2018}).

\bibitem{aaronson2007learnability}
\bibinfo{author}{Aaronson, S.}
\newblock \bibinfo{title}{The learnability of quantum states}.
\newblock \emph{\bibinfo{journal}{Proceedings of the Royal Society A:
  Mathematical, Physical and Engineering Sciences}}
  \textbf{\bibinfo{volume}{463}}, \bibinfo{pages}{3089--3114}
  (\bibinfo{year}{2007}).

\bibitem{rocchetto2019experimental}
\bibinfo{author}{Rocchetto, A.} \emph{et~al.}
\newblock \bibinfo{title}{Experimental learning of quantum states}.
\newblock \emph{\bibinfo{journal}{Science advances}}
  \textbf{\bibinfo{volume}{5}}, \bibinfo{pages}{eaau1946}
  (\bibinfo{year}{2019}).

\bibitem{lloyd2018quantum}
\bibinfo{author}{Lloyd, S.} \& \bibinfo{author}{Weedbrook, C.}
\newblock \bibinfo{title}{Quantum generative adversarial learning}.
\newblock \emph{\bibinfo{journal}{Physical review letters}}
  \textbf{\bibinfo{volume}{121}}, \bibinfo{pages}{040502}
  (\bibinfo{year}{2018}).

\bibitem{carrasquilla2019reconstructing}
\bibinfo{author}{Carrasquilla, J.}, \bibinfo{author}{Torlai, G.},
  \bibinfo{author}{Melko, R.~G.} \& \bibinfo{author}{Aolita, L.}
\newblock \bibinfo{title}{Reconstructing quantum states with generative
  models}.
\newblock \emph{\bibinfo{journal}{Nature Machine Intelligence}}
  \textbf{\bibinfo{volume}{1}}, \bibinfo{pages}{155--161}
  (\bibinfo{year}{2019}).

\bibitem{chakrabarti2019quantum}
\bibinfo{author}{Chakrabarti, S.}, \bibinfo{author}{Yiming, H.},
  \bibinfo{author}{Li, T.}, \bibinfo{author}{Feizi, S.} \& \bibinfo{author}{Wu,
  X.}
\newblock \bibinfo{title}{Quantum wasserstein generative adversarial networks}.
\newblock In \emph{\bibinfo{booktitle}{Advances in Neural Information
  Processing Systems}}, \bibinfo{pages}{6781--6792} (\bibinfo{year}{2019}).

\bibitem{beer2020training}
\bibinfo{author}{Beer, K.} \emph{et~al.}
\newblock \bibinfo{title}{Training deep quantum neural networks}.
\newblock \emph{\bibinfo{journal}{Nature communications}}
  \textbf{\bibinfo{volume}{11}}, \bibinfo{pages}{1--6} (\bibinfo{year}{2020}).

\bibitem{kiani2020learning}
\bibinfo{author}{Kiani, B.~T.}, \bibinfo{author}{Lloyd, S.} \&
  \bibinfo{author}{Maity, R.}
\newblock \bibinfo{title}{Learning unitaries by gradient descent}.
\newblock \emph{\bibinfo{journal}{arXiv preprint arXiv:2001.11897}}
  (\bibinfo{year}{2020}).

\bibitem{mitarai2018quantum}
\bibinfo{author}{Mitarai, K.}, \bibinfo{author}{Negoro, M.},
  \bibinfo{author}{Kitagawa, M.} \& \bibinfo{author}{Fujii, K.}
\newblock \bibinfo{title}{Quantum circuit learning}.
\newblock \emph{\bibinfo{journal}{Physical Review A}}
  \textbf{\bibinfo{volume}{98}}, \bibinfo{pages}{032309}
  (\bibinfo{year}{2018}).

\bibitem{bisio2010optimal}
\bibinfo{author}{Bisio, A.}, \bibinfo{author}{Chiribella, G.},
  \bibinfo{author}{D’Ariano, G.~M.}, \bibinfo{author}{Facchini, S.} \&
  \bibinfo{author}{Perinotti, P.}
\newblock \bibinfo{title}{Optimal quantum learning of a unitary
  transformation}.
\newblock \emph{\bibinfo{journal}{Physical Review A}}
  \textbf{\bibinfo{volume}{81}}, \bibinfo{pages}{032324}
  (\bibinfo{year}{2010}).

\bibitem{quintino2019reversing}
\bibinfo{author}{Quintino, M.~T.}, \bibinfo{author}{Dong, Q.},
  \bibinfo{author}{Shimbo, A.}, \bibinfo{author}{Soeda, A.} \&
  \bibinfo{author}{Murao, M.}
\newblock \bibinfo{title}{Reversing unknown quantum transformations: Universal
  quantum circuit for inverting general unitary operations}.
\newblock \emph{\bibinfo{journal}{Physical Review Letters}}
  \textbf{\bibinfo{volume}{123}}, \bibinfo{pages}{210502}
  (\bibinfo{year}{2019}).

\bibitem{lloyd2020quantum}
\bibinfo{author}{Lloyd, S.} \emph{et~al.}
\newblock \bibinfo{title}{Quantum polar decomposition algorithm}.
\newblock \emph{\bibinfo{journal}{arXiv preprint arXiv:2006.00841}}
  (\bibinfo{year}{2020}).

\bibitem{carolan2020variational}
\bibinfo{author}{Carolan, J.} \emph{et~al.}
\newblock \bibinfo{title}{Variational quantum unsampling on a quantum photonic
  processor}.
\newblock \emph{\bibinfo{journal}{Nature Physics}}
  \textbf{\bibinfo{volume}{16}}, \bibinfo{pages}{322--327}
  (\bibinfo{year}{2020}).

\bibitem{sharma2020noise}
\bibinfo{author}{Sharma, K.}, \bibinfo{author}{Khatri, S.},
  \bibinfo{author}{Cerezo, M.} \& \bibinfo{author}{Coles, P.~J.}
\newblock \bibinfo{title}{Noise resilience of variational quantum compiling}.
\newblock \emph{\bibinfo{journal}{New Journal of Physics}}
  \textbf{\bibinfo{volume}{22}}, \bibinfo{pages}{043006}
  (\bibinfo{year}{2020}).

\bibitem{benedetti2019generative}
\bibinfo{author}{Benedetti, M.} \emph{et~al.}
\newblock \bibinfo{title}{A generative modeling approach for benchmarking and
  training shallow quantum circuits}.
\newblock \emph{\bibinfo{journal}{npj Quantum Information}}
  \textbf{\bibinfo{volume}{5}}, \bibinfo{pages}{1--9} (\bibinfo{year}{2019}).

\bibitem{liu2018differentiable}
\bibinfo{author}{Liu, J.-G.} \& \bibinfo{author}{Wang, L.}
\newblock \bibinfo{title}{Differentiable learning of quantum circuit born
  machines}.
\newblock \emph{\bibinfo{journal}{Physical Review A}}
  \textbf{\bibinfo{volume}{98}}, \bibinfo{pages}{062324}
  (\bibinfo{year}{2018}).

\bibitem{coyle2020born}
\bibinfo{author}{Coyle, B.}, \bibinfo{author}{Mills, D.},
  \bibinfo{author}{Danos, V.} \& \bibinfo{author}{Kashefi, E.}
\newblock \bibinfo{title}{The born supremacy: Quantum advantage and training of
  an ising born machine}.
\newblock \emph{\bibinfo{journal}{npj Quantum Information}}
  \textbf{\bibinfo{volume}{6}}, \bibinfo{pages}{1--11} (\bibinfo{year}{2020}).

\bibitem{mcclean2018barren}
\bibinfo{author}{McClean, J.~R.}, \bibinfo{author}{Boixo, S.},
  \bibinfo{author}{Smelyanskiy, V.~N.}, \bibinfo{author}{Babbush, R.} \&
  \bibinfo{author}{Neven, H.}
\newblock \bibinfo{title}{Barren plateaus in quantum neural network training
  landscapes}.
\newblock \emph{\bibinfo{journal}{Nature communications}}
  \textbf{\bibinfo{volume}{9}}, \bibinfo{pages}{1--6} (\bibinfo{year}{2018}).

\bibitem{wang2020noise}
\bibinfo{author}{Wang, S.} \emph{et~al.}
\newblock \bibinfo{title}{Noise-induced barren plateaus in variational quantum
  algorithms}.
\newblock \emph{\bibinfo{journal}{arXiv preprint arXiv:2007.14384}}
  (\bibinfo{year}{2020}).

\bibitem{cerezo2020cost}
\bibinfo{author}{Cerezo, M.}, \bibinfo{author}{Sone, A.},
  \bibinfo{author}{Volkoff, T.}, \bibinfo{author}{Cincio, L.} \&
  \bibinfo{author}{Coles, P.~J.}
\newblock \bibinfo{title}{Cost-function-dependent barren plateaus in shallow
  quantum neural networks}.
\newblock \emph{\bibinfo{journal}{arXiv preprint arXiv:2001.00550}}
  (\bibinfo{year}{2020}).

\bibitem{pechen2011there}
\bibinfo{author}{Pechen, A.~N.} \& \bibinfo{author}{Tannor, D.~J.}
\newblock \bibinfo{title}{Are there traps in quantum control landscapes?}
\newblock \emph{\bibinfo{journal}{Physical review letters}}
  \textbf{\bibinfo{volume}{106}}, \bibinfo{pages}{120402}
  (\bibinfo{year}{2011}).

\bibitem{moore2012exploring}
\bibinfo{author}{Moore, K.~W.} \& \bibinfo{author}{Rabitz, H.}
\newblock \bibinfo{title}{Exploring constrained quantum control landscapes}.
\newblock \emph{\bibinfo{journal}{The Journal of chemical physics}}
  \textbf{\bibinfo{volume}{137}}, \bibinfo{pages}{134113}
  (\bibinfo{year}{2012}).

\bibitem{cerezo2020variational}
\bibinfo{author}{Cerezo, M.} \emph{et~al.}
\newblock \bibinfo{title}{Variational quantum algorithms}.
\newblock \emph{\bibinfo{journal}{arXiv preprint arXiv:2012.09265}}
  (\bibinfo{year}{2020}).

\bibitem{de2020quantum}
\bibinfo{author}{De~Palma, G.}, \bibinfo{author}{Marvian, M.},
  \bibinfo{author}{Trevisan, D.} \& \bibinfo{author}{Lloyd, S.}
\newblock \bibinfo{title}{The {Q}uantum {W}asserstein {D}istance of {O}rder 1}.
\newblock \emph{\bibinfo{journal}{IEEE Transactions on Information Theory}}
  \textbf{\bibinfo{volume}{67}}, \bibinfo{pages}{6627--6643}
  (\bibinfo{year}{2021}).

\bibitem{arjovsky2017wasserstein}
\bibinfo{author}{Arjovsky, M.}, \bibinfo{author}{Chintala, S.} \&
  \bibinfo{author}{Bottou, L.}
\newblock \bibinfo{title}{Wasserstein gan}.
\newblock \emph{\bibinfo{journal}{arXiv preprint arXiv:1701.07875}}
  (\bibinfo{year}{2017}).

\bibitem{chen2018adversarial}
\bibinfo{author}{Chen, L.} \emph{et~al.}
\newblock \bibinfo{title}{Adversarial text generation via feature-mover's
  distance}.
\newblock In \emph{\bibinfo{booktitle}{Advances in Neural Information
  Processing Systems}}, \bibinfo{pages}{4666--4677} (\bibinfo{year}{2018}).

\bibitem{rubner1998metric}
\bibinfo{author}{Rubner, Y.}, \bibinfo{author}{Tomasi, C.} \&
  \bibinfo{author}{Guibas, L.~J.}
\newblock \bibinfo{title}{A metric for distributions with applications to image
  databases}.
\newblock In \emph{\bibinfo{booktitle}{Sixth International Conference on
  Computer Vision (IEEE Cat. No. 98CH36271)}}, \bibinfo{pages}{59--66}
  (\bibinfo{organization}{IEEE}, \bibinfo{year}{1998}).

\bibitem{villani2008optimal}
\bibinfo{author}{Villani, C.}
\newblock \emph{\bibinfo{title}{Optimal transport: old and new}}, vol.
  \bibinfo{volume}{338} (\bibinfo{publisher}{Springer Science \& Business
  Media}, \bibinfo{year}{2008}).

\bibitem{gulrajani2017improved}
\bibinfo{author}{Gulrajani, I.}, \bibinfo{author}{Ahmed, F.},
  \bibinfo{author}{Arjovsky, M.}, \bibinfo{author}{Dumoulin, V.} \&
  \bibinfo{author}{Courville, A.~C.}
\newblock \bibinfo{title}{Improved training of wasserstein gans}.
\newblock In \emph{\bibinfo{booktitle}{Advances in neural information
  processing systems}}, \bibinfo{pages}{5767--5777} (\bibinfo{year}{2017}).

\bibitem{goodfellow2014generative}
\bibinfo{author}{Goodfellow, I.} \emph{et~al.}
\newblock \bibinfo{title}{Generative adversarial nets}.
\newblock In \emph{\bibinfo{booktitle}{Advances in neural information
  processing systems}}, \bibinfo{pages}{2672--2680} (\bibinfo{year}{2014}).

\bibitem{hu2019quantum}
\bibinfo{author}{Hu, L.} \emph{et~al.}
\newblock \bibinfo{title}{Quantum generative adversarial learning in a
  superconducting quantum circuit}.
\newblock \emph{\bibinfo{journal}{Science advances}}
  \textbf{\bibinfo{volume}{5}}, \bibinfo{pages}{eaav2761}
  (\bibinfo{year}{2019}).

\bibitem{campos2021abrupt}
\bibinfo{author}{Campos, E.}, \bibinfo{author}{Nasrallah, A.} \&
  \bibinfo{author}{Biamonte, J.}
\newblock \bibinfo{title}{Abrupt transitions in variational quantum circuit
  training}.
\newblock \emph{\bibinfo{journal}{Physical Review A}}
  \textbf{\bibinfo{volume}{103}}, \bibinfo{pages}{032607}
  (\bibinfo{year}{2021}).

\bibitem{skolik2020layerwise}
\bibinfo{author}{Skolik, A.}, \bibinfo{author}{McClean, J.~R.},
  \bibinfo{author}{Mohseni, M.}, \bibinfo{author}{van~der Smagt, P.} \&
  \bibinfo{author}{Leib, M.}
\newblock \bibinfo{title}{Layerwise learning for quantum neural networks}.
\newblock \emph{\bibinfo{journal}{arXiv preprint arXiv:2006.14904}}
  (\bibinfo{year}{2020}).

\bibitem{brandao2017quantum}
\bibinfo{author}{Brandao, F.~G.} \& \bibinfo{author}{Svore, K.~M.}
\newblock \bibinfo{title}{Quantum speed-ups for solving semidefinite programs}.
\newblock In \emph{\bibinfo{booktitle}{2017 IEEE 58th Annual Symposium on
  Foundations of Computer Science (FOCS)}}, \bibinfo{pages}{415--426}
  (\bibinfo{organization}{IEEE}, \bibinfo{year}{2017}).

\bibitem{van2020quantum}
\bibinfo{author}{Van~Apeldoorn, J.}, \bibinfo{author}{Gily{\'e}n, A.},
  \bibinfo{author}{Gribling, S.} \& \bibinfo{author}{de~Wolf, R.}
\newblock \bibinfo{title}{Quantum sdp-solvers: Better upper and lower bounds}.
\newblock \emph{\bibinfo{journal}{Quantum}} \textbf{\bibinfo{volume}{4}},
  \bibinfo{pages}{230} (\bibinfo{year}{2020}).

\bibitem{bertsimas1997introduction}
\bibinfo{author}{Bertsimas, D.} \& \bibinfo{author}{Tsitsiklis, J.~N.}
\newblock \emph{\bibinfo{title}{Introduction to linear optimization}},
  vol.~\bibinfo{volume}{6} (\bibinfo{publisher}{Athena Scientific Belmont, MA},
  \bibinfo{year}{1997}).

\bibitem{huang2020predicting}
\bibinfo{author}{Huang, H.-Y.}, \bibinfo{author}{Kueng, R.} \&
  \bibinfo{author}{Preskill, J.}
\newblock \bibinfo{title}{Predicting many properties of a quantum system from
  very few measurements}.
\newblock \emph{\bibinfo{journal}{Nature Physics}}
  \textbf{\bibinfo{volume}{16}}, \bibinfo{pages}{1050--1057}
  (\bibinfo{year}{2020}).

\bibitem{huang2021efficient}
\bibinfo{author}{Huang, H.-Y.}, \bibinfo{author}{Kueng, R.} \&
  \bibinfo{author}{Preskill, J.}
\newblock \bibinfo{title}{Efficient estimation of pauli observables by
  derandomization}.
\newblock \emph{\bibinfo{journal}{arXiv preprint arXiv:2103.07510}}
  (\bibinfo{year}{2021}).

\bibitem{krizhevsky2017imagenet}
\bibinfo{author}{Krizhevsky, A.}, \bibinfo{author}{Sutskever, I.} \&
  \bibinfo{author}{Hinton, G.~E.}
\newblock \bibinfo{title}{Imagenet classification with deep convolutional
  neural networks}.
\newblock \emph{\bibinfo{journal}{Communications of the ACM}}
  \textbf{\bibinfo{volume}{60}}, \bibinfo{pages}{84--90}
  (\bibinfo{year}{2017}).

\bibitem{gu2018recent}
\bibinfo{author}{Gu, J.} \emph{et~al.}
\newblock \bibinfo{title}{Recent advances in convolutional neural networks}.
\newblock \emph{\bibinfo{journal}{Pattern Recognition}}
  \textbf{\bibinfo{volume}{77}}, \bibinfo{pages}{354--377}
  (\bibinfo{year}{2018}).

\bibitem{lecun2015deep}
\bibinfo{author}{LeCun, Y.}, \bibinfo{author}{Bengio, Y.} \&
  \bibinfo{author}{Hinton, G.}
\newblock \bibinfo{title}{Deep learning}.
\newblock \emph{\bibinfo{journal}{nature}} \textbf{\bibinfo{volume}{521}},
  \bibinfo{pages}{436--444} (\bibinfo{year}{2015}).

\bibitem{vaswani2017attention}
\bibinfo{author}{Vaswani, A.} \emph{et~al.}
\newblock \bibinfo{title}{Attention is all you need}.
\newblock In \emph{\bibinfo{booktitle}{Advances in neural information
  processing systems}}, \bibinfo{pages}{5998--6008} (\bibinfo{year}{2017}).

\bibitem{brown2020language}
\bibinfo{author}{Brown, T.~B.} \emph{et~al.}
\newblock \bibinfo{title}{Language models are few-shot learners}.
\newblock \emph{\bibinfo{journal}{arXiv preprint arXiv:2005.14165}}
  (\bibinfo{year}{2020}).

\bibitem{devlin2018bert}
\bibinfo{author}{Devlin, J.}, \bibinfo{author}{Chang, M.-W.},
  \bibinfo{author}{Lee, K.} \& \bibinfo{author}{Toutanova, K.}
\newblock \bibinfo{title}{Bert: Pre-training of deep bidirectional transformers
  for language understanding}.
\newblock \emph{\bibinfo{journal}{arXiv preprint arXiv:1810.04805}}
  (\bibinfo{year}{2018}).

\bibitem{benedetti2019parameterized}
\bibinfo{author}{Benedetti, M.}, \bibinfo{author}{Lloyd, E.},
  \bibinfo{author}{Sack, S.} \& \bibinfo{author}{Fiorentini, M.}
\newblock \bibinfo{title}{Parameterized quantum circuits as machine learning
  models}.
\newblock \emph{\bibinfo{journal}{Quantum Science and Technology}}
  \textbf{\bibinfo{volume}{4}}, \bibinfo{pages}{043001} (\bibinfo{year}{2019}).

\bibitem{du2018expressive}
\bibinfo{author}{Du, Y.}, \bibinfo{author}{Hsieh, M.-H.}, \bibinfo{author}{Liu,
  T.} \& \bibinfo{author}{Tao, D.}
\newblock \bibinfo{title}{The expressive power of parameterized quantum
  circuits}.
\newblock \emph{\bibinfo{journal}{arXiv preprint arXiv:1810.11922}}
  (\bibinfo{year}{2018}).

\bibitem{sharma2020trainability}
\bibinfo{author}{Sharma, K.}, \bibinfo{author}{Cerezo, M.},
  \bibinfo{author}{Cincio, L.} \& \bibinfo{author}{Coles, P.~J.}
\newblock \bibinfo{title}{Trainability of dissipative perceptron-based quantum
  neural networks}.
\newblock \emph{\bibinfo{journal}{arXiv preprint arXiv:2005.12458}}
  (\bibinfo{year}{2020}).

\bibitem{schuld2014quest}
\bibinfo{author}{Schuld, M.}, \bibinfo{author}{Sinayskiy, I.} \&
  \bibinfo{author}{Petruccione, F.}
\newblock \bibinfo{title}{The quest for a quantum neural network}.
\newblock \emph{\bibinfo{journal}{Quantum Information Processing}}
  \textbf{\bibinfo{volume}{13}}, \bibinfo{pages}{2567--2586}
  (\bibinfo{year}{2014}).

\bibitem{killoran2019continuous}
\bibinfo{author}{Killoran, N.} \emph{et~al.}
\newblock \bibinfo{title}{Continuous-variable quantum neural networks}.
\newblock \emph{\bibinfo{journal}{Physical Review Research}}
  \textbf{\bibinfo{volume}{1}}, \bibinfo{pages}{033063} (\bibinfo{year}{2019}).

\bibitem{cong2019quantum}
\bibinfo{author}{Cong, I.}, \bibinfo{author}{Choi, S.} \&
  \bibinfo{author}{Lukin, M.~D.}
\newblock \bibinfo{title}{Quantum convolutional neural networks}.
\newblock \emph{\bibinfo{journal}{Nature Physics}}
  \textbf{\bibinfo{volume}{15}}, \bibinfo{pages}{1273--1278}
  (\bibinfo{year}{2019}).

\bibitem{schuld2019evaluating}
\bibinfo{author}{Schuld, M.}, \bibinfo{author}{Bergholm, V.},
  \bibinfo{author}{Gogolin, C.}, \bibinfo{author}{Izaac, J.} \&
  \bibinfo{author}{Killoran, N.}
\newblock \bibinfo{title}{Evaluating analytic gradients on quantum hardware}.
\newblock \emph{\bibinfo{journal}{Physical Review A}}
  \textbf{\bibinfo{volume}{99}}, \bibinfo{pages}{032331}
  (\bibinfo{year}{2019}).

\bibitem{huembeli2020characterizing}
\bibinfo{author}{Huembeli, P.} \& \bibinfo{author}{Dauphin, A.}
\newblock \bibinfo{title}{Characterizing the loss landscape of variational
  quantum circuits}.
\newblock \emph{\bibinfo{journal}{arXiv preprint arXiv:2008.02785}}
  (\bibinfo{year}{2020}).

\bibitem{fingerhuth2018quantum}
\bibinfo{author}{Fingerhuth, M.}, \bibinfo{author}{Babej, T.} \emph{et~al.}
\newblock \bibinfo{title}{A quantum alternating operator ansatz with hard and
  soft constraints for lattice protein folding}.
\newblock \emph{\bibinfo{journal}{arXiv preprint arXiv:1810.13411}}
  (\bibinfo{year}{2018}).

\bibitem{hadfield2019quantum}
\bibinfo{author}{Hadfield, S.} \emph{et~al.}
\newblock \bibinfo{title}{From the quantum approximate optimization algorithm
  to a quantum alternating operator ansatz}.
\newblock \emph{\bibinfo{journal}{Algorithms}} \textbf{\bibinfo{volume}{12}},
  \bibinfo{pages}{34} (\bibinfo{year}{2019}).

\bibitem{farhi2014quantum}
\bibinfo{author}{Farhi, E.}, \bibinfo{author}{Goldstone, J.} \&
  \bibinfo{author}{Gutmann, S.}
\newblock \bibinfo{title}{A quantum approximate optimization algorithm}.
\newblock \emph{\bibinfo{journal}{arXiv preprint arXiv:1411.4028}}
  (\bibinfo{year}{2014}).

\bibitem{kingma2014adam}
\bibinfo{author}{Kingma, D.~P.} \& \bibinfo{author}{Ba, J.}
\newblock \bibinfo{title}{Adam: A method for stochastic optimization}.
\newblock \emph{\bibinfo{journal}{arXiv preprint arXiv:1412.6980}}
  (\bibinfo{year}{2014}).

\bibitem{ge2020robust}
\bibinfo{author}{Ge, X.}, \bibinfo{author}{Ding, H.}, \bibinfo{author}{Rabitz,
  H.} \& \bibinfo{author}{Wu, R.-B.}
\newblock \bibinfo{title}{Robust quantum control in games: An adversarial
  learning approach}.
\newblock \emph{\bibinfo{journal}{Physical Review A}}
  \textbf{\bibinfo{volume}{101}}, \bibinfo{pages}{052317}
  (\bibinfo{year}{2020}).

\bibitem{palittapongarnpim2017learning}
\bibinfo{author}{Palittapongarnpim, P.}, \bibinfo{author}{Wittek, P.},
  \bibinfo{author}{Zahedinejad, E.}, \bibinfo{author}{Vedaie, S.} \&
  \bibinfo{author}{Sanders, B.~C.}
\newblock \bibinfo{title}{Learning in quantum control: High-dimensional global
  optimization for noisy quantum dynamics}.
\newblock \emph{\bibinfo{journal}{Neurocomputing}}
  \textbf{\bibinfo{volume}{268}}, \bibinfo{pages}{116--126}
  (\bibinfo{year}{2017}).

\bibitem{schuld2020circuit}
\bibinfo{author}{Schuld, M.}, \bibinfo{author}{Bocharov, A.},
  \bibinfo{author}{Svore, K.~M.} \& \bibinfo{author}{Wiebe, N.}
\newblock \bibinfo{title}{Circuit-centric quantum classifiers}.
\newblock \emph{\bibinfo{journal}{Physical Review A}}
  \textbf{\bibinfo{volume}{101}}, \bibinfo{pages}{032308}
  (\bibinfo{year}{2020}).

\bibitem{romero2017quantum}
\bibinfo{author}{Romero, J.}, \bibinfo{author}{Olson, J.~P.} \&
  \bibinfo{author}{Aspuru-Guzik, A.}
\newblock \bibinfo{title}{Quantum autoencoders for efficient compression of
  quantum data}.
\newblock \emph{\bibinfo{journal}{Quantum Science and Technology}}
  \textbf{\bibinfo{volume}{2}}, \bibinfo{pages}{045001} (\bibinfo{year}{2017}).

\bibitem{jones2018quantum}
\bibinfo{author}{Jones, T.} \& \bibinfo{author}{Benjamin, S.~C.}
\newblock \bibinfo{title}{Quantum compilation and circuit optimisation via
  energy dissipation}.
\newblock \emph{\bibinfo{journal}{arXiv preprint arXiv:1811.03147}}
  (\bibinfo{year}{2018}).

\bibitem{nautrup2019optimizing}
\bibinfo{author}{Nautrup, H.~P.}, \bibinfo{author}{Delfosse, N.},
  \bibinfo{author}{Dunjko, V.}, \bibinfo{author}{Briegel, H.~J.} \&
  \bibinfo{author}{Friis, N.}
\newblock \bibinfo{title}{Optimizing quantum error correction codes with
  reinforcement learning}.
\newblock \emph{\bibinfo{journal}{Quantum}} \textbf{\bibinfo{volume}{3}},
  \bibinfo{pages}{215} (\bibinfo{year}{2019}).

\bibitem{baireuther2018machine}
\bibinfo{author}{Baireuther, P.}, \bibinfo{author}{O'Brien, T.~E.},
  \bibinfo{author}{Tarasinski, B.} \& \bibinfo{author}{Beenakker, C.~W.}
\newblock \bibinfo{title}{Machine-learning-assisted correction of correlated
  qubit errors in a topological code}.
\newblock \emph{\bibinfo{journal}{Quantum}} \textbf{\bibinfo{volume}{2}},
  \bibinfo{pages}{48} (\bibinfo{year}{2018}).

\bibitem{bausch2020quantum}
\bibinfo{author}{Bausch, J.} \& \bibinfo{author}{Leditzky, F.}
\newblock \bibinfo{title}{Quantum codes from neural networks}.
\newblock \emph{\bibinfo{journal}{New Journal of Physics}}
  \textbf{\bibinfo{volume}{22}}, \bibinfo{pages}{023005}
  (\bibinfo{year}{2020}).

\bibitem{johnson2017qvector}
\bibinfo{author}{Johnson, P.~D.}, \bibinfo{author}{Romero, J.},
  \bibinfo{author}{Olson, J.}, \bibinfo{author}{Cao, Y.} \&
  \bibinfo{author}{Aspuru-Guzik, A.}
\newblock \bibinfo{title}{Qvector: an algorithm for device-tailored quantum
  error correction}.
\newblock \emph{\bibinfo{journal}{arXiv preprint arXiv:1711.02249}}
  (\bibinfo{year}{2017}).

\bibitem{zhao2021analyzing}
\bibinfo{author}{Zhao, C.} \& \bibinfo{author}{Gao, X.-S.}
\newblock \bibinfo{title}{Analyzing the barren plateau phenomenon in training
  quantum neural networks with the zx-calculus}.
\newblock \emph{\bibinfo{journal}{Quantum}} \textbf{\bibinfo{volume}{5}},
  \bibinfo{pages}{466} (\bibinfo{year}{2021}).

\bibitem{cerezo2021variational}
\bibinfo{author}{Cerezo, M.} \emph{et~al.}
\newblock \bibinfo{title}{Variational quantum algorithms}.
\newblock \emph{\bibinfo{journal}{Nature Reviews Physics}}
  \textbf{\bibinfo{volume}{3}}, \bibinfo{pages}{625--644}
  (\bibinfo{year}{2021}).

\bibitem{larocca2020navigating}
\bibinfo{author}{Larocca, M.}, \bibinfo{author}{Calzetta, E.~A.} \&
  \bibinfo{author}{Wisniacki, D.~A.}
\newblock \bibinfo{title}{Navigating on quantum control solution subspaces}.
\newblock \emph{\bibinfo{journal}{arXiv preprint arXiv:2001.05941}}
  (\bibinfo{year}{2020}).

\bibitem{grant2019initialization}
\bibinfo{author}{Grant, E.}, \bibinfo{author}{Wossnig, L.},
  \bibinfo{author}{Ostaszewski, M.} \& \bibinfo{author}{Benedetti, M.}
\newblock \bibinfo{title}{An initialization strategy for addressing barren
  plateaus in parametrized quantum circuits}.
\newblock \emph{\bibinfo{journal}{Quantum}} \textbf{\bibinfo{volume}{3}},
  \bibinfo{pages}{214} (\bibinfo{year}{2019}).

\bibitem{zhou2020quantum}
\bibinfo{author}{Zhou, L.}, \bibinfo{author}{Wang, S.-T.},
  \bibinfo{author}{Choi, S.}, \bibinfo{author}{Pichler, H.} \&
  \bibinfo{author}{Lukin, M.~D.}
\newblock \bibinfo{title}{Quantum approximate optimization algorithm:
  Performance, mechanism, and implementation on near-term devices}.
\newblock \emph{\bibinfo{journal}{Physical Review X}}
  \textbf{\bibinfo{volume}{10}}, \bibinfo{pages}{021067}
  (\bibinfo{year}{2020}).

\bibitem{pesah2020absence}
\bibinfo{author}{Pesah, A.} \emph{et~al.}
\newblock \bibinfo{title}{Absence of barren plateaus in quantum convolutional
  neural networks}.
\newblock \emph{\bibinfo{journal}{arXiv preprint arXiv:2011.02966}}
  (\bibinfo{year}{2020}).

\bibitem{bharti2020quantum}
\bibinfo{author}{Bharti, K.} \& \bibinfo{author}{Haug, T.}
\newblock \bibinfo{title}{Quantum assisted simulator}.
\newblock \emph{\bibinfo{journal}{arXiv preprint arXiv:2011.06911}}
  (\bibinfo{year}{2020}).

\bibitem{stokes2020quantum}
\bibinfo{author}{Stokes, J.}, \bibinfo{author}{Izaac, J.},
  \bibinfo{author}{Killoran, N.} \& \bibinfo{author}{Carleo, G.}
\newblock \bibinfo{title}{Quantum natural gradient}.
\newblock \emph{\bibinfo{journal}{Quantum}} \textbf{\bibinfo{volume}{4}},
  \bibinfo{pages}{269} (\bibinfo{year}{2020}).

\bibitem{zhang2019self}
\bibinfo{author}{Zhang, H.}, \bibinfo{author}{Goodfellow, I.},
  \bibinfo{author}{Metaxas, D.} \& \bibinfo{author}{Odena, A.}
\newblock \bibinfo{title}{Self-attention generative adversarial networks}.
\newblock In \emph{\bibinfo{booktitle}{International conference on machine
  learning}}, \bibinfo{pages}{7354--7363} (\bibinfo{organization}{PMLR},
  \bibinfo{year}{2019}).

\bibitem{miyato2018spectral}
\bibinfo{author}{Miyato, T.}, \bibinfo{author}{Kataoka, T.},
  \bibinfo{author}{Koyama, M.} \& \bibinfo{author}{Yoshida, Y.}
\newblock \bibinfo{title}{Spectral normalization for generative adversarial
  networks}.
\newblock \emph{\bibinfo{journal}{arXiv preprint arXiv:1802.05957}}
  (\bibinfo{year}{2018}).

\bibitem{karras2017progressive}
\bibinfo{author}{Karras, T.}, \bibinfo{author}{Aila, T.},
  \bibinfo{author}{Laine, S.} \& \bibinfo{author}{Lehtinen, J.}
\newblock \bibinfo{title}{Progressive growing of gans for improved quality,
  stability, and variation}.
\newblock \emph{\bibinfo{journal}{arXiv preprint arXiv:1710.10196}}
  (\bibinfo{year}{2017}).

\bibitem{roth2017stabilizing}
\bibinfo{author}{Roth, K.}, \bibinfo{author}{Lucchi, A.},
  \bibinfo{author}{Nowozin, S.} \& \bibinfo{author}{Hofmann, T.}
\newblock \bibinfo{title}{Stabilizing training of generative adversarial
  networks through regularization}.
\newblock In \emph{\bibinfo{booktitle}{Advances in neural information
  processing systems}}, \bibinfo{pages}{2018--2028} (\bibinfo{year}{2017}).

\bibitem{petzka2017regularization}
\bibinfo{author}{Petzka, H.}, \bibinfo{author}{Fischer, A.} \&
  \bibinfo{author}{Lukovnicov, D.}
\newblock \bibinfo{title}{On the regularization of wasserstein gans}.
\newblock \emph{\bibinfo{journal}{arXiv preprint arXiv:1709.08894}}
  (\bibinfo{year}{2017}).

\bibitem{gao2017wasserstein}
\bibinfo{author}{Gao, R.}, \bibinfo{author}{Chen, X.} \&
  \bibinfo{author}{Kleywegt, A.~J.}
\newblock \bibinfo{title}{Wasserstein distributional robustness and
  regularization in statistical learning}.
\newblock \emph{\bibinfo{journal}{arXiv preprint arXiv:1712.06050}}
  (\bibinfo{year}{2017}).

\bibitem{li2018machine}
\bibinfo{author}{Li, Z.}, \bibinfo{author}{Meier, M.-A.},
  \bibinfo{author}{Hauksson, E.}, \bibinfo{author}{Zhan, Z.} \&
  \bibinfo{author}{Andrews, J.}
\newblock \bibinfo{title}{Machine learning seismic wave discrimination:
  Application to earthquake early warning}.
\newblock \emph{\bibinfo{journal}{Geophysical Research Letters}}
  \textbf{\bibinfo{volume}{45}}, \bibinfo{pages}{4773--4779}
  (\bibinfo{year}{2018}).

\bibitem{xuan2018multiview}
\bibinfo{author}{Xuan, Q.} \emph{et~al.}
\newblock \bibinfo{title}{Multiview generative adversarial network and its
  application in pearl classification}.
\newblock \emph{\bibinfo{journal}{IEEE Transactions on Industrial Electronics}}
  \textbf{\bibinfo{volume}{66}}, \bibinfo{pages}{8244--8252}
  (\bibinfo{year}{2018}).

\bibitem{yi2017dualgan}
\bibinfo{author}{Yi, Z.}, \bibinfo{author}{Zhang, H.}, \bibinfo{author}{Tan,
  P.} \& \bibinfo{author}{Gong, M.}
\newblock \bibinfo{title}{Dualgan: Unsupervised dual learning for
  image-to-image translation}.
\newblock In \emph{\bibinfo{booktitle}{Proceedings of the IEEE international
  conference on computer vision}}, \bibinfo{pages}{2849--2857}
  (\bibinfo{year}{2017}).

\bibitem{elgammal2017can}
\bibinfo{author}{Elgammal, A.}, \bibinfo{author}{Liu, B.},
  \bibinfo{author}{Elhoseiny, M.} \& \bibinfo{author}{Mazzone, M.}
\newblock \bibinfo{title}{Can: Creative adversarial networks, generating" art"
  by learning about styles and deviating from style norms}.
\newblock \emph{\bibinfo{journal}{arXiv preprint arXiv:1706.07068}}
  (\bibinfo{year}{2017}).

\bibitem{wang2018intelligent}
\bibinfo{author}{Wang, Z.}, \bibinfo{author}{Wang, J.} \&
  \bibinfo{author}{Wang, Y.}
\newblock \bibinfo{title}{An intelligent diagnosis scheme based on generative
  adversarial learning deep neural networks and its application to planetary
  gearbox fault pattern recognition}.
\newblock \emph{\bibinfo{journal}{Neurocomputing}}
  \textbf{\bibinfo{volume}{310}}, \bibinfo{pages}{213--222}
  (\bibinfo{year}{2018}).

\bibitem{anand2020experimental}
\bibinfo{author}{Anand, A.}, \bibinfo{author}{Romero, J.},
  \bibinfo{author}{Degroote, M.} \& \bibinfo{author}{Aspuru-Guzik, A.}
\newblock \bibinfo{title}{Experimental demonstration of a quantum generative
  adversarial network for continuous distributions}.
\newblock \emph{\bibinfo{journal}{arXiv preprint arXiv:2006.01976}}
  (\bibinfo{year}{2020}).

\bibitem{ahmed2020quantum}
\bibinfo{author}{Ahmed, S.}, \bibinfo{author}{Mu{\~n}oz, C.~S.},
  \bibinfo{author}{Nori, F.} \& \bibinfo{author}{Kockum, A.~F.}
\newblock \bibinfo{title}{Quantum state tomography with conditional generative
  adversarial networks}.
\newblock \emph{\bibinfo{journal}{arXiv preprint arXiv:2008.03240}}
  (\bibinfo{year}{2020}).

\bibitem{lu2020quantum}
\bibinfo{author}{Lu, S.}, \bibinfo{author}{Duan, L.-M.} \&
  \bibinfo{author}{Deng, D.-L.}
\newblock \bibinfo{title}{Quantum adversarial machine learning}.
\newblock \emph{\bibinfo{journal}{Physical Review Research}}
  \textbf{\bibinfo{volume}{2}}, \bibinfo{pages}{033212} (\bibinfo{year}{2020}).

\bibitem{zeng2019learning}
\bibinfo{author}{Zeng, J.}, \bibinfo{author}{Wu, Y.}, \bibinfo{author}{Liu,
  J.-G.}, \bibinfo{author}{Wang, L.} \& \bibinfo{author}{Hu, J.}
\newblock \bibinfo{title}{Learning and inference on generative adversarial
  quantum circuits}.
\newblock \emph{\bibinfo{journal}{Physical Review A}}
  \textbf{\bibinfo{volume}{99}}, \bibinfo{pages}{052306}
  (\bibinfo{year}{2019}).

\bibitem{romero2019variational}
\bibinfo{author}{Romero, J.} \& \bibinfo{author}{Aspuru-Guzik, A.}
\newblock \bibinfo{title}{Variational quantum generators: Generative
  adversarial quantum machine learning for continuous distributions}.
\newblock \emph{\bibinfo{journal}{arXiv preprint arXiv:1901.00848}}
  (\bibinfo{year}{2019}).

\bibitem{zoufal2019quantum}
\bibinfo{author}{Zoufal, C.}, \bibinfo{author}{Lucchi, A.} \&
  \bibinfo{author}{Woerner, S.}
\newblock \bibinfo{title}{Quantum generative adversarial networks for learning
  and loading random distributions}.
\newblock \emph{\bibinfo{journal}{npj Quantum Information}}
  \textbf{\bibinfo{volume}{5}}, \bibinfo{pages}{1--9} (\bibinfo{year}{2019}).

\bibitem{nakaji2020quantum}
\bibinfo{author}{Nakaji, K.} \& \bibinfo{author}{Yamamoto, N.}
\newblock \bibinfo{title}{Quantum semi-supervised generative adversarial
  network for enhanced data classification}.
\newblock \emph{\bibinfo{journal}{arXiv preprint arXiv:2010.13727}}
  (\bibinfo{year}{2020}).

\bibitem{herr2020anomaly}
\bibinfo{author}{Herr, D.}, \bibinfo{author}{Obert, B.} \&
  \bibinfo{author}{Rosenkranz, M.}
\newblock \bibinfo{title}{Anomaly detection with variational quantum generative
  adversarial networks}.
\newblock \emph{\bibinfo{journal}{arXiv preprint arXiv:2010.10492}}
  (\bibinfo{year}{2020}).

\bibitem{huang2020quantum}
\bibinfo{author}{Huang, B.}, \bibinfo{author}{Symonds, N.~O.} \&
  \bibinfo{author}{von Lilienfeld, O.~A.}
\newblock \bibinfo{title}{Quantum machine learning in chemistry and materials}.
\newblock \emph{\bibinfo{journal}{Handbook of Materials Modeling: Methods:
  Theory and Modeling}} \bibinfo{pages}{1883--1909} (\bibinfo{year}{2020}).

\bibitem{stamatopoulos2020option}
\bibinfo{author}{Stamatopoulos, N.} \emph{et~al.}
\newblock \bibinfo{title}{Option pricing using quantum computers}.
\newblock \emph{\bibinfo{journal}{Quantum}} \textbf{\bibinfo{volume}{4}},
  \bibinfo{pages}{291} (\bibinfo{year}{2020}).

\bibitem{orus2019quantum}
\bibinfo{author}{Orus, R.}, \bibinfo{author}{Mugel, S.} \&
  \bibinfo{author}{Lizaso, E.}
\newblock \bibinfo{title}{Quantum computing for finance: overview and
  prospects}.
\newblock \emph{\bibinfo{journal}{Reviews in Physics}}
  \textbf{\bibinfo{volume}{4}}, \bibinfo{pages}{100028} (\bibinfo{year}{2019}).

\bibitem{kiani2020quantum}
\bibinfo{author}{Kiani, B.~T.}, \bibinfo{author}{Villanyi, A.} \&
  \bibinfo{author}{Lloyd, S.}
\newblock \bibinfo{title}{Quantum medical imaging algorithms}.
\newblock \emph{\bibinfo{journal}{arXiv preprint arXiv:2004.02036}}
  (\bibinfo{year}{2020}).

\bibitem{kiani2020diffeq}
\bibinfo{author}{Kiani, B.~T.} \emph{et~al.}
\newblock \bibinfo{title}{Quantum advantage for differential equation
  analysis}.
\newblock \emph{\bibinfo{journal}{arXiv preprint arXiv:2010.15776}}
  (\bibinfo{year}{2020}).

\bibitem{yao2017quantum}
\bibinfo{author}{Yao, X.-W.} \emph{et~al.}
\newblock \bibinfo{title}{Quantum image processing and its application to edge
  detection: theory and experiment}.
\newblock \emph{\bibinfo{journal}{Physical Review X}}
  \textbf{\bibinfo{volume}{7}}, \bibinfo{pages}{031041} (\bibinfo{year}{2017}).

\bibitem{lloyd2018qaoa}
\bibinfo{author}{Lloyd, S.}
\newblock \bibinfo{title}{Quantum approximate optimization is computationally
  universal}.
\newblock \emph{\bibinfo{journal}{arXiv preprint arXiv:1812.11075}}
  (\bibinfo{year}{2018}).

\bibitem{zhang2020qed}
\bibinfo{author}{Zhang, Y.}, \bibinfo{author}{Zhang, R.} \&
  \bibinfo{author}{Potter, A.~C.}
\newblock \bibinfo{title}{Qed driven qaoa for network-flow optimization}.
\newblock \emph{\bibinfo{journal}{arXiv preprint arXiv:2006.09418}}
  (\bibinfo{year}{2020}).

\bibitem{kandala2017hardware}
\bibinfo{author}{Kandala, A.} \emph{et~al.}
\newblock \bibinfo{title}{Hardware-efficient variational quantum eigensolver
  for small molecules and quantum magnets}.
\newblock \emph{\bibinfo{journal}{Nature}} \textbf{\bibinfo{volume}{549}},
  \bibinfo{pages}{242--246} (\bibinfo{year}{2017}).

\bibitem{parrish2019quantum}
\bibinfo{author}{Parrish, R.~M.}, \bibinfo{author}{Hohenstein, E.~G.},
  \bibinfo{author}{McMahon, P.~L.} \& \bibinfo{author}{Mart{\'\i}nez, T.~J.}
\newblock \bibinfo{title}{Quantum computation of electronic transitions using a
  variational quantum eigensolver}.
\newblock \emph{\bibinfo{journal}{Physical review letters}}
  \textbf{\bibinfo{volume}{122}}, \bibinfo{pages}{230401}
  (\bibinfo{year}{2019}).

\bibitem{monge1781memoire}
\bibinfo{author}{Monge, G.}
\newblock \bibinfo{title}{M{\'e}moire sur la th{\'e}orie des d{\'e}blais et des
  remblais}.
\newblock \emph{\bibinfo{journal}{Histoire de l'Acad{\'e}mie Royale des
  Sciences de Paris}}  (\bibinfo{year}{1781}).

\bibitem{kantorovich1942translocation}
\bibinfo{author}{Kantorovich, L.~V.}
\newblock \bibinfo{title}{On the translocation of masses}.
\newblock In \emph{\bibinfo{booktitle}{Dokl. Akad. Nauk. USSR (NS)}},
  vol.~\bibinfo{volume}{37}, \bibinfo{pages}{199--201} (\bibinfo{year}{1942}).

\bibitem{ambrosio2008gradient}
\bibinfo{author}{Ambrosio, L.}, \bibinfo{author}{Gigli, N.} \&
  \bibinfo{author}{Savar{\'e}, G.}
\newblock \emph{\bibinfo{title}{Gradient flows: in metric spaces and in the
  space of probability measures}} (\bibinfo{publisher}{Springer Science \&
  Business Media}, \bibinfo{year}{2008}).

\bibitem{peyre2019computational}
\bibinfo{author}{Peyr{\'e}, G.} \& \bibinfo{author}{Cuturi, M.}
\newblock \bibinfo{title}{Computational {O}ptimal {T}ransport: {W}ith
  {A}pplications to {D}ata {S}cience}.
\newblock \emph{\bibinfo{journal}{Foundations and Trends{\textregistered} in
  Machine Learning}} \textbf{\bibinfo{volume}{11}}, \bibinfo{pages}{355--607}
  (\bibinfo{year}{2019}).

\bibitem{vershik2013long}
\bibinfo{author}{Vershik, A.~M.}
\newblock \bibinfo{title}{Long history of the {M}onge-{K}antorovich
  transportation problem}.
\newblock \emph{\bibinfo{journal}{The Mathematical Intelligencer}}
  \textbf{\bibinfo{volume}{35}}, \bibinfo{pages}{1--9} (\bibinfo{year}{2013}).

\bibitem{ornstein1973application}
\bibinfo{author}{Ornstein, D.~S.}
\newblock \bibinfo{title}{An application of ergodic theory to probability
  theory}.
\newblock \emph{\bibinfo{journal}{The Annals of Probability}}
  \textbf{\bibinfo{volume}{1}}, \bibinfo{pages}{43--58} (\bibinfo{year}{1973}).

\bibitem{nielsen2002quantum}
\bibinfo{author}{Nielsen, M.~A.} \& \bibinfo{author}{Chuang, I.}
\newblock \bibinfo{title}{Quantum computation and quantum information}
  (\bibinfo{year}{2002}).

\bibitem{carlen2014analog}
\bibinfo{author}{Carlen, E.~A.} \& \bibinfo{author}{Maas, J.}
\newblock \bibinfo{title}{An analog of the 2-{W}asserstein metric in
  non-commutative probability under which the {F}ermionic {F}okker--{P}lanck
  equation is gradient flow for the entropy}.
\newblock \emph{\bibinfo{journal}{Communications in Mathematical Physics}}
  \textbf{\bibinfo{volume}{331}}, \bibinfo{pages}{887--926}
  (\bibinfo{year}{2014}).

\bibitem{carlen2017gradient}
\bibinfo{author}{Carlen, E.~A.} \& \bibinfo{author}{Maas, J.}
\newblock \bibinfo{title}{Gradient flow and entropy inequalities for quantum
  {M}arkov semigroups with detailed balance}.
\newblock \emph{\bibinfo{journal}{Journal of Functional Analysis}}
  \textbf{\bibinfo{volume}{273}}, \bibinfo{pages}{1810--1869}
  (\bibinfo{year}{2017}).

\bibitem{carlen2020non}
\bibinfo{author}{Carlen, E.~A.} \& \bibinfo{author}{Maas, J.}
\newblock \bibinfo{title}{Non-commutative calculus, optimal transport and
  functional inequalities in dissipative quantum systems}.
\newblock \emph{\bibinfo{journal}{Journal of Statistical Physics}}
  \textbf{\bibinfo{volume}{178}}, \bibinfo{pages}{319--378}
  (\bibinfo{year}{2020}).

\bibitem{rouze2019concentration}
\bibinfo{author}{Rouz{\'e}, C.} \& \bibinfo{author}{Datta, N.}
\newblock \bibinfo{title}{Concentration of quantum states from quantum
  functional and transportation cost inequalities}.
\newblock \emph{\bibinfo{journal}{Journal of Mathematical Physics}}
  \textbf{\bibinfo{volume}{60}}, \bibinfo{pages}{012202}
  (\bibinfo{year}{2019}).

\bibitem{datta2020relating}
\bibinfo{author}{Datta, N.} \& \bibinfo{author}{Rouz{\'e}, C.}
\newblock \bibinfo{title}{Relating relative entropy, optimal transport and
  {F}isher information: {A} quantum {H}{W}{I} inequality}.
\newblock \emph{\bibinfo{journal}{Annales Henri Poincar{\'e}}}
  \textbf{\bibinfo{volume}{21}}, \bibinfo{pages}{2115--2150}
  (\bibinfo{year}{2020}).

\bibitem{van2020geometrical}
\bibinfo{author}{Van~Vu, T.} \& \bibinfo{author}{Hasegawa, Y.}
\newblock \bibinfo{title}{Geometrical {B}ounds of the {I}rreversibility in
  {M}arkovian {S}ystems}.
\newblock \emph{\bibinfo{journal}{arXiv preprint arXiv:2005.02871}}
  (\bibinfo{year}{2020}).

\bibitem{de2018conditional}
\bibinfo{author}{De~Palma, G.} \& \bibinfo{author}{Huber, S.}
\newblock \bibinfo{title}{The conditional {E}ntropy {P}ower {I}nequality for
  quantum additive noise channels}.
\newblock \emph{\bibinfo{journal}{Journal of Mathematical Physics}}
  \textbf{\bibinfo{volume}{59}}, \bibinfo{pages}{122201}
  (\bibinfo{year}{2018}).

\bibitem{gao2020fisher}
\bibinfo{author}{Gao, L.}, \bibinfo{author}{Junge, M.} \&
  \bibinfo{author}{LaRacuente, N.}
\newblock \bibinfo{title}{Fisher information and logarithmic sobolev inequality
  for matrix-valued functions}.
\newblock \emph{\bibinfo{journal}{Annales Henri Poincar{\'e}}}
  \textbf{\bibinfo{volume}{21}}, \bibinfo{pages}{3409--3478}
  (\bibinfo{year}{2020}).

\bibitem{chen2017matricial}
\bibinfo{author}{Chen, Y.}, \bibinfo{author}{Georgiou, T.~T.},
  \bibinfo{author}{Ning, L.} \& \bibinfo{author}{Tannenbaum, A.}
\newblock \bibinfo{title}{Matricial {W}asserstein-1 distance}.
\newblock \emph{\bibinfo{journal}{IEEE control systems letters}}
  \textbf{\bibinfo{volume}{1}}, \bibinfo{pages}{14--19} (\bibinfo{year}{2017}).

\bibitem{ryu2018vector}
\bibinfo{author}{Ryu, E.~K.}, \bibinfo{author}{Chen, Y.}, \bibinfo{author}{Li,
  W.} \& \bibinfo{author}{Osher, S.}
\newblock \bibinfo{title}{Vector and matrix optimal mass transport: theory,
  algorithm, and applications}.
\newblock \emph{\bibinfo{journal}{SIAM Journal on Scientific Computing}}
  \textbf{\bibinfo{volume}{40}}, \bibinfo{pages}{A3675--A3698}
  (\bibinfo{year}{2018}).

\bibitem{chen2018matrix}
\bibinfo{author}{Chen, Y.}, \bibinfo{author}{Georgiou, T.~T.} \&
  \bibinfo{author}{Tannenbaum, A.}
\newblock \bibinfo{title}{Matrix optimal mass transport: a quantum mechanical
  approach}.
\newblock \emph{\bibinfo{journal}{IEEE Transactions on Automatic Control}}
  \textbf{\bibinfo{volume}{63}}, \bibinfo{pages}{2612--2619}
  (\bibinfo{year}{2018}).

\bibitem{chen2018wasserstein}
\bibinfo{author}{Chen, Y.}, \bibinfo{author}{Georgiou, T.~T.} \&
  \bibinfo{author}{Tannenbaum, A.}
\newblock \bibinfo{title}{Wasserstein geometry of quantum states and optimal
  transport of matrix-valued measures}.
\newblock In \emph{\bibinfo{booktitle}{Emerging Applications of Control and
  Systems Theory}}, \bibinfo{pages}{139--150} (\bibinfo{publisher}{Springer},
  \bibinfo{year}{2018}).

\bibitem{agredo2013wasserstein}
\bibinfo{author}{Agredo, J.}
\newblock \bibinfo{title}{A {W}asserstein-type distance to measure deviation
  from equilibrium of quantum {M}arkov semigroups}.
\newblock \emph{\bibinfo{journal}{Open Systems \& Information Dynamics}}
  \textbf{\bibinfo{volume}{20}}, \bibinfo{pages}{1350009}
  (\bibinfo{year}{2013}).

\bibitem{agredo2016exponential}
\bibinfo{author}{Agredo, J.}
\newblock \bibinfo{title}{On exponential convergence of generic quantum
  {M}arkov semigroups in a {W}asserstein-type distance}.
\newblock \emph{\bibinfo{journal}{International Journal of Pure and Applied
  Mathematics}} \textbf{\bibinfo{volume}{107}}, \bibinfo{pages}{909--925}
  (\bibinfo{year}{2016}).

\bibitem{ikeda2020foundation}
\bibinfo{author}{Ikeda, K.}
\newblock \bibinfo{title}{Foundation of quantum optimal transport and
  applications}.
\newblock \emph{\bibinfo{journal}{Quantum Information Processing}}
  \textbf{\bibinfo{volume}{19}}, \bibinfo{pages}{25} (\bibinfo{year}{2020}).

\bibitem{golse2016mean}
\bibinfo{author}{Golse, F.}, \bibinfo{author}{Mouhot, C.} \&
  \bibinfo{author}{Paul, T.}
\newblock \bibinfo{title}{On the mean field and classical limits of quantum
  mechanics}.
\newblock \emph{\bibinfo{journal}{Communications in Mathematical Physics}}
  \textbf{\bibinfo{volume}{343}}, \bibinfo{pages}{165--205}
  (\bibinfo{year}{2016}).

\bibitem{caglioti2018towards}
\bibinfo{author}{Caglioti, E.}, \bibinfo{author}{Golse, F.} \&
  \bibinfo{author}{Paul, T.}
\newblock \bibinfo{title}{Towards optimal transport for quantum densities}
  (\bibinfo{year}{2018}).
\newblock \bibinfo{note}{Preprint}.

\bibitem{golse2018quantum}
\bibinfo{author}{Golse, F.}
\newblock \bibinfo{title}{The quantum {N}-body problem in the mean-field and
  semiclassical regime}.
\newblock \emph{\bibinfo{journal}{Philosophical Transactions of the Royal
  Society A: Mathematical, Physical and Engineering Sciences}}
  \textbf{\bibinfo{volume}{376}}, \bibinfo{pages}{20170229}
  (\bibinfo{year}{2018}).

\bibitem{golse2017schrodinger}
\bibinfo{author}{Golse, F.} \& \bibinfo{author}{Paul, T.}
\newblock \bibinfo{title}{The {S}chr{\"o}dinger equation in the mean-field and
  semiclassical regime}.
\newblock \emph{\bibinfo{journal}{Archive for Rational Mechanics and Analysis}}
  \textbf{\bibinfo{volume}{223}}, \bibinfo{pages}{57--94}
  (\bibinfo{year}{2017}).

\bibitem{golse2018wave}
\bibinfo{author}{Golse, F.} \& \bibinfo{author}{Paul, T.}
\newblock \bibinfo{title}{Wave packets and the quadratic {M}onge--{K}antorovich
  distance in quantum mechanics}.
\newblock \emph{\bibinfo{journal}{Comptes Rendus Mathematique}}
  \textbf{\bibinfo{volume}{356}}, \bibinfo{pages}{177--197}
  (\bibinfo{year}{2018}).

\bibitem{caglioti2019quantum}
\bibinfo{author}{Caglioti, E.}, \bibinfo{author}{Golse, F.} \&
  \bibinfo{author}{Paul, T.}
\newblock \bibinfo{title}{{Q}uantum {O}ptimal {T}ransport is {C}heaper}.
\newblock \emph{\bibinfo{journal}{Journal of Statistical Physics}}
  (\bibinfo{year}{2020}).

\bibitem{de2019quantum}
\bibinfo{author}{De~Palma, G.} \& \bibinfo{author}{Trevisan, D.}
\newblock \bibinfo{title}{Quantum optimal transport with quantum channels}.
\newblock \emph{\bibinfo{journal}{Annales Henri Poincar{\'e}}}
  \textbf{\bibinfo{volume}{22}}, \bibinfo{pages}{3199--3234}
  (\bibinfo{year}{2021}).

\bibitem{duvenhage2018balance}
\bibinfo{author}{Duvenhage, R.} \& \bibinfo{author}{Snyman, M.}
\newblock \bibinfo{title}{Balance between quantum {M}arkov semigroups}.
\newblock \emph{\bibinfo{journal}{Annales Henri Poincar{\'e}}}
  \textbf{\bibinfo{volume}{19}}, \bibinfo{pages}{1747--1786}
  (\bibinfo{year}{2018}).

\bibitem{agredo2017quantum}
\bibinfo{author}{Agredo, J.} \& \bibinfo{author}{Fagnola, F.}
\newblock \bibinfo{title}{On quantum versions of the classical {W}asserstein
  distance}.
\newblock \emph{\bibinfo{journal}{Stochastics}} \textbf{\bibinfo{volume}{89}},
  \bibinfo{pages}{910--922} (\bibinfo{year}{2017}).

\bibitem{zyczkowski1998monge}
\bibinfo{author}{Zyczkowski, K.} \& \bibinfo{author}{Slomczynski, W.}
\newblock \bibinfo{title}{The {M}onge distance between quantum states}.
\newblock \emph{\bibinfo{journal}{Journal of Physics A: Mathematical and
  General}} \textbf{\bibinfo{volume}{31}}, \bibinfo{pages}{9095}
  (\bibinfo{year}{1998}).

\bibitem{zyczkowski2001monge}
\bibinfo{author}{Zyczkowski, K.} \& \bibinfo{author}{Slomczynski, W.}
\newblock \bibinfo{title}{The {M}onge metric on the sphere and geometry of
  quantum states}.
\newblock \emph{\bibinfo{journal}{Journal of Physics A: Mathematical and
  General}} \textbf{\bibinfo{volume}{34}}, \bibinfo{pages}{6689}
  (\bibinfo{year}{2001}).

\bibitem{bengtsson2017geometry}
\bibinfo{author}{Bengtsson, I.} \& \bibinfo{author}{{\.Z}yczkowski, K.}
\newblock \emph{\bibinfo{title}{Geometry of {Q}uantum {S}tates: {A}n
  {I}ntroduction to {Q}uantum {E}ntanglement}} (\bibinfo{publisher}{Cambridge
  University Press}, \bibinfo{year}{2017}).

\bibitem{mathieu2014fast}
\bibinfo{author}{Mathieu, M.} \& \bibinfo{author}{LeCun, Y.}
\newblock \bibinfo{title}{Fast approximation of rotations and hessians
  matrices}.
\newblock \emph{\bibinfo{journal}{arXiv preprint arXiv:1404.7195}}
  (\bibinfo{year}{2014}).

\bibitem{jing2017tunable}
\bibinfo{author}{Jing, L.} \emph{et~al.}
\newblock \bibinfo{title}{Tunable efficient unitary neural networks (eunn) and
  their application to rnns}.
\newblock In \emph{\bibinfo{booktitle}{International Conference on Machine
  Learning}}, \bibinfo{pages}{1733--1741} (\bibinfo{organization}{PMLR},
  \bibinfo{year}{2017}).

\bibitem{dao2019learning}
\bibinfo{author}{Dao, T.}, \bibinfo{author}{Gu, A.}, \bibinfo{author}{Eichhorn,
  M.}, \bibinfo{author}{Rudra, A.} \& \bibinfo{author}{R{\'e}, C.}
\newblock \bibinfo{title}{Learning fast algorithms for linear transforms using
  butterfly factorizations}.
\newblock \emph{\bibinfo{journal}{Proceedings of machine learning research}}
  \textbf{\bibinfo{volume}{97}}, \bibinfo{pages}{1517} (\bibinfo{year}{2019}).

\bibitem{clements2016optimal}
\bibinfo{author}{Clements, W.~R.}, \bibinfo{author}{Humphreys, P.~C.},
  \bibinfo{author}{Metcalf, B.~J.}, \bibinfo{author}{Kolthammer, W.~S.} \&
  \bibinfo{author}{Walmsley, I.~A.}
\newblock \bibinfo{title}{Optimal design for universal multiport
  interferometers}.
\newblock \emph{\bibinfo{journal}{Optica}} \textbf{\bibinfo{volume}{3}},
  \bibinfo{pages}{1460--1465} (\bibinfo{year}{2016}).

\bibitem{shen2017deep}
\bibinfo{author}{Shen, Y.} \emph{et~al.}
\newblock \bibinfo{title}{Deep learning with coherent nanophotonic circuits}.
\newblock \emph{\bibinfo{journal}{Nature Photonics}}
  \textbf{\bibinfo{volume}{11}}, \bibinfo{pages}{441} (\bibinfo{year}{2017}).

\bibitem{verdon2017quantum}
\bibinfo{author}{Verdon, G.}, \bibinfo{author}{Broughton, M.} \&
  \bibinfo{author}{Biamonte, J.}
\newblock \bibinfo{title}{A quantum algorithm to train neural networks using
  low-depth circuits}.
\newblock \emph{\bibinfo{journal}{arXiv preprint arXiv:1712.05304}}
  (\bibinfo{year}{2017}).

\bibitem{wang2018quantum}
\bibinfo{author}{Wang, Z.}, \bibinfo{author}{Hadfield, S.},
  \bibinfo{author}{Jiang, Z.} \& \bibinfo{author}{Rieffel, E.~G.}
\newblock \bibinfo{title}{Quantum approximate optimization algorithm for
  maxcut: A fermionic view}.
\newblock \emph{\bibinfo{journal}{Physical Review A}}
  \textbf{\bibinfo{volume}{97}}, \bibinfo{pages}{022304}
  (\bibinfo{year}{2018}).

\bibitem{hodson2019portfolio}
\bibinfo{author}{Hodson, M.}, \bibinfo{author}{Ruck, B.}, \bibinfo{author}{Ong,
  H.}, \bibinfo{author}{Garvin, D.} \& \bibinfo{author}{Dulman, S.}
\newblock \bibinfo{title}{Portfolio rebalancing experiments using the quantum
  alternating operator ansatz}.
\newblock \emph{\bibinfo{journal}{arXiv preprint arXiv:1911.05296}}
  (\bibinfo{year}{2019}).

\bibitem{chancellor2019domain}
\bibinfo{author}{Chancellor, N.}
\newblock \bibinfo{title}{Domain wall encoding of discrete variables for
  quantum annealing and qaoa}.
\newblock \emph{\bibinfo{journal}{Quantum Science and Technology}}
  \textbf{\bibinfo{volume}{4}}, \bibinfo{pages}{045004} (\bibinfo{year}{2019}).

\bibitem{bergholm2018pennylane}
\bibinfo{author}{Bergholm, V.} \emph{et~al.}
\newblock \bibinfo{title}{Pennylane: Automatic differentiation of hybrid
  quantum-classical computations}.
\newblock \emph{\bibinfo{journal}{arXiv preprint arXiv:1811.04968}}
  (\bibinfo{year}{2018}).

\bibitem{tensorflow2015-whitepaper}
\bibinfo{author}{Abadi, M.} \emph{et~al.}
\newblock \bibinfo{title}{{TensorFlow}: Large-scale machine learning on
  heterogeneous systems} (\bibinfo{year}{2015}).
\newblock \urlprefix\url{https://www.tensorflow.org/}.
\newblock \bibinfo{note}{Software available from tensorflow.org}.

\bibitem{paszke2019pytorch}
\bibinfo{author}{Paszke, A.} \emph{et~al.}
\newblock \bibinfo{title}{Pytorch: An imperative style, high-performance deep
  learning library}.
\newblock In \emph{\bibinfo{booktitle}{Advances in neural information
  processing systems}}, \bibinfo{pages}{8026--8037} (\bibinfo{year}{2019}).

\end{thebibliography}

\clearpage
\begin{appendices}

\section{Related Works}\label{app:rw}

\subsection{Loss landscape in quantum machine learning}
Prior research has theoretically and numerically analyzed the typical properties of loss landscapes in quantum machine learning and control settings. Notably, for commonly used cost functions, prior work has proved and numerically shown the existence of ``barren plateaus'' characterized by exponentially decaying gradients for large depth quantum parameterized circuits \cite{mcclean2018barren,zhao2021analyzing,cerezo2021variational}, cost functions with global observables \cite{cerezo2020cost}, and noisy circuits  \cite{wang2020noise}. Furthermore, research in quantum control theory has identified the presence of traps when the control landscape is constrained \cite{moore2012exploring,pechen2011there,larocca2020navigating}. 

Various attempts have been made to potentially avoid these roadblocks. These include methods for initializing circuit parameters \cite{grant2019initialization,zhou2020quantum}, algorithms for layer-wise training \cite{skolik2020layerwise}, and choosing ansatzes that can avoid decaying gradients \cite{pesah2020absence,bharti2020quantum}. Despite these efforts, prior work \cite{cerezo2020cost} has shown that barren plateaus are unavoidable unless cost functions are appropriately chosen to be local. Thus more pertinent to our work are prior studies that have specifically designed loss metrics or gradient-based algorithms that help avoid issues with barren plateaus. This includes Ref. \cite{cerezo2020cost}, which considers local operator loss functions, but not necessarily distance metrics. The quantum EM distance, when used as a cost function, is biased towards local operators, gaining the benefits of using local cost functions outlined in \cite{cerezo2020cost}. Beyond direct changes to the cost functions, Ref. \cite{huembeli2020characterizing} includes second order derivatives in optimizing the loss function to better navigate flat loss landscapes, and Ref. \cite{stokes2020quantum} constructs an algorithm to optimize over the Fubini-study metric tensor. Though not analyzed here, these methods can be used in tandem with our qWGAN to improve optimization of circuit parameters and increase rates of convergence.

\subsection{Classical and quantum generative adversarial networks}
Generative models, as their name indicates, aim to generate a target object or produce samples from a target distribution by approximating the given target through a learning procedure. In the quantum setting, one popular variant for generative models are Born machines whose cost function is measured by comparing a target classical distribution to the sample distribution of a measurement from a variational quantum circuit \cite{liu2018differentiable,benedetti2019generative,coyle2020born}. Ref. \cite{benedetti2019generative} considers the classical EM distance in their evaluation of the cost function which compares the sampled distribution of the quantum computer to the target distribution.

One commonly used generative algorithm is the generative adversarial network (GAN), a classical algorithm first introduced in \cite{goodfellow2014generative}. 
Most relevant to the current work, Ref. \cite{arjovsky2017wasserstein} constructed the first classical Wasserstein GAN employing an earth mover's distance. Later work improved upon the stability and training of the original Wasserstein GAN by, for example, constructing improved discriminators and generators \cite{gulrajani2017improved,zhang2019self,miyato2018spectral}, progressively adding layers during training \cite{karras2017progressive}, and employing various regularization techniques \cite{roth2017stabilizing,petzka2017regularization,gao2017wasserstein}. In the classical literature, GANs have been extensively used in many real-world applications \cite{li2018machine,xuan2018multiview,yi2017dualgan,elgammal2017can,wang2018intelligent}.

In the quantum setting, quantum GANs were first proposed by Refs. \cite{dallaire2018quantum,lloyd2018quantum}. Simple experiments were performed showing the power of quantum GANs in learning quantum data for relatively small systems that can be simulated or experimentally analyzed \cite{benedetti2019adversarial,hu2019quantum,anand2020experimental,ahmed2020quantum,lu2020quantum}. Hybrid classical-quantum GANs, fully classical in the discriminator, generator, and/or loss function, were proposed in Refs. \cite{zeng2019learning,romero2019variational,zoufal2019quantum,nakaji2020quantum}. Ref. \cite{chakrabarti2019quantum} proposed a version of a quantum Wasserstein GAN (qWGAN), though the employed earth mover's distance is unitarily invariant (see \autoref{sec:QW} for details). Ref. \cite{herr2020anomaly} also proposed a qWGAN structure with a classical discriminator and the classical EM distance as their loss function. Our work differs from both of these prior qWGAN papers in that it implements the first qWGAN with a quantum EM distance. Some early experimental demonstrations of quantum GANs have also been performed on various different systems \cite{huang2020quantum,anand2020experimental,stamatopoulos2020option,hu2019quantum}.

\subsection{Applications of quantum machine learning}
Most of the work in quantum machine learning has focused on finding useful applications for quantum machine learning. These include applications in finance \cite{orus2019quantum,stamatopoulos2020option}, chemistry \cite{huang2020quantum}, and post-processing quantum outputs \cite{kiani2020quantum,kiani2020diffeq,yao2017quantum}. Beyond quantum GANs, there has been a focus in recent years on developing near term quantum algorithms potentially implementable on quantum computers with around 100 qubits. Among the most promising candidates include the quantum approximate optimization algorithm \cite{farhi2014quantum,lloyd2018qaoa,zhang2020qed}, the variational quantum eigensolver \cite{kandala2017hardware,parrish2019quantum}, and quantum GANs discussed earlier.

\section{The Classical Earth Mover's Distance}
\label{app:clEM}
The classical earth mover's (EM) distance, also called Monge-Kantorovich distance, is a distance between probability distributions on a metric space which dates back to Monge \cite{monge1781memoire} and has its roots in the theory of optimal mass transport.
Let $p,\,q$ be probability distributions on the metric space $\mathcal{X}$, which for simplicity we will assume to be finite, and let $d$ be the distance on $\mathcal{X}$.
Following the Kantorovich's formulation of the EM distance \cite{kantorovich1942translocation}, we define the set of the \emph{couplings} between $p$ and $q$ as the set of the probability distributions on two copies of $\mathcal{X}$ with marginals equal to $p$ and $q$, respectively.
In the interpretation of mass transport, $p$ and $q$ are considered as distributions of a unit amount of mass, and any coupling $\pi$ prescribes a plan to transform the distribution $p$ into the distribution $q$, in the sense that $\pi(x,y)$
is the amount of mass that is moved from $x$ to $y$.
Assuming that the cost of moving a unit of mass from $x$ to $y$ is equal to $d(x,y)$, the cost of the coupling $\pi$ is equal to $\sum_{x,\,y\in\mathcal{X}}\pi(x,y)\,d(x,y)$, \emph{i.e.}, to the expectation value of the distance with respect to $\pi$.
The EM distance between $p$ and $q$ is given by the minimum cost among all the couplings between $p$ and $q$.
The EM distance has been generalized to a transport cost equal to a power of $d$, leading to the family of the Wasserstein distances of order $\alpha$, of which the $\alpha=1$ case recovers the EM distance.
The exploration of the Wasserstein distances has led to the creation of an extremely fruitful field in mathematical analysis, with applications ranging from differential geometry and partial differential equations to machine learning \cite{villani2008optimal, ambrosio2008gradient, peyre2019computational,vershik2013long}.

The EM distance can be considered as a generalization of the total variation distance.
Indeed, the EM distance recovers the total variation distance when the distance $d$ on $\mathcal{X}$ is the trivial distance for which all the elements of $\mathcal{X}$ are equivalent, \emph{i.e.}, $d(x,y)=1$ for any $x\neq y\in\mathcal{X}$.

When $\mathcal{X}$ is a set of the strings of $n$ bits, the natural choice for $d$ is the Hamming distance, given by the number of different bits.
In this case, the EM distance is also known as Ornstein's $\bar{d}$ distance \cite{ornstein1973application}.

\section{Quantum Mechanics and Qubits}\label{app:QM}
Any quantum system has an associated Hilbert space.
If the Hilbert space has finite dimension $N$, it is always isomorphic to $\mathbb{C}^N$.
For the sake of simplicity, we restrict our discussion to this case.

We denote a column vector in $\mathbb{C}^N$ with $\ket{\cdot}$, where $\cdot$ is a label for the vector.
We will mostly consider vectors with unit norm.
For any $\ket{\psi}\in\mathbb{C}^N$, we denote with $\bra{\psi}\in\left(\mathbb{C}^N\right)^*$ the row vector whose entries are the complex conjugates of the entries of $\ket{\psi}$.
Following the usual rule for matrix multiplication, $\braket{\cdot|\cdot}$ denotes the canonical Hermitian inner product of $\mathbb{C}^N$, defined to be antilinear in the first entry and linear in the second.

A \emph{quantum state} is the quantum counterpart of a probability distribution on a set of $N$ elements, and is a positive semidefinite Hermitian matrix in $\mathbb{C}^{N\times N}$ with unit trace.
A quantum state is \emph{pure} if it cannot be expressed as a nontrivial convex combination of quantum states.
This is the case iff the quantum state is an orthogonal projector with rank one, \emph{i.e.}, if it can be expressed as $|\psi\rangle\langle\psi|$ for some unit vector $\ket{\psi}\in\mathbb{C}^N$.
With some abuse of notation, we call also the unit vectors in $\mathbb{C}^N$ quantum states, formally meaning the associated orthogonal projectors.
Similarly, we call the inner product between two pure quantum states the inner product between the associated unit vectors.
A quantum state is called \emph{mixed} if it is not pure.
Two quantum states are called \emph{orthogonal} if the corresponding supports are orthogonal.
Any mixed quantum state can be expressed as a convex combination of mutually orthogonal pure quantum states.

An \emph{observable} is the quantum counterpart of a real-valued function on a set of $N$ elements, and is given by an $N\times N$ Hermitian matrix.
The expectation value of the observable $H$ on the quantum state $\rho$ is given by $\mathrm{Tr}\left[\rho\,H\right]$.

The Hilbert space associated to a composite quantum system is the tensor product of the Hilbert spaces associated to each subsystem.
Let $\rho$ be a quantum state of the composite quantum system with Hilbert space $\mathbb{C}^{N_1}\otimes\mathbb{C}^{N_2}$, \emph{i.e.}, a Hermitian matrix in $\mathbb{C}^{N_1\times N_1}\otimes\mathbb{C}^{N_2\times N_2}$.
We denote with $\rho_1$ the \emph{marginal} state of $\rho$ on the first subsystem, \emph{i.e.}, the quantum state in $\mathbb{C}^{N_1\times N_1}$ such that $\mathrm{Tr}\left[\rho_1\,H\right] = \mathrm{Tr}\left[\rho\left(H\otimes\mathbb{I}_{N_2}\right)\right]$ for any quantum observable $H$ of $\mathbb{C}^{N_1}$.
$\rho_1$ is equal to the partial trace of $\rho$ over the second subsystem: $\rho_1 = \mathrm{Tr}_2\rho$.

In this paper, we focus on a quantum system composed of $n$ qubits.
A \emph{qubit} is the quantum system associated to the Hilbert space $\mathbb{C}^2$.
We denote with $\ket{0},\,\ket{1}$ the vectors of its canonical basis, which is also called the computational basis.
The Hilbert space of $n$ qubits is $\left(\mathbb{C}^2\right)^{\otimes n}$, and is isomorphic to $\mathbb{C}^N$ with $N=2^n$.
The computational basis of $\left(\mathbb{C}^2\right)^{\otimes n}$ is $\left\{\ket{x_1}\otimes\ldots\otimes\ket{x_n}:x\in\{0,1\}^n\right\}$.
By the sake of a simpler notation, we denote each vector $\ket{x_1}\otimes\ldots\otimes\ket{x_n}$ with $\ket{x}$, and we set $\ket{0}^{\otimes n} = \ket{0_n},\,\ket{1}^{\otimes n} = \ket{1_n}$.
We denote with $\mathcal{O}_n$ the set of the observables of $\left(\mathbb{C}^2\right)^{\otimes n}$.
We say that a linear operator on $\left(\mathbb{C}^2\right)^{\otimes n}$ acts on the $i$-th qubit if it is equal to a $2\times2$ matrix acting on the $i$-th qubit tensored with the identity operator acting on the remaining $n-1$ qubits.
The definition of a linear operator acting on a subset of qubits is analogous.

Perhaps the most important observables used and studied in quantum computation are the Pauli matrices. Together with the identity matrix, the Pauli matrices shown below form a basis for the observables on $\mathbb{C}^2$ (\textit{i.e.,} one qubit). 
\begin{align}
\sigma_X = &
\begin{pmatrix}0&1\\1&0\end{pmatrix} \\
\sigma_Y = &
\begin{pmatrix}0&-i\\i&0\end{pmatrix} \\
\sigma_Z = &
\begin{pmatrix}1&0\\0&-1\end{pmatrix} 
\end{align}
A single Pauli observable can act as a measurement on one qubit; however, multiple qubits can be measured by a ``Pauli string" represented by a set of Pauli matrices placed in tensor product form (\textit{e.g.,} $\sigma_X \otimes \sigma_Y \otimes I \otimes \sigma_X$ or equivalently the string 'XYIX').

Pauli operators are often used as parameterized quantum gates. In their parameterized form:
\begin{equation}
    R_P(t) = e^{-it \sigma_P/2} = \cos{\left(\frac{t}{2}\right)} I - i \sin{\left(\frac{t}{2}\right)} \sigma_P
\end{equation}
where subscript $P \in \{X,Y,Z\}$ indicates the specific Pauli operator chosen. 

In our exposition, we outline the computations performed on a quantum computer as quantum circuits, which are models representing a computation as a sequence of reversible quantum gates and measurement operators. Quantum circuits contain $n$-bit registers and the sequence of gates are applied accordingly to the qubits in the register. For further details on how to read quantum circuits, the reader is referred to the book \cite{nielsen2002quantum}.

\section{Quantum Generalizations of Wasserstein Distances}\label{sec:QW}
Several quantum generalizations of optimal transport distances have been proposed.
One line of research by Carlen, Maas, Datta and Rouz\'e \cite{carlen2014analog,carlen2017gradient,carlen2020non,rouze2019concentration,datta2020relating,van2020geometrical} defines a quantum Wasserstein distance of order 2 from a Riemannian metric on the space of quantum states based on a quantum analog of a differential structure.
This quantum Wasserstein distance is intimately linked to both entropy and Fisher information \cite{datta2020relating}, and has led to determine the rate of convergence of the quantum Ornstein-Uhlenbeck semigroup \cite{carlen2017gradient,de2018conditional}.
Exploiting their quantum differential structure, Refs. \cite{rouze2019concentration,carlen2020non,gao2020fisher} also define a quantum generalization of the Lipschitz constant and of the earth mover's distance.
Alternative definitions of quantum earth mover's distances based on a quantum differential structure are proposed in Refs. \cite{chen2017matricial,ryu2018vector,chen2018matrix,chen2018wasserstein}.
Refs. \cite{agredo2013wasserstein,agredo2016exponential,ikeda2020foundation} propose quantum earth mover's distances based on a distance between the vectors of the canonical basis.

Another line of research by Golse, Mouhot, Paul and Caglioti \cite{golse2016mean,caglioti2018towards,golse2018quantum,golse2017schrodinger,golse2018wave,caglioti2019quantum} arose in the context of the study of the semiclassical limit of quantum mechanics and defines a family of quantum Wasserstein distances of order 2 built on the notion of couplings.
A coupling between the quantum states $\rho$ and $\sigma$ of $\mathbb{C}^N$ is a quantum state $\Pi$ of $\left(\mathbb{C}^N\right)^{\otimes2}$ whose marginal states on the first and on the second subsystems are equal to $\rho$ and $\sigma$, respectively.
The transport cost of the coupling $\Pi$ is $\mathrm{Tr}\left[\Pi\,C\right]$, where $C$ is a suitable positive semidefinite $N^2\times N^2$ cost matrix.
Different choices of $C$ will lead to different distances.
The square distance between $\rho$ and $\sigma$ is defined as the minimum cost among all the couplings between $\rho$ and $\sigma$.
Refs. \cite{golse2016mean,caglioti2018towards,golse2018quantum,golse2017schrodinger,golse2018wave,caglioti2019quantum} consider the case of a quantum harmonic oscillator, which is actually infinite dimensional, and choose as cost matrix the quantum analog of the square Euclidean distance:
\begin{equation}
    C = \left(Q_1 - Q_2\right)^2 + \left(P_1 - P_2\right)^2\,,
\end{equation}
where $Q_{1,2}$ and $P_{1,2}$ are the position and momentum operators of the two subsystems, respectively.
However, the resulting distance has the undesirable property that the distance between a quantum state and itself may not be zero.
Ref. \cite{chakrabarti2019quantum} notices that the distance between any quantum state and itself is zero whenever the support of the cost matrix $C$ is contained in the antisymmetric subspace with respect to the swap of the two subsystems of $\left(\mathbb{C}^N\right)^{\otimes2}$.
Therefore, Ref. \cite{chakrabarti2019quantum} chooses the orthogonal projector onto the antisymmetric subspace as cost matrix, and employs the resulting distance as a cost function for quantum GANs.
We stress that this distance is unitarily invariant.
Indeed, for any $N\times N$ unitary matrix $U$, if $\Pi$ is a coupling between the quantum states $\rho$ and $\sigma$, then $U^{\otimes2}\,\Pi\,U^{\dag\otimes2}$ is a coupling between $U\,\rho\,U^\dag$ and $U\,\sigma\,U^\dag$, and these two couplings have the same cost since the projector onto the antisymmetric subspace commutes with $U^{\otimes2}$.
Moreover, the only coupling between the pure quantum states $\ket{\psi}$ and $\ket{\phi}$ is the product state $\Pi = \ket{\psi}\!\bra{\psi}\otimes\ket{\phi}\!\bra{\phi}$, whose cost is equal to $\left(1-|\braket{\phi|\psi}|^2\right)/2$.
Therefore, the distance between pure quantum states is a function of their overlap.

Ref. \cite{de2019quantum} proposes another quantum Wasserstein distance of order 2 based on couplings, with the property that each quantum coupling is associated to a quantum channel.
The relation between quantum couplings and quantum channels in the framework of von Neumann algebras has been explored in \cite{duvenhage2018balance}.
The problem of defining a quantum earth mover's distance through quantum couplings has been explored in Ref. \cite{agredo2017quantum}.

The quantum Wasserstein distance between two quantum states can be defined as the classical Wasserstein distance between the probability distributions of the outcomes of an informationally complete measurement performed on the states, which is a measurement whose probability distribution completely determines the state.
This definition has been explored for Gaussian quantum systems with the heterodyne measurement in Refs. \cite{zyczkowski1998monge,zyczkowski2001monge,bengtsson2017geometry}.

\section{Toy Model Details}
\label{app:toy_model_details}
Our toy model (\autoref{sec:toy_model}) analyzes the learnability of the GHZ state when using a loss function either corresponding to fidelity (function of inner product between target GHZ state and generated state) or the quantum EM distance. For optimizing over the fidelity, we have a loss function, copied below, that is easily evaluated as a function of $\boldsymbol{\theta}$.

\begin{equation}
\begin{split}
    F & =  |\braket{GHZ_n| \psi(\boldsymbol{\theta})} |^2 = \biggl( \frac{\cos(\theta_1)}{\sqrt{2}} + \frac{\prod_{i=1}^n \sin(\theta_i)}{\sqrt{2}} \biggr)^2
\end{split}
\label{eq:toy_loss_function_appendix}
\end{equation}

Here, to perform optimization, we simply perform gradient based updates on the parameters $\theta_i$ in the equation above. For all experiments, the Adam optimizer was used to perform gradient updates with a learning rate of $0.2$ \cite{kingma2014adam}. For each simulation, 100000 steps of optimization were performed before stopping. Learning is considered successful if $1-F < 0.02$. 

In our toy model, to efficiently approximate the quantum EM distance, we construct a loss function $\tilde{D}_{EM} \approx D_{EM}\bigl(\ket{\psi(\boldsymbol{\theta})} \bra{\psi(\boldsymbol{\theta})}, \ket{GHZ_n} \bra{GHZ_n} \bigr)$ which takes the maximum over $O(n)$ expectations of Pauli operators. We first note that the state  $\ket{\psi(\boldsymbol{\theta})}$ is spanned by up to $n+1$ computational basis states.

\begin{equation}
\begin{split}
    \ket{\psi(\boldsymbol{\theta})} & = \cos \theta_1 \ket{0_n} + i \sin \theta_1 \cos \theta_2 \ket{1} \ket{0_{n-1}} - \sin \theta_1 \sin \theta_2 \cos \theta_3 \ket{1_2} \ket{0_{n-2}} + \dots \\
    & = \cos \theta_1  \ket{0_n}  + \sum_{k=1}^{n-1} i^k \biggl[ \prod_{j=1}^{k} \sin \theta_j \biggr] \cos \theta_{k+1} \ket{1_k} \ket{0_{n-k}}  + i^n \biggl[ \prod_{j=1}^{n} \sin \theta_j \biggr] \ket{1_n}
\end{split}
\end{equation}

We can write the state above in a vector of length $n+1$ only including the terms in the above span:
\begin{equation}
    \ket{\psi(\boldsymbol{\theta})} = \begin{bmatrix} \cos \theta_1  \\ i \sin \theta_1 \cos \theta_2 \\ \vdots \\ i^n  \prod_{j=1}^{n} \sin \theta_j  \end{bmatrix} ,
    \label{eq:psi_compact_form}
\end{equation}
where the above vector can be easily stored in the memory of a classical computer. 

To calculate $\tilde{D}_{EM}$, we first measure the expectation of $\ket{\psi(\boldsymbol{\theta})}$ with respect to the following $2n$ Pauli operators $P_i$:
\begin{equation}
    \begin{split}
        P_i \in \{ & \sigma_Z^{(1)}, \; \sigma_Z^{(2)}, \dots, \; \sigma_Z^{(n)}, \\ 
        & \sigma_Y^{(1)}, \; \sigma_X^{(1)} \otimes \sigma_X^{(2)}, \; \sigma_X^{(1)} \otimes \sigma_X^{(2)} \otimes \sigma_Y^{(3)} , \dots , \; \sigma_X^{(1)} \otimes \sigma_X^{(2)} \otimes \cdots \otimes \sigma_X^{(n)} \} ,
    \end{split}
\end{equation}
which is equivalent to the complete set of single qubit Pauli Z operators combined with a multi-qubit Pauli operator for each qubit $k$ consisting of the Pauli X operator or Pauli Y operator acting on qubit $k$ if $k$ is even or odd respectively (to handle relative phases) and Pauli X operators acting on all qubits $j<k$. In the above, we use the notation $\sigma_L^{(i)}$ to indicate Pauli $L \in \{X,Y,Z\}$ acting on qubit $i$.

Since as mentioned earlier, $\ket{\psi(\boldsymbol{\theta})}$ is written compactly in vector form, expectations for each of the above operators can be efficiently evaluated using a classical computer. As discussed in \autoref{sec:EM_evaluation}, an optimal Hamiltonian whose expectation approximates the quantum EM distance can be efficiently constructed as a parameterized sum of the above operators. Since all Pauli Z operators act on individual qubits, $\tilde{D}_{EM}$ can be calculated as the maximum amongst the following $n+1$ parameterized sums of expectations of operators:
\begin{equation}
    \begin{split}
         \tilde{D}_{EM} = \max \biggl\{ & \bigl| \mathbb{E}[\sigma_Z^{(1)}]\bigr| + \bigl| \mathbb{E}[ \sigma_Z^{(2)}]\bigr| + \cdots + \bigl|\mathbb{E}[ \sigma_Z^{(n)}]\bigr|, \\ 
        & \bigl|\mathbb{E}[\sigma_Y^{(1)}]\bigr| + \bigl| \mathbb{E}[ \sigma_Z^{(2)}]\bigr| + \cdots + \bigl|\mathbb{E}[ \sigma_Z^{(n)}]\bigr|, \\
        & \bigl|\mathbb{E}[\sigma_X^{(1)} \otimes \sigma_Y^{(2)}] \bigr| + \bigl| \mathbb{E}[ \sigma_Z^{(3)}]\bigr| + \cdots + \bigl|\mathbb{E}[ \sigma_Z^{(n)}]\bigr|, \\
        & \dots, \\
        &  \bigl|\mathbb{E}[\sigma_X^{(1)} \otimes \sigma_X^{(2)} \otimes \cdots \otimes \sigma_X^{n}] \bigr| \biggr\} ,
    \end{split}
\end{equation}
where $\mathbb{E}[\cdot]$ indicates the difference in expectation of the operator $\cdot$ on the generated state versus the target GHZ state. For faster simulation, we actually consider the maximum over a simpler set of operators that is equally effective at learning the GHZ state:
\begin{equation}
    \begin{split}
         \tilde{D}_{EM} = \max \biggl\{ & \bigl| \mathbb{E}[\sigma_Z^{(1)}]\bigr| + \bigl| \mathbb{E}[ \sigma_Z^{(2)}]\bigr| + \cdots + \bigl|\mathbb{E}[ \sigma_Z^{(n)}]\bigr|, \\ 
        & \bigl|\mathbb{E}[\sigma_Y^{(1)}]\bigr| , \\
        & \bigl|\mathbb{E}[\sigma_X^{(1)} \otimes \sigma_Y^{(2)}] \bigr| , \\
        & \dots, \\
        &  \bigl|\mathbb{E}[\sigma_X^{(1)} \otimes \sigma_X^{(2)} \otimes \cdots \otimes \sigma_X^{n}] \bigr| \biggr\} .  
    \end{split}
\end{equation}

Using the equation for $\tilde{D}_{EM}$ above, gradient updates can efficiently be performed on the parameters of the circuit. As with the fidelity loss function, we perform optimization with the Adam optimizer at a learning rate of $0.2$ \cite{kingma2014adam}. Only up to 10000 steps of optimization were performed since convergence was almost always achieved within about 1000 steps. Learning is considered successful if $| \braket{GHZ_n | \psi(\boldsymbol{\theta})} |^2 > 0.98 $ after the optimization. Success was achieved in virtually all instances when using $\tilde{D}_{EM}$.

\section{Proof of \autoref{prop:GHZ}}\label{app:GHZ}
For any $k=0,\,\ldots,\,n-1$, let
\begin{equation}
\Delta_k = |\Psi_k\rangle\langle \Psi_k| - |GHZ_n\rangle\langle GHZ_n|\,,
\end{equation}
and let $\mathcal{D}_1$ be the completely dephasing channel acting on the first qubit.
From \cite[Proposition 3]{de2020quantum}, the quantum EM distance is contractive with respect to a quantum channel acting on a single qubit.
We then have on the one hand
\begin{equation}
\left\|\Delta_k\right\|_{EM} \ge \left\|\mathcal{D}_1(\Delta_k)\right\|_{EM} = \left\||1_k\rangle\langle1_k|\otimes\frac{|0_{n-k}\rangle\langle0_{n-k}| - |1_{n-k}\rangle\langle1_{n-k}|}{2}\right\|_{EM} = \frac{n-k}{2}\,.
\end{equation}
On the other hand, we have
\begin{align}
\left\|\Delta_k\right\|_{EM} &\le \left\|\mathcal{D}_1(\Delta_k)\right\|_{EM} + \left\|\Delta_k - \mathcal{D}_1(\Delta_k)\right\|_{EM} = \frac{n-k}{2} + \frac{1}{2}\left\|\Delta_k - \mathcal{D}_1(\Delta_k)\right\|_1\nonumber\\
&= \frac{n-k}{2} + \frac{1}{2}\left\{
                     \begin{array}{cl}
                       1& \qquad k=0 \\
                       \sqrt{2}& \qquad k=1,\,\ldots,\,n-1 \\
                     \end{array}
                   \right.\,,
\end{align}
where the first equality follows from \cite[Proposition 2]{de2020quantum}, stating that $\|X\|_{EM} = \|X\|_1/2$ for any $X\in\mathcal{O}_n$ with $\mathrm{Tr}_1X=0$.
The claim follows.

\section{Bias towards local operators}\label{app:bias}
Consider the maximization problem \eqref{eq:linear_program} copied below:

\begin{equation*}
\begin{array}{ll@{}ll}
\text{maximize}  & \displaystyle\sum\limits_{j=1}^{|W|} c_{j}w_{j} &\\
\text{subject to}& \displaystyle\sum\limits_{j:i \in \mathcal{I}_j}   |w_{j}| \leq 1,  & \qquad i=1 ,..., n
\end{array}
\end{equation*}

Here we prove that the optimal Hamiltonian for the maximization problem contains only terms with few qubits whenever all the coefficients $c_j$ associated to single Pauli operators are $\Omega(1)$.

\begin{prop}
Let $w^*:\{I,X,Y,Z\}^n\to\mathbb{R}$ be the set of parameters that achieve the maximum in \eqref{eq:linear_program}, and let
\begin{equation}
a = \min_{i=1,\,\ldots,\,n}\max_{P=X,Y,Z}\left|\mathrm{Tr}\left[\left(G -\rho_{\mathrm{tar}}\right)\sigma_P^{(i)}\right]\right|\,.
\end{equation}
Then, $w^*_{P_1\ldots P_n} = 0$ for any $P_1,\,\ldots,\,P_n\in\{I,X,Y,Z\}$ such that
\begin{equation}\label{eq:hypca}
|c_{P_1\ldots P_n}| < a\left|\left\{i=1,\,\ldots,\,n:P_i\neq I\right\}\right|\,.
\end{equation}
In particular, $w^*_{P_1\ldots P_n} = 0$ for any Pauli string that acts nontrivially on more than $2/a$ qubits.
\begin{proof}
The maximization problem \eqref{eq:linear_program} is a linear program with dual
\begin{equation}\label{eq:min}
\min_{z\in\mathbb{R}_{\ge0}^n}\sum_{i=1}^n z_i\quad  : \quad |c_{P_1\ldots P_n}| \le \sum_{i\in[n]:P_i\neq I} z_i \quad \forall\,P_1,\,\ldots,\,P_n\in\{I,X,Y,Z\}\,.
\end{equation}
Let $z^*\in\mathbb{R}^n_{\ge0}$ achieve the minimum in \eqref{eq:min}.
For any $P_1,\,\ldots,\,P_n\in\{I,X,Y,Z\}^n$ such that $w^*_{P_1\ldots P_n}\neq0$ we have
\begin{equation}\label{eq:cz*}
|c_{P_1\ldots P_n}| = \sum_{i\in[n]:P_i\neq I} z_i^*\,.
\end{equation}
From \eqref{eq:min} we have $a \le z_i^*$ for any $i=1,\,\ldots,\,n$.
Let $P_1,\,\ldots,\,P_n\in\{I,X,Y,Z\}$ satisfy \eqref{eq:hypca}, and let us assume that $w^*_{P_1\ldots P_n}\neq0$.
We get from \eqref{eq:cz*}
\begin{equation}
|c_{P_1\ldots P_n}| = \sum_{i\in[n]:P_i\neq I} z_i^* \ge a\left|\left\{i\in[n]:P_i\neq I\right\}\right|\,,
\end{equation}
which contradicts \eqref{eq:hypca}, and the claim follows.
\end{proof}
\end{prop}

\section{Supplementary details of estimated EM distance}
\label{app:correlations}

\paragraph{Linear relaxation example over product states}
To help aid intuition for the linear relaxation, consider the simple setting of estimating the quantum EM distance between two $n$-qubit product states $\ket{\psi_1}$ and $\ket{\psi_2}$ using only single qubit Pauli terms in the linear relaxation of \eqref{eq:linear_program} copied below:
\begin{equation}
\begin{array}{ll@{}ll}
\text{maximize}  & \displaystyle\sum\limits_{j=1}^{|W|} c_{j}w_{j} &\\
\text{subject to}& \displaystyle\sum\limits_{j:i \in \mathcal{I}_j}   |w_{j}| \leq 1,  & \qquad i=1 ,..., n
\end{array}
\end{equation}
where $c_{j}= \bra{\psi_1}P_j\ket{\psi_1} - \bra{\psi_2}P_j\ket{\psi_2} $ and $P_j$ is one of the single qubit Pauli terms that we use in estimating the EM distance. Solving the above outputs an optimal Hamiltonian that takes the form
\begin{equation}
    H_{\max} = \sum_{i=0}^{n_{\mathrm{active}}} w_i^* H_i^* ,
\end{equation}
where $w_{i}^*$ and $H_i^*$ are the weights and active operators respectively.

In this setting, for each qubit $i$, the linear program above will set $w_j^*=\operatorname{sign}(c_j)$ for the $c_j$ with largest magnitude in the set $\{j:i \in \mathcal{I}_j\}$ and $w_j^*=0$ for all other elements in the set. Thus, the linear program will select the Pauli terms that makes the largest contribution in the difference in expectations between the two states. For example, if $\ket{\psi_1} = \ket{+}\ket{+}$ and $\ket{\psi_2}=\ket{-}\ket{-}$, where $\ket{+}$ and $\ket{-}$ are the $+1$ and $-1$ eigenvectors of the $X$ basis respectively, then the optimal Hamiltonian unsurprisingly equals $H_{\max} = X \otimes I + I \otimes X$ since these are the single qubit Pauli terms that differ between the two states. Note, that $H_{\max} = X \otimes I + I \otimes X$ also has a Lipschitz constant of one since any single qubit change can only change the value of the Hamiltonian by at most one.

As an aside, adding higher order Pauli terms in the above linear program makes no difference in the outcome. Note, that for any higher order Pauli term, the magnitude of the difference between expectations of the two states $\ket{\psi_1}$ and $\ket{\psi_2}$ will only be captured by differences in their individual qubits. Since the difference in expectation for this higher order Pauli term cannot exceed that of the single qubit Pauli terms within it, these terms will never appear in the optimal Hamiltonian calculated above (see \autoref{app:bias} for further details).

\paragraph{Correlation of estimated and actual quantum EM distance}
Exact calculations of the quantum EM distance would require computational resources that grow exponentially with the number of qubits. However, as discussed in the main text, efficient estimates that lower bound the quantum EM distance can be obtained by formulating a new metric $D_{EM}^{(k)}$ optimizing over Hamiltonians of local operators (also see prior section for further motivation for this formalism). This estimated distance can be efficiently calculated as a linear program that requires computational resources that grow polynomially with the number of qubits. 

\begin{figure}[ht]
    \centering
    \begin{subfigure}[]{0.5\textwidth}
         \centering
         \includegraphics[]{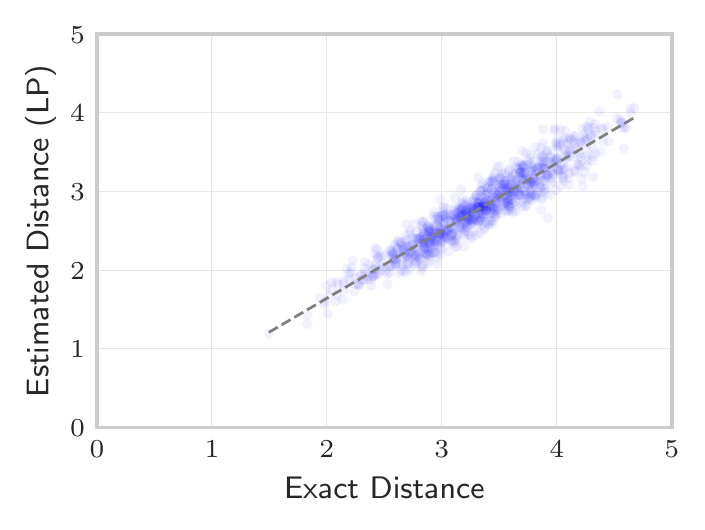}
         \caption{}
         \label{fig:product_correlation_exact_estimate}
    \end{subfigure}
    \hfill
    \begin{subfigure}[]{0.49\textwidth}
         \centering
         \includegraphics[]{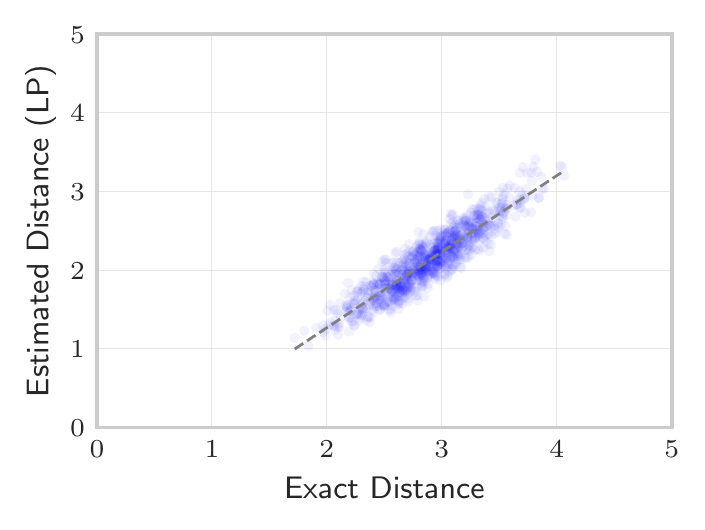}
         \caption{}
         \label{fig:bp_correlation_exact_estimate}
     \end{subfigure}
     \caption{Randomly drawn values of the exact distance $D_{EM}$ and estimated distance $D_{EM}^{(2)}$ (calculated from linear program over 2-local Pauli operators) are highly correlated. Individual points are drawn from (a) random product states and (b) states drawn from randomly parameterizing three layers of the barren plateau mixing circuit (\autoref{fig:mixing_circuit_appendix}). All simulations are for a system of five qubits where exact calculations of the quantum EM distance can be performed efficiently on a classical computer. Each plot contains 1000 randomly drawn data points. A linear fit is plotted over the data as a dotted line. }
     \label{fig:correlations}
\end{figure}

\autoref{fig:correlations} shows that the exact distance $D_{EM}$ and estimated distance $D_{EM}^{(2)}$ are well correlated for a system of five qubits where calculation of exact distances is possible on a classical computer. In \autoref{fig:correlations}, exact and estimated distances are compared for random product states (\autoref{fig:product_correlation_exact_estimate}) and states drawn randomly from a depth 3 circuit (\autoref{fig:bp_correlation_exact_estimate}). Though correlations are strong for random product states (\autoref{fig:product_correlation_exact_estimate}), the estimated distance is not as strong of an approximator for the depth 3 circuit (\autoref{fig:bp_correlation_exact_estimate}) where the slope of the correlation is slightly below one. This situation is one where many of the qubits have interacted with each other and we do not expect learning with the estimated distance to perform well at all times since the approximation is clearly not optimal (\textit{i.e.,} higher order Pauli operators may be needed to approximate the distance better). Nevertheless, the results here lend further support to the proof in the prior section showing that the optimal Hamiltonian in the quantum EM distance is biased towards local operators.

\section{Gradients of qWGAN}\label{app:grad}
For the generic version of our generator (equation \eqref{eq:generic_generator}), optimization is performed over probability parameters $p_i$ and gate parameters in each unitary $U_i$.
The generator optimizes the parameters $\theta$ to minimize $\mathrm{Tr}\left[G(\theta)\,H\right]$, where $H$ is the Hamiltonian provided by the discriminator.
The following \autoref{prop:grad} proves that the gradient of $\mathrm{Tr}\left[G(\theta)\,H\right]$ coincides with the gradient of the EM distance between $G(\theta)$ and $\rho_{\mathrm{tar}}$ if $H$ is the optimal Hamiltonian that achieves the EM distance in \eqref{eq:qWass1}. Therefore, our learning algorithm decreases the EM distance between $G(\theta)$ and $\rho_{\mathrm{tar}}$.
The proof is in \autoref{sec:gradproof}.

\begin{prop}\label{prop:grad}
For any target quantum state $\sigma$, any parametric family of quantum states $\rho(t),\,0\le t\le T$ that is differentiable in $t=0$ and any $k=1,\,\ldots,\,n$,
\begin{align}\label{eq:dW1}
\left.\frac{\mathrm{d}}{\mathrm{d}t}D_{EM}(\rho(t),\sigma)\right|_{t=0} &= \max\left(\mathrm{Tr}\left[\rho'(0)\,H\right]:H\in\mathcal{O}_n\,,\;\left\|H\right\|_L\le1\,,\;\mathrm{Tr}\left[\left(\rho(0) - \sigma\right)H\right] = D_{EM}(\rho(0),\sigma)\right)\,,\nonumber\\
\left.\frac{\mathrm{d}}{\mathrm{d}t}D_{EM}^{(k)}(\rho(t),\sigma)\right|_{t=0} &= \max\left(\mathrm{Tr}\left[\rho'(0)\,H\right]:H\in\mathcal{O}_n^{(k)}\,,\;\left\|H\right\|_{\tilde{L}}\le1\,,\;\mathrm{Tr}\left[\left(\rho(0) - \sigma\right)H\right] = D_{EM}^{(k)}(\rho(0),\sigma)\right)\,.
\end{align}
If $\rho(t)$ admits a differentiable extension to negative values of $t$, \eqref{eq:dW1} provides the right derivative of $D_{EM}(\rho(t) - \sigma)$, which can be different from the left derivative if the $\max$ in \eqref{eq:dW1} is nontrivial.
\end{prop}

For parameters $p_i$, the gradient of $\mathrm{Tr}\left[G(\theta)\,H\right]$ can be evaluated using $U_i$:
\begin{equation}
     \frac{\partial D_{EM}}{\partial p_i} = \Tr (U_i \rho_0 U_i^\dagger H_{\max} ) \,,
\end{equation}
where $H_{\max}$ is the optimal Hamiltonian outputted by the discriminator (equation \eqref{eq:discriminator_max_hamiltonian}). Note, that the above is simply the average measured value of $H_{\max}$ for the quantum state $U_i \rho_0 U_i^\dagger$.
For gate parameters, we can use standard techniques \cite{schuld2019evaluating} for evaluating gradients with respect to gate parameters.

\subsection{Proof of \autoref{prop:grad}}\label{sec:gradproof}
We prove the claim for the exact quantum EM distance.
The proof for the approximated quantum EM distance is completely analogous.

On the one hand, we have for any $H$ as in \eqref{eq:dW1}
\begin{equation}
\liminf_{t\to0^+}\frac{\left\|\rho(t) - \sigma\right\|_{EM} - \left\|\rho(0) - \sigma\right\|_{EM}}{t} \ge \liminf_{t\to0^+}\mathrm{Tr}\left[\frac{\rho(t) - \rho(0)}{t}\,H\right] = \mathrm{Tr}\left[\rho'(0)\,H\right]\,.
\end{equation}
On the other hand, for any $0<t<T$, let $H(t)\in\mathcal{O}_n$ be traceless and such that $\left\|H(t)\right\|_L\le1$ and $\mathrm{Tr}\left[\left(\rho(t) - \sigma\right)H(t)\right] = \left\|\rho(t) - \sigma\right\|_{EM}$.
We have
\begin{equation}\label{eq:supder}
\limsup_{t\to0^+}\frac{\left\|\rho(t) - \sigma\right\|_{EM} - \left\|\rho(0) - \sigma\right\|_{EM}}{t} \le \limsup_{t\to0^+}\mathrm{Tr}\left[\frac{\rho(t) - \rho(0)}{t}\,H(t)\right]\,.
\end{equation}
Let $t_k\downarrow0$ be a sequence that achieves the $\limsup$ in the right-hand side of \eqref{eq:supder} and such that
\begin{equation}
\lim_{k\to\infty}H(t_k) = H_0 \in \mathcal{O}_n\,.
\end{equation}
We have
\begin{align}
\left\|H_0\right\|_L &= \lim_{k\to\infty}\left\|H(t_k)\right\|_L \le 1\,,\nonumber\\
\mathrm{Tr}\left[\left(\rho(0) - \sigma\right)H_0\right] &= \lim_{k\to\infty}\mathrm{Tr}\left[\left(\rho(t_k) - \sigma\right)H(t_k)\right] = \lim_{k\to\infty}\left\|\rho(t_k) - \sigma\right\|_{EM} = \left\|\rho(0) - \sigma\right\|_{EM}\,,
\end{align}
and
\begin{equation}
\limsup_{t\to0^+}\frac{\left\|\rho(t) - \sigma\right\|_{EM} - \left\|\rho(0) - \sigma\right\|_{EM}}{t} \le \lim_{k\to\infty}\mathrm{Tr}\left[\frac{\rho(t_k) - \rho(0)}{t_k}\,H(t_k)\right] = \mathrm{Tr}\left[\rho'(0)\,H_0\right]\,,
\end{equation}
and the claim follows.

\section{Additional Simulations and Figures}
\label{app:other_simulations}

\subsection{Learning the GHZ state}

\begin{figure}[ht]
    \centering
    \includegraphics[center]{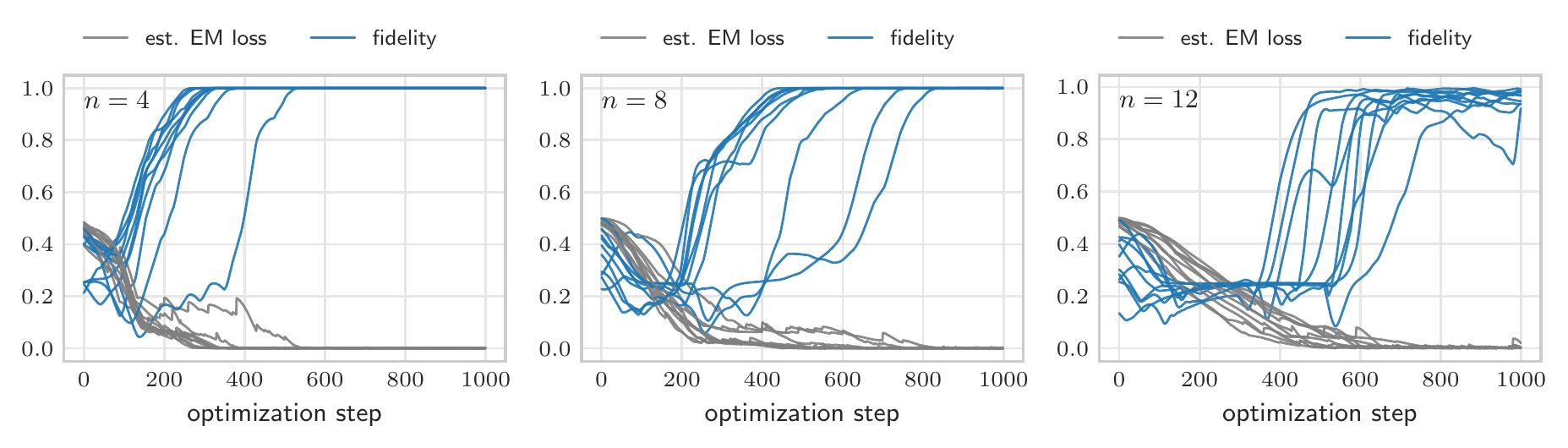}
    \caption{The qWGAN consistently generates the GHZ state in simulations with circuits of 4, 8, and 12 qubits. Estimated $EM$ loss (quantum EM distance estimated by active operators) is also plotted in above chart, normalized by dividing by the number of qubits. Plots contain one line for each of 10 simulations for each circuit size.}
    \label{fig:app_GHZ_loss_plot}
\end{figure}

The analysis in \autoref{subsec:GHZ_simulations} showed that the qWGAN is especially effective at learning the GHZ state. In addition to the results shown in \autoref{subsec:GHZ_simulations}, \autoref{fig:app_GHZ_loss_plot} shows the typical dynamics of learning the GHZ state of 4, 8, and 12 qubits. In all cases, the GHZ state is learned within 1000 steps of optimization.

\subsection{Teacher-student learning}
\begin{figure}[ht]
    \centering
    \includegraphics[center]{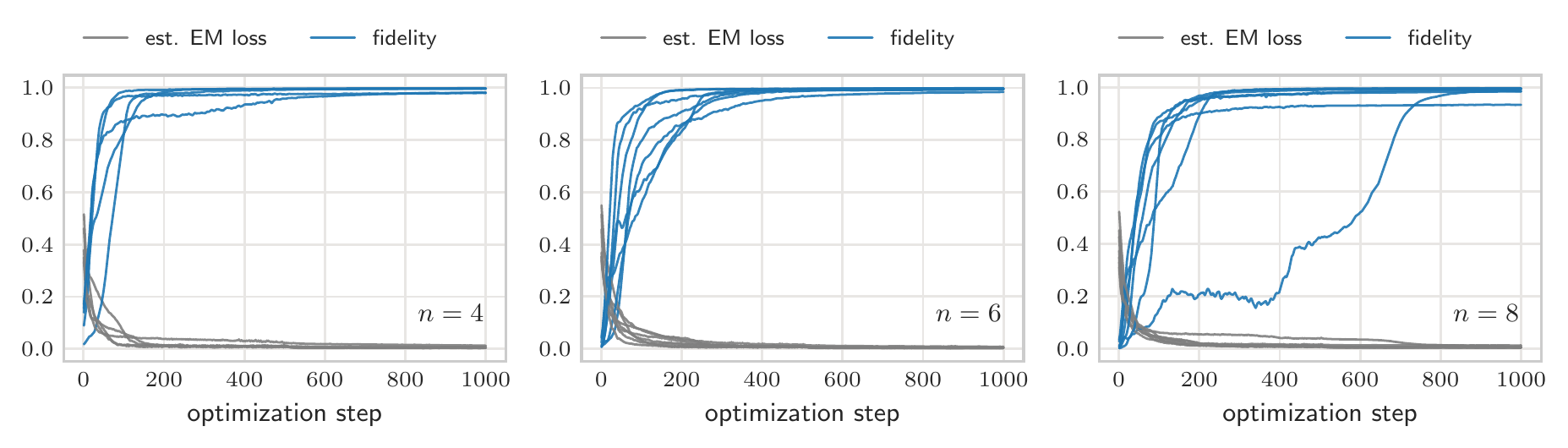}
    \caption{The student circuit is able to approximate well the state generated by the teacher circuit. Here, the target is constructed by randomly setting the parameters of a depth $2$ mixing circuit (teacher circuit). The qWGAN, equipped with a generator circuit of depth $4$, successfully learns the target state generated by the teacher circuit. For each plot, $5$ simulations are performed. }
    \label{fig:app_mixing_loss_plot}
\end{figure}

Supplementary to the results in \autoref{subsec:teach_student}, we include \autoref{fig:app_mixing_loss_plot} which shows the typical profile of learning in the teacher-student setup. In almost all instances, learning of the state generated by the teacher circuit was achieved.

\subsection{Gradients of qWGAN vs. conventional GANs}
\begin{figure}[ht]
    \centering
    \includegraphics[center]{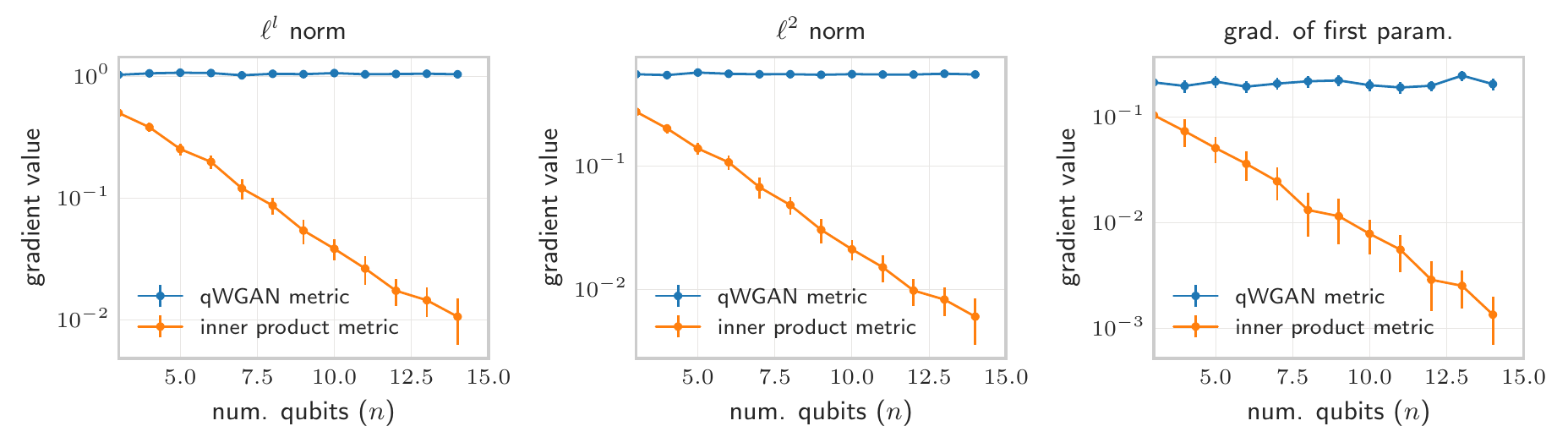}
    \caption{Comparison of gradients between the quantum EM loss metric and a conventional loss metric that is a function of the inner product. Here the average L1 norm (left), L2 norm (center), and the  absolute value of the gradient of the first parameter (right) are shown. Decaying gradients observed in the inner product loss metric. In contrast, regardless of the number of qubits, the gradients of the qWGAN remain stable. Gradients are calculated at first step of optimization. $L_1$ norm and $L_2$ norm are divided by $n$ and $\sqrt{n}$ respectively to normalize based on the number of parameters in the circuit. Results are averaged across 100 simulations for each data point }
    \label{fig:app_all_gradient_norms}
\end{figure}

Supplementary to \autoref{fig:mixing_gradients}, we include further details of the gradients of the quantum EM loss metric and its comparison to a inner product loss metric in \autoref{fig:app_all_gradient_norms}. As a reminder, the inner product loss metric is $F = 1- \left| \braket{ \phi_{\mathrm{tar}} | \phi(\boldsymbol{\theta}) } \right|^2$.

\subsection{Butterfly circuit learning}
\label{subsec:butterfly_simulations}

In this section, we consider learning the parameters of a  ``butterfly" circuit which constructs interactions between all qubits in $O(\log_2 n)$ layers. The general form of this circuit is shown in \autoref{app:circuits} and is motivated by prior work in classical machine learning and photonics where similar parameterizations of unitary transformations produced interesting results \cite{mathieu2014fast,jing2017tunable,dao2019learning,clements2016optimal,shen2017deep}. Here, the generator takes the form of $r_{\mathrm{gen}}$ copies of the parameterized butterfly circuit. The generator aims to learn a target density matrix $\rho_{\mathrm{tar}}$ of rank $r_{\mathrm{tar}}$ which is generated from a circuit of the same form as the generator but with randomly chosen parameters. In other words,

\begin{equation}
    \rho_{\mathrm{tar}} = \frac{1}{r_{\mathrm{tar}}} \sum_{i=1}^{r_{\mathrm{tar}}} U_{b}(\theta_{\mathrm{ran}}^{(i)}) \rho_0 U_{b}(\theta_{\mathrm{ran}}^{(i)})^\dagger
\end{equation}
where $U_{b}(\theta_{\mathrm{ran}}^{(i)})$ is the unitary transformation associated to the butterfly circuit with parameters $\theta_{\mathrm{ran}}^{(i)}$ chosen randomly (we choose each parameter uniformly from $[0,2\pi)$).

\begin{figure}[ht]
    \centering
    \includegraphics[center]{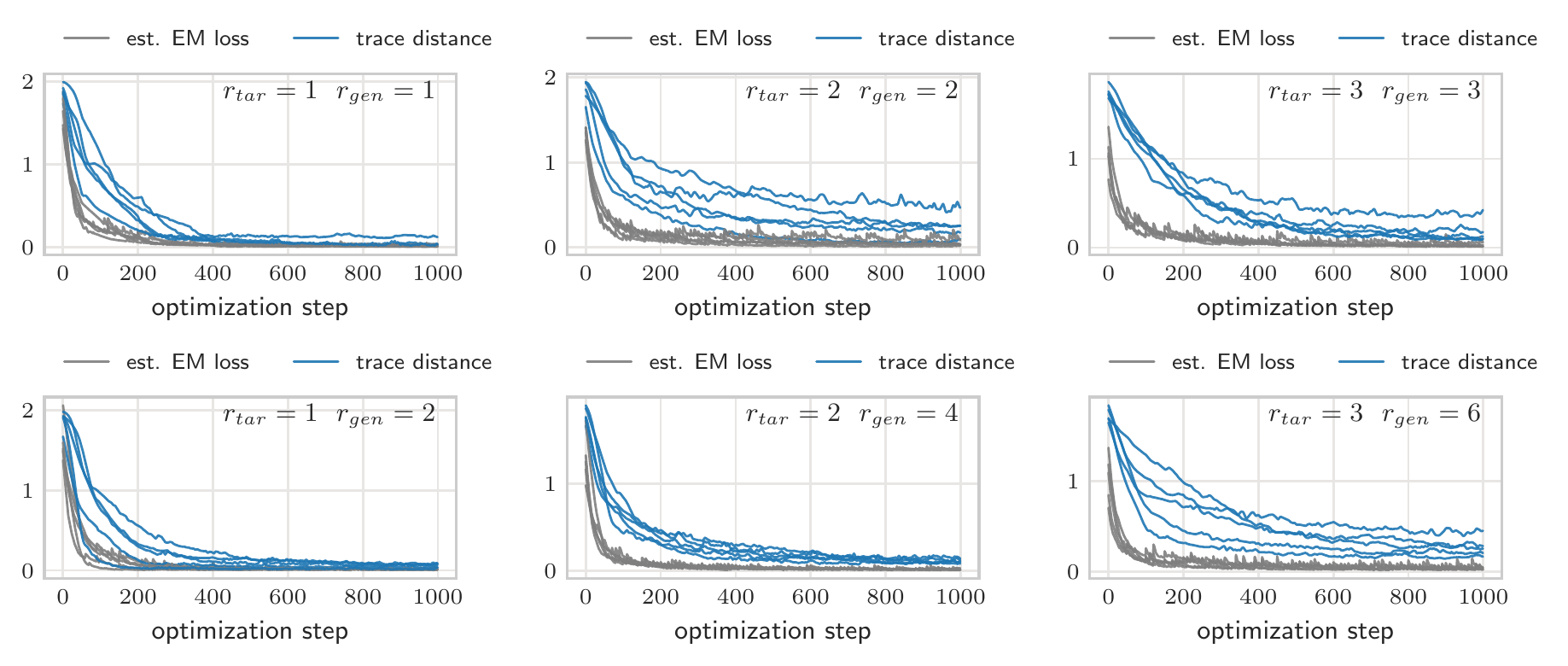}
    \caption{The qWGAN is also able to generate mixed states that approximate well their given target (also a mixed state) in both trace distance and quantum EM distance. Here, the generator circuit takes the form of a butterfly circuit of $4$ qubits (see \autoref{app:circuits}), and the target is constructed by randomly setting the parameters of the generator circuit. The qWGAN aims to learn the target density matrix of rank $r_{\mathrm{tar}}$ with either $r_{\mathrm{gen}}=r_{\mathrm{tar}}$ or $r_{\mathrm{gen}}=2r_{\mathrm{tar}}$ parameterized circuits of the same form. For each plot, $5$ simulations are performed. }
    \label{fig:butterfly_loss_plot}
\end{figure}

\autoref{fig:butterfly_loss_plot} shows that the qWGAN is effective at learning mixed states of $4$ qubits, though learning is clearly more challenging as the rank of the target density matrix increases. We recognize that the form of the generator \eqref{eq:generic_generator} may not be well suited to optimization over mixed states. For example, it is often the case that different circuits in the generator optimize to the same critical point in the loss landscape, thus outputting the same state. Future improvements to the design of generators can improve the results shown here.

\subsection{QAOA learning}
\label{subsec:QAOA_simulations}
The quantum approximate optimization algorithm and its related extension to the quantum alternating operator ansatz, both given the acronym QAOA, are promising candidates for achieving quantum speedups in classical optimization problems \cite{fingerhuth2018quantum,hadfield2019quantum,farhi2014quantum}. Recent work has shown that QAOA is computationally universal \cite{lloyd2018qaoa} and potentially an effective algorithm in a wide range of quantum machine learning settings \cite{kiani2020learning,verdon2017quantum,wang2018quantum,zhang2020qed,hodson2019portfolio,chancellor2019domain}. 

Here, we use a QAOA circuit as the generator for our qWGAN to learn the ground state of a simple translationally invariant Ising Hamiltonian cost function $C$:
\begin{equation}
    C = B \sum_{i=1}^N \sigma_Z^{(i)} \sigma_Z^{(i+1)},
\end{equation}
where $B$ is a constant assumed to be positive and $\sigma_Z^{(i)}$ is the Pauli Z operator acting on qubit $i$. Given the simple translationally invariant form of $C$, its ground state is spanned by the states $\ket{01}^{\otimes \frac{1}{2}n}$ and $\ket{10}^{\otimes \frac{1}{2}n}$. Note that this setting is different from more traditional QAOA settings since here we do not aim to find the ground state but instead are given the ground state and aim to construct that state from a parameterized circuit.

For our experiments, we attempt to learn the ground state of $C$: $\frac{1}{\sqrt{2}} \bigl(\ket{01}^{\otimes \frac{1}{2}n} + \ket{10}^{\otimes \frac{1}{2}n} \bigr)$. We use a QAOA circuit which applies, repeating for a depth of $L$ times, a mixing Hamiltonian $e^{-i \alpha_l H_{\mathrm{mix}}}$ and the cost Hamiltonian $e^{-i \beta_l H_C}$ where $l \in \{1, \dots, L\}$ indicates the layer of the QAOA circuit. In total, the circuit has $2L$ trainable parameters $\alpha_l$ and $\beta_l$ (see \autoref{app:circuits} for details of circuit).

\begin{equation}
    \begin{split}
        H_{\mathrm{mix}} = \sum_{i=1}^N \sigma_X^{(i)} \; \; \; \; \; \; \; \; \;
        H_{C} = \sum_{i=1}^N \sigma_Z^{(i)} \sigma_Z^{(i+1)}
    \end{split}
\end{equation}

\begin{figure}[ht]
    \centering
    \includegraphics[center]{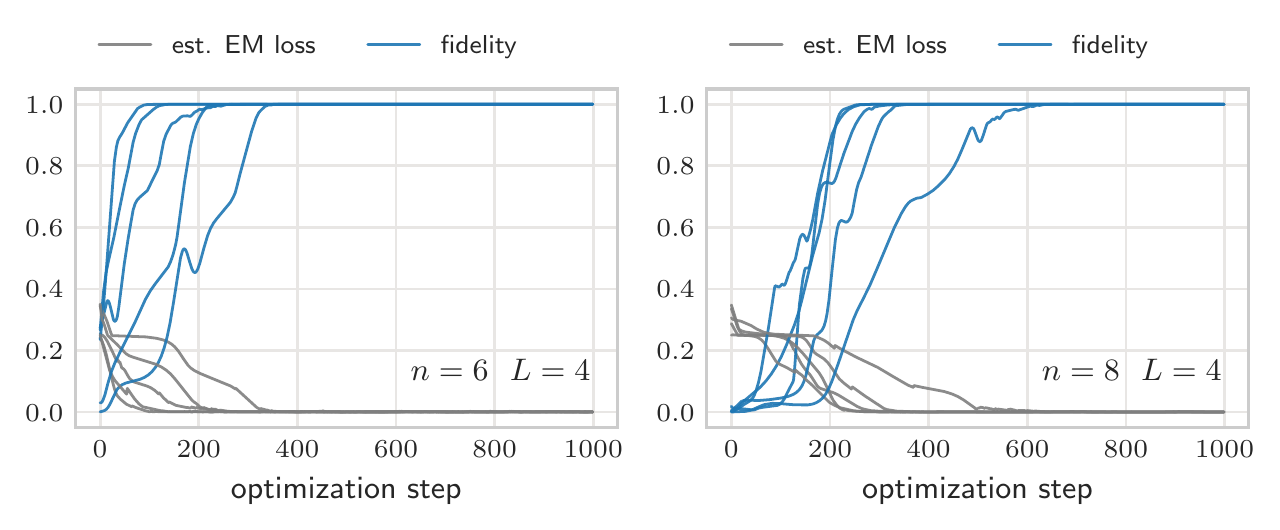}
    \caption{The qWGAN is effective at learning the ground state of a translationally invariant Ising Hamiltonian. Here, the generator is a QAOA circuit (see \autoref{app:circuits}) of depth $L=4$. Estimated $W_1$ loss (quantum EM distance estimated by active operators) is also plotted in above chart, normalized by dividing by the number of qubits. For each plot, $5$ simulations are performed. }
    \label{fig:qaoa}
\end{figure}

\autoref{fig:qaoa} shows that our qWGAN is very effective at learning the ground state using the QAOA circuit as the generator. Convergence to the ground state is achieved within a few hundred steps of optimization.

\section{Circuits Used in Experiments}
\label{app:circuits}

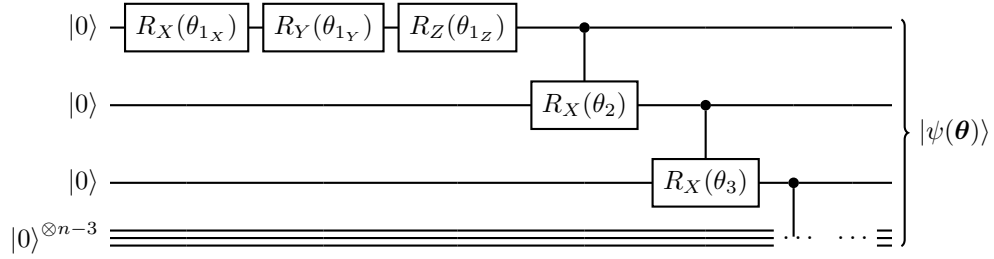
\begin{figure}
    \centering
        \begin{tikzpicture}
        \node[scale=1.] {
        \begin{quantikz}[row sep=0.4cm, column sep=0.2cm]
            \lstick{$\ket{0}$} & \gate{R_X(\theta_{1_X})} & \gate{R_Y(\theta_{1_Y})} & \gate{R_Z(\theta_{1_Z})}   & \ctrl{1} & \qw & \qw & \qw & \qw \rstick[wires = 4]{$\ket{\psi(\boldsymbol{\theta})}$}  \\     
            \lstick{$\ket{0}$} & \qw & \qw & \qw & \gate{R_X(\theta_2)} & \ctrl{1} & \qw & \qw & \qw  \\
            \lstick{$\ket{0}$} & \qw & \qw & \qw & \qw & \gate{R_X(\theta_3)} & \ctrl{1} & \qw & \qw  \\
            \lstick{$\ket{0}^{\otimes n-3}$} & \qwbundle[alternate]{} & \qwbundle[alternate]{} & \qwbundle[alternate]{} & \qwbundle[alternate]{} & \qwbundle[alternate]{} & \ \ldots \qwbundle[alternate]{} & \ \ldots \ & \qwbundle[alternate]{} 
        \end{quantikz}
        };
        \end{tikzpicture}
        \caption{Circuit for generator in GHZ simulations (\autoref{subsec:GHZ_simulations}).}
        \label{fig:GHZ_sim_circuit}
\end{figure}

\begin{figure}
    \centering
        \begin{tikzpicture}
        \node[scale=1.] {
        \begin{quantikz}[row sep=0.4cm, column sep=0.5cm]
            \lstick{$\ket{0}$} & \gate{R_Y(\theta_{1,1})} & \phase{\theta_{2,1}} & 
            \gate{R_Y(\theta_{3,1})} & \qw & \phase{\theta_{4,3}} & \qw \\     
            \lstick{$\ket{0}$} & \gate{R_Y(\theta_{1,2})} & \ctrl{-1} &
            \gate{R_Y(\theta_{3,2})} & \phase{\theta_{4,1}} & \qw & \qw  \\
            \lstick{$\ket{0}$} & \gate{R_Y(\theta_{1,3})} & \phase{\theta_{2,2}} &
            \gate{R_Y(\theta_{3,3})} &  \ctrl{-1} & \qw & \qw \\
            \lstick{$\ket{0}$} & \gate{R_Y(\theta_{1,4})} & \ctrl{-1} &
            \gate{R_Y(\theta_{3,4})} & \phase{\theta_{4,2}} & \qw & \qw \\
            \lstick{$\ket{0}$} & \gate{R_Y(\theta_{1,5})} & \phase{\theta_{2,3}} &
            \gate{R_Y(\theta_{3,5})} & \ctrl{-1} & \qw & \qw \\
            \lstick{$\ket{0}$} & \gate{R_Y(\theta_{1,6})} & \ctrl{-1} &  
            \gate{R_Y(\theta_{3,6})} & \qw & \ctrl{-5} & \qw
        \end{quantikz}
        };
        \end{tikzpicture}
        \caption{Single layer of mixing circuit used to perform learning and generate targets in \autoref{subsec:GHZ_simulations}. This circuit consists of alternating layers of parameterized Pauli Y rotations and parameterized Pauli Z-Z rotations. The circuit above may be repeated to construct deeper circuits for simulations.}
        \label{fig:mixing_circuit_appendix}
\end{figure}
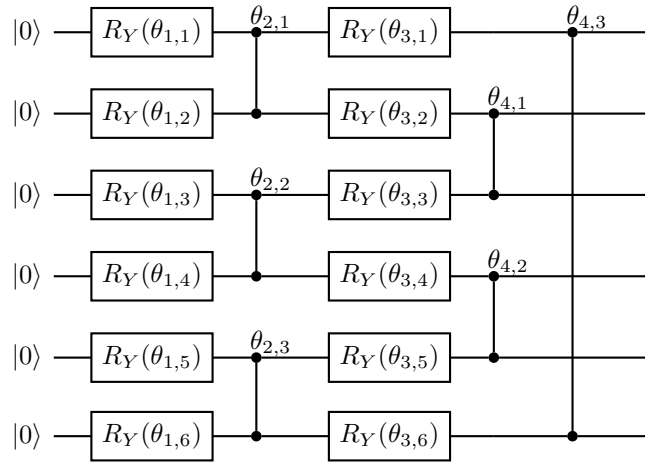

\begin{figure}
    \centering
        \begin{subfigure}[]{1\textwidth}
         \centering
         \begin{tikzpicture}
            \node[scale=1.] {
            \begin{quantikz}[row sep=0.4cm, column sep=0.2cm]
                \lstick{$\ket{0}$}
                & \gate{H} & \gate[3]{e^{-i \beta_1 H_{C}}} & \gate[3]{e^{-i \alpha_1 H_{\mathrm{mix}}}} & \gate[3]{e^{-i \beta_2 H_{C}}} & \gate[3]{e^{-i \alpha_2 H_{\mathrm{mix}}}} & \gate[3]{\cdots} & \gate[3]{e^{-i \beta_L H_{C}}} & \gate[3]{e^{-i \alpha_L H_{\mathrm{mix}}}} & \qw \\
                \lstick{$\ket{0}^{\otimes n-2}$} & \gate{H} \qwbundle[alternate]{} & & & & & & & & \qwbundle[alternate]{} \\
                \lstick{$\ket{0}$} & \gate{H} & & & & & & & & \qw
            \end{quantikz}
            };
        \end{tikzpicture}
         \caption{}
         \label{fig:qaoa_circuit_general}
    \end{subfigure}
    \hfill
    \begin{subfigure}[]{1.0\textwidth}
        \centering
        \begin{tikzpicture}
        \node[scale=1.] {
        \begin{quantikz}[row sep=0.4cm, column sep=0.2cm]
            \lstick{$\ket{0}$}  & \gate{H}             & \gate{R_Z(2\beta_1)}  & \qw                    & 
            \qw                 & \gate{R_Z(2\beta_1)} & \qw      & \qw                   & \gate{H} &
            \gate{R_Z(2\alpha_1)} & \gate{H} & \qw \\     
            \lstick{$\ket{0}$}  & \gate{H}             & \gate{R_Z(2\beta_1)}  & \gate{R_Z(2\beta_1)}   &  
            \qw                 & \qw                  & \gate{H} & \gate{R_Z(2\alpha_1)} & \gate{H} &
            \qw                   & \qw & \qw \\
            \lstick{$\ket{0}$}  & \gate{H}             & \qw                   & \gate{R_Z(2\beta_1)}   & 
            \gate{R_Z(2\beta_1)}& \qw                  & \gate{H} & \gate{R_Z(2\alpha_1)} & \gate{H} & 
            \qw                   & \qw & \qw \\
            \lstick{$\ket{0}$}  & \gate{H}             & \qw                   & \qw                    & 
            \gate{R_Z(2\beta_1)}& \gate{R_Z(2\beta_1)} & \qw      & \qw                   & \gate{H} &
            \gate{R_Z(2\alpha_1)} & \gate{H} & \qw
        \end{quantikz}
        };
        \end{tikzpicture}
        \caption{}
        \label{fig:qaoa_circuit_4_qubit}
    \end{subfigure}
     \caption{(a) General form for QAOA circuit. (b) QAOA circuit for $4$ qubits and $L=1$. }
\end{figure}
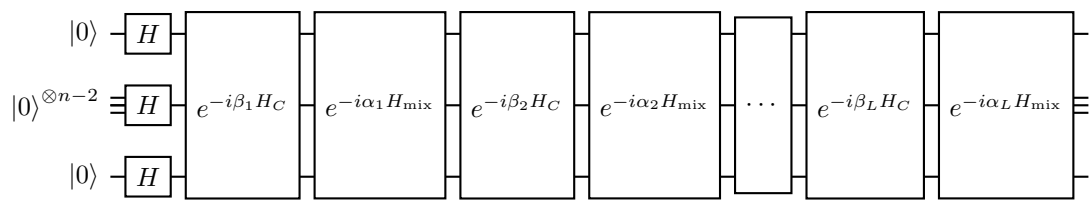
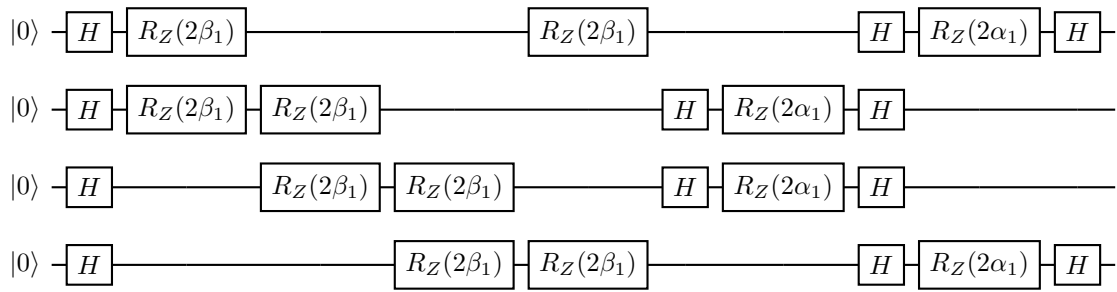

In all our experiments, the generators are parameterized circuits. The form of those circuits are listed below.
\begin{itemize}
    \item GHZ circuit (\autoref{subsec:GHZ_simulations}): circuit is shown in \autoref{fig:GHZ_sim_circuit}. This circuit differs from that used in the toy model (\autoref{fig:simple_circuit}) only in the first qubit. Here, three parameterized Pauli rotations are applied to the first qubit to allow for complete control over the relative phase of the first qubit.
    \item Mixing circuit (\autoref{subsec:teach_student}): circuit is shown in \autoref{fig:mixing_circuit_appendix}. This circuit is commonly used in prior literature to show the existence of barren plateaus in the loss landscape \cite{mcclean2018barren,cerezo2020cost}. This circuit contains alternating layers of parameterized Pauli Y rotations and pairwise Pauli Z-Z rotations.
    \item Butterfly circuit (\autoref{subsec:butterfly_simulations}): The butterfly circuit takes the form of alternating layers of single qubit Pauli $X$ rotations followed by controlled Pauli $X$ rotations applied in the order of the butterfly pattern (\autoref{fig:butterfly_pattern}). The form of the circuit for $4$ qubits shown in \autoref{fig:4_qubit_butterfly}.
    \item QAOA circuit (\autoref{subsec:QAOA_simulations}): general form of circuit is shown in \autoref{fig:qaoa_circuit_general} consisting of alternating applications of a mixing Hamiltonian $H_{\mathrm{mix}}$ and cost Hamiltonian $H_C$. An initial layer of Hadamard gates is also included. At a given layer $l$, Trotterized time evolution circuits are used to apply $H_{\mathrm{mix}}$ and $H_C$ for times $\alpha_l$ and $\beta_l$ respectively \cite{bergholm2018pennylane}. The form of the circuit for $4$ qubits and a single QAOA layer ($L=1$) is shown in \autoref{fig:qaoa_circuit_4_qubit}.
\end{itemize}

\begin{figure}
    \centering
        \begin{subfigure}[]{1\textwidth}
         \centering
         \includegraphics[]{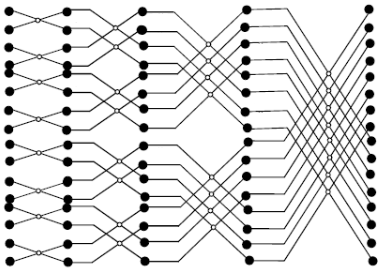}
         \caption{}
         \label{fig:butterfly_pattern}
    \end{subfigure}
    \hfill
    \begin{subfigure}[]{1.0\textwidth}
        \centering
        \begin{tikzpicture}
        \node[scale=1.] {
        \begin{quantikz}[row sep=0.4cm, column sep=0.2cm]
            \lstick{$\ket{0}$} & \gate{R_X(\theta_{1,1})} & \ctrl{1} & \gate{R_X(\theta_{3,1})} & \ctrl{2} & \qw  & \qw \rstick[wires = 4]{$\ket{\psi(\boldsymbol{\theta})}$}  \\     
            \lstick{$\ket{0}$} & \gate{R_X(\theta_{1,2})} & \gate{R_X(\theta_{2,1})} & \gate{R_X(\theta_{3,2})} &  \qw & \ctrl{2} & \qw  \\
            \lstick{$\ket{0}$} & \gate{R_X(\theta_{1,3})} & \ctrl{1} & \gate{R_X(\theta_{3,3})} & \gate{R_X(\theta_{4,1})}  & \qw & \qw   \\
            \lstick{$\ket{0}$} & \gate{R_X(\theta_{1,4})} & \gate{R_X(\theta_{2,2})} & \gate{R_X(\theta_{3,4})} & \qw & \gate{R_X(\theta_{4,2})} & \qw
        \end{quantikz}
        };
        \end{tikzpicture}
        \caption{}
        \label{fig:4_qubit_butterfly}
    \end{subfigure}
     \caption{(a) Butterfly pattern of interactions, here shown for a system of $16$ qubits. (b) Circuit for generator in butterfly circuit simulations (\autoref{subsec:butterfly_simulations}) here shown for $4$ qubits. }
\end{figure}
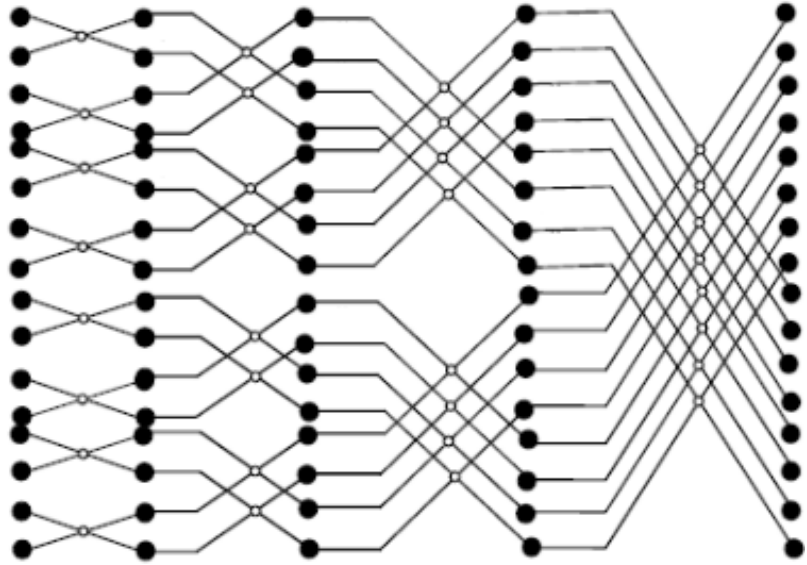
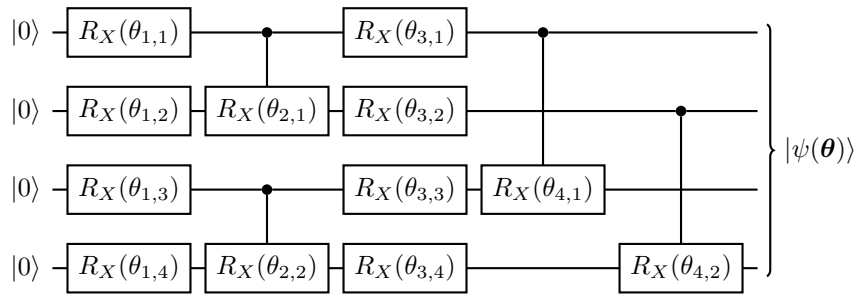

\section{Computational Details}
\label{app:computational_details}

All code used for this paper is available here: \url{https://github.com/bkiani/Quantum-EM-distance-and-qWGAN}

Quantum circuit simulations were performed using PennyLane \cite{bergholm2018pennylane} with a backend of Tensorflow \cite{tensorflow2015-whitepaper} or Pytorch \cite{paszke2019pytorch}. Unless specified otherwise, the Adam optimizer is used for performing gradient-based updates on a generator \cite{kingma2014adam}. The default Adam optimizer was set to a learning rate of $0.01$. In some cases, learning was performed in two phases, first with a learning rate of $0.02$ decreased to $0.007$ for a second phase.

All parameters of the generator are initialized according to a standard normal distribution unless otherwise stated. In its default setting, we cycle the operators of the discriminator every ten optimization steps. When operators are cycled, a cycling threshold of $P=0.8$ is used (see \autoref{par:cycling}). Discriminators are initialized with the set of $2$-local Pauli operators.

% \bibliographyappref{main.bib}
% \bibliographystyleappref{naturemag}

\end{appendices}

\end{document}